\documentclass[]{sapthesis} 

\usepackage{anysize}
\marginsize{3cm}{3cm}{1cm}{1cm}
\usepackage{amsmath}
\usepackage[english]{babel}
\usepackage{graphicx}
\usepackage{hyperref}
\usepackage{color}
\usepackage{longtable}
\usepackage{tabu}

\def\neigGroupI{j \in \partial a^R_{i}}
\def\neigOppositeI{j \in \partial a^R_{\neg i}}
\def\neigIn{j \in \partial a^R_{in}}
\def\neigOut{j \in \partial a^R_{out}}

\def\etheta{e^{\theta}}

\newcommand{\avg}[1]{\left\langle{#1}\right\rangle}

\newcommand{\bvec}[1]{\ensuremath{\boldsymbol{#1}}}

\newcommand{\Z}[2]{\ensuremath{Z^{#1\rightarrow #2}}}
\newcommand{\Zt}[2]{\ensuremath{\tilde{Z}^{#1\rightarrow #2}}}
\newcommand{\mess}[3]{\ensuremath{#1^{#2 \rightarrow #3}}}
\newcommand{\neigMeta}[1][e]{\ensuremath{\partial #1^M}}
\newcommand{\neigRea}[1][a]{\ensuremath{\partial #1^R}}

\newcommand{\inn}{{\rm in}}
\newcommand{\outt}{{\rm out}}

\newcommand{\cc}{\ensuremath{\lambda}}

\author{Alessandro Seganti} 
\title{Boolean constraint satisfaction problems for reaction networks}
\submitdate{October 2013} \cycle{XXV} \coordinator{Prof. Massimo Testa} \firstadvisor{Prof. Federico Ricci-Tersenghi}
\secondadvisor{Dr. Andrea De Martino}

\begin{document}

\frontmatter 
\maketitle 
\begin{abstract} 
In this Thesis we will present a work at the boundary between Physics and Biology. On the one hand we studied theoretically, on a newly defined class of random networks, a constraint satisfaction problem (CSP) devised to understand the capabilities of a metabolic network. On the other hand we showed that it is possible to apply the same techniques in real networks by obtaining preliminary results on the real metabolic network of E.Coli. We will thus show in this Thesis that it is possible to analyze biological problems on metabolic networks both theoretically and practically by computer simulations of a simplified model.
\end{abstract} 

\tableofcontents

\mainmatter 

\addcontentsline{toc}{chapter}{Introduction}

\chapter*{Introduction}
In this Thesis we will present a work at the boundary between Physics and Biology. On the one hand we studied theoretically, on a newly defined class of random networks, a constraint satisfaction problem (CSP) devised to understand the capabilities of a metabolic network. On the other hand we showed that it is possible to apply the same techniques in real networks by obtaining preliminary results on the real metabolic network of E.Coli. We will thus show in this Thesis that it is possible to analyze biological problems on metabolic networks both theoretically and practically by computer simulations of a simplified model.

Biological networks map out the complex set of interactions that may occur among different units (genes, proteins, signalling molecules, enzymes, etc.) in cells \cite{Barabasi_2004,Tkacik_Bialek_2007}. In recent years, the use of high-throughput sequencing and gene expression profiling techniques has allowed researchers to map the structure of different of these networks for several organisms (see \cite{Kanehisa2012kegg}). Their structure is thought to reflect, at least in part, the specific physiologic function(s) they are meant to control and many important questions can be formulated about the optimality, robustness and evolvability of these systems \cite{Barabasi,Palsson_Theo,Alon,Crombach_2008,Berkhout_2012,Wagner_2008,Wagner_2005,Ciliberti_2007}. Still the topology-to-function mapping remains unclear and it is not fully understood how to link the network's architecture to the biological function it is meant to carry out.

On the other hand,  functional control in cells is achieved through the physical dynamics that takes place {\it on} the networks, which is usually much more complicated than the network structure by itself would suggest. To make an example, consider transcriptional regulatory networks. While their structure only encodes for the possible protein-DNA interactions by which the transcription of RNA can be turned on or off, regulation results from the reciprocal adjustment of transcriptional activity and protein levels. This process however involves a variety of regulated steps, like DNA-binding and unbinding events by multiple proteins (possibly preceded by the formation of protein complexes), RNA polymerization (by specifically recruited molecular machinery) and transport, post-transcriptional modification events and, finally, translation. Each node in this network therefore lumps together a number of molecular species and elementary processes, all of which are subject to noise. In such a complex interacting environment, the overall patterns of activity may be hard to uncover even if one is only interested in steady states.

In Chapter \ref{chap:biological_back}, we will show how using the available data it is possible to develop models to simulate a biological system and predict its behavior. This has been possible by using an holistic approach to biological problem e.g. a \textit{Systems biology} point of view. In this approach, it is considered that it should be possible to understand the behavior of complex biological systems as a whole. These approaches have brought many insights on the functioning of biological systems (e.g. FBA for metabolism, see Section \ref{bio:FBA}) and are still one of the most active field of research in computational biology. Interestingly the paradigmatic shift introduced by these new approaches in biology has many similarities to the ideas at the foundation of Statistical Physics. It is thus not surprising that many techniques developed in this field are suitable to study complex biological problems. In Chapter \ref{sec:stat_mech} we will briefly present the most important for our studies and show by some examples what has been understood of the characteristics of complex random systems in the last 30 years. Hopefully we will show in this Chapter that these approaches are suitable to describe and understand systems with an exponential number of states.

In Chapter \ref{chap:CSP_for_FBA_VN} we will present the coarse-grained approach to characterize the operation of biological networks through elementary feasibility constraints that we developed. This kind of approach has been widely used \cite{Kauffman,Leone_Zecchina,Martin,MartinII,Francois,Samal} and is especially suited (a) to identify robust attractors of the dynamics and/or groups of nodes that are likely to behave in a highly correlated way (viz. the emergence of network motifs discussed in \cite{Martin}), and (b) to evaluate `degrees of activity' for the different nodes, by which one may, for example, guide more refined techniques that simulate the full dynamics of the system towards physiologically relevant states. Thus in Section \ref{sec:boolean_problem} we will present the rationale behind the use of boolean variables in the case of metabolic networks. We will then define a novel Boolean constraint-satisfaction problem designed to represent minimal operational and stability requirements for the non-equilibrium steady states (NESS) of biochemical reaction networks. In essence, we shall enforce feasibility constraints that link enzyme activity to substrate and product availability, and vice-versa, similarly to the approach defined in \cite{Ebenhoh,Handorf_2007,Ebenhoh_2004} to characterize the productive capabilities of a metabolic network. From a physical viewpoint, the model describes, in different limits, different types of NESS, and therefore different physiological scenarios. The corresponding CSPs, on the other hand, turn out to be of a novel type, requiring {\it ad hoc} message-passing methods to be analyzed in detail. Then in Sections \ref{sec:real_net} and \ref{sec:RRN} we will define the networks we have studied: the random reaction network (RRN) and the metabolic network of Escherichia Coli (E.Coli). The RRN is a network we defined in order to have a random network that could represent a real metabolic network on which to test the properties of the equations we developed. Furthermore, we have chosen the metabolic network because it is one of the best known system and in the specific, the metabolic network of E.Coli is known to an impressive precision.

Finally we will show in Chapter \ref{sec:random_case} the results of our analysis on RRNs, showing that the structure of the solution space is non trivial. Then in Chapter \ref{sec:real_case} we will show how to apply the equations on the real metabolic network of E.Coli showing what it has been so far possible to infer on this network.

%%% Local Variables: 
%%% mode: latex
%%% TeX-master: "thesis"
%%% End: 

\chapter{Biological background}
\label{chap:biological_back}
\section{Metabolism: theory and modeling}
\label{sec:metabolism}
\begin{figure}[h]
\center
\includegraphics[scale=0.45]{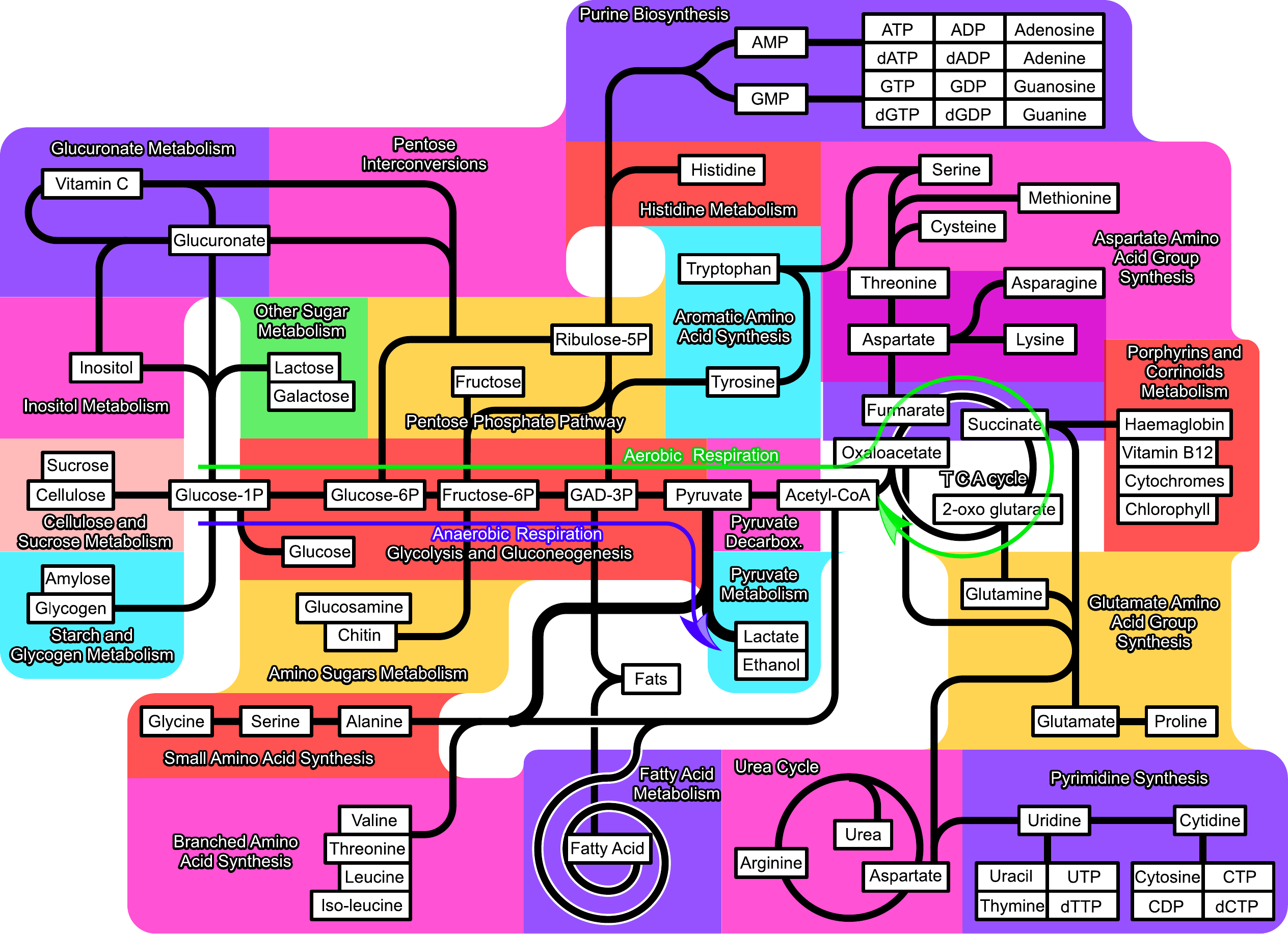}
  \caption{Schematic representation of the Metabolism divided in pathways. In each square there is the pathway name and the key metabolites involved in this pathway. \cite{Img_metabolism}}
  \label{fig:metabolism}
\end{figure}

 In this section we will briefly summarize the most important characteristics of metabolism, for a comprehensive approach of this argument see \cite{Brock_2012}.

Metabolism is the combination of processes through which cells make and use the energy needed for all their physiological functions. It is generally divided in two parts: Catabolism and Anabolism. In the latter part, molecules used by the cell are synthesized while in the other part macromolecules (e.g. glucose) are broken down in smaller components. Furthermore, catabolic reactions produce energy that is then used for the anabolic reactions, for mechanical work inside the cell and for the active transport of molecules against the pressure gradient. The energy is transported and stored in the cell by the means of key metabolites as ATP or NADPH that form an active link between anabolism and Catabolism. Thus the cell can be considered as a ``machine'' that uses the energy produced by the catabolic part to accomplish cellular functions (e.g. homeostasis to maintain the cellular pressure) and to produce and maintain the biological structure of which it is made (e.g. production of amino acids).

\begin{figure}[h]
\center
\includegraphics[scale=0.55]{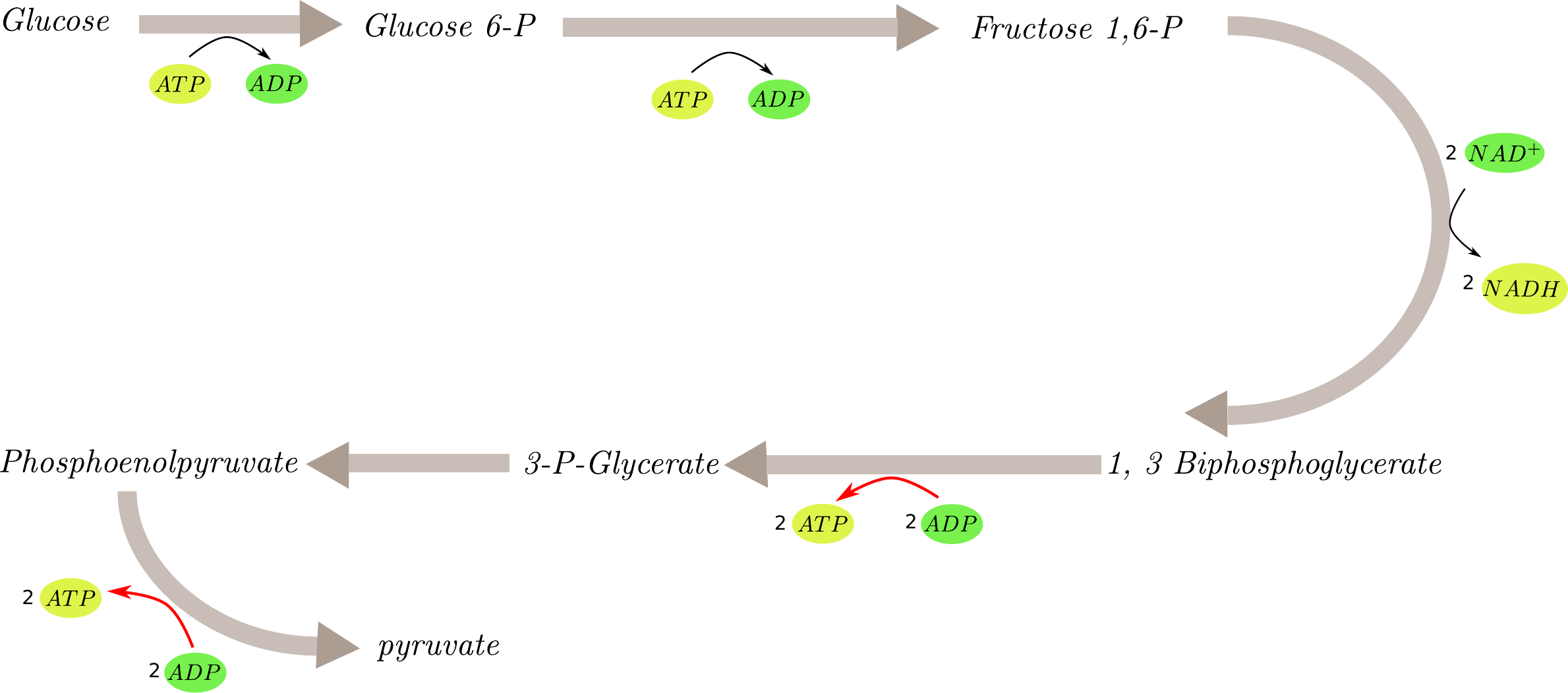}
  \caption{Schematic representation of the Glycolysis.}
  \label{fig:glyc}
\end{figure}

It is possible to divide reactions in {\it metabolic pathways} that are non overlapping functional groups of reactions (see Figure \ref{fig:metabolism} for a schematic view of the typical pathways of cellular metabolism). Generally reactions inside a pathway work together in certain parts of the cell producing substrates needed by the cell. The main catabolic pathways are: the Glycolysis and the Krebs cycle (or TCA cycle) that we will detail further in the following. Whereas anabolic pathways can be divided in four main groups: {\bf the fatty acid metabolism}, for the synthesis of the elements of the membrane of the cell, {\bf the nucleotide synthesis pathways}, for the synthesis of the base elements of DNA and RNA, {\bf the pathways for amino acids synthesis} (protein synthesis) and {\bf the penthose phosphate pathway} that is an alternative to Glycolysis for the degradation of the {\it glucose} but without energy yield. Pathways are strongly interconnected between them also if not all the groups of reactions are localized in the same part of the cell.

\begin{figure}[h]
\center
\includegraphics[scale=0.25]{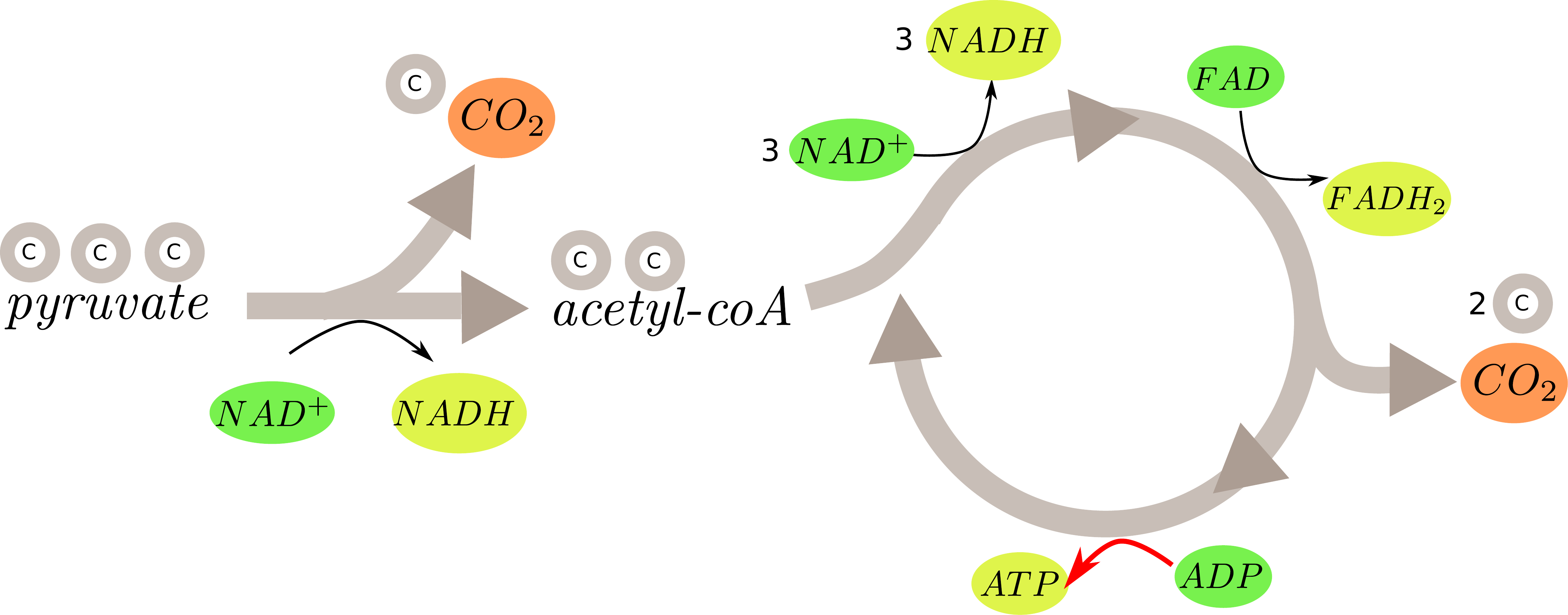}
  \caption{Schematic representation of the Krebs cycle.}
  \label{fig:krebs}
\end{figure}

{\bf Glycolysis} is the pathway by which the cell decomposes {\it glucose} in smaller molecules. The net energy yield of this pathway are {\it two molecules of {\it ATP} per molecule of {\it glucose}} as it uses two ATP molecules and produces four, see Figure \ref{fig:glyc}. Furthermore the by-product of the pathway is the {\it pyruvate}. This molecule is then used by the cell mainly in two ways: it can be reduced by {\it fermentation} or in {\it aerobic respiration} (Figure \ref{fig:metabolism}).

In fermentation pyruvate is reduced in various types of fermentation products (in E.Coli the possible products are: {\it Succinate}, {\it Ethanol}, {\it Lactate}, {\it Formate}, {\it Acetate},...). The common characteristic of all these processes of reduction is that {\it NADH} is oxidized back to $NAD^+$, thus permitting the Glycolysis to continue. The combination of fermentation and Glycolysis is often called {\it anaerobic respiration}. In this setting, the only active catabolic reaction is the Glycolysis, hence anaerobic respiration has a net yield of 2 molecules of ATP per molecule of glucose that is a very low yield. Nevertheless in environments poor of oxygen this is the main source of energy.

In presence of oxygen, it is possible to activate the {\bf The Krebs cycle} by oxydating {\it pyruvate} into {\it acetyl-CoA}. This cycle produces an impressive net energy yield of 32 ATP molecules per molecule of glucose (Figure \ref{fig:krebs}) being thus much more efficient from an energetic point of view than fermentation. The combination of Glycolysis and Krebs Cycle is called {\it aerobic respiration}. This is clearly the most efficient process to produce energy and it is the preferred pathway in environments with oxygen.

\begin{figure} 
  \center
  \includegraphics[scale=0.45]{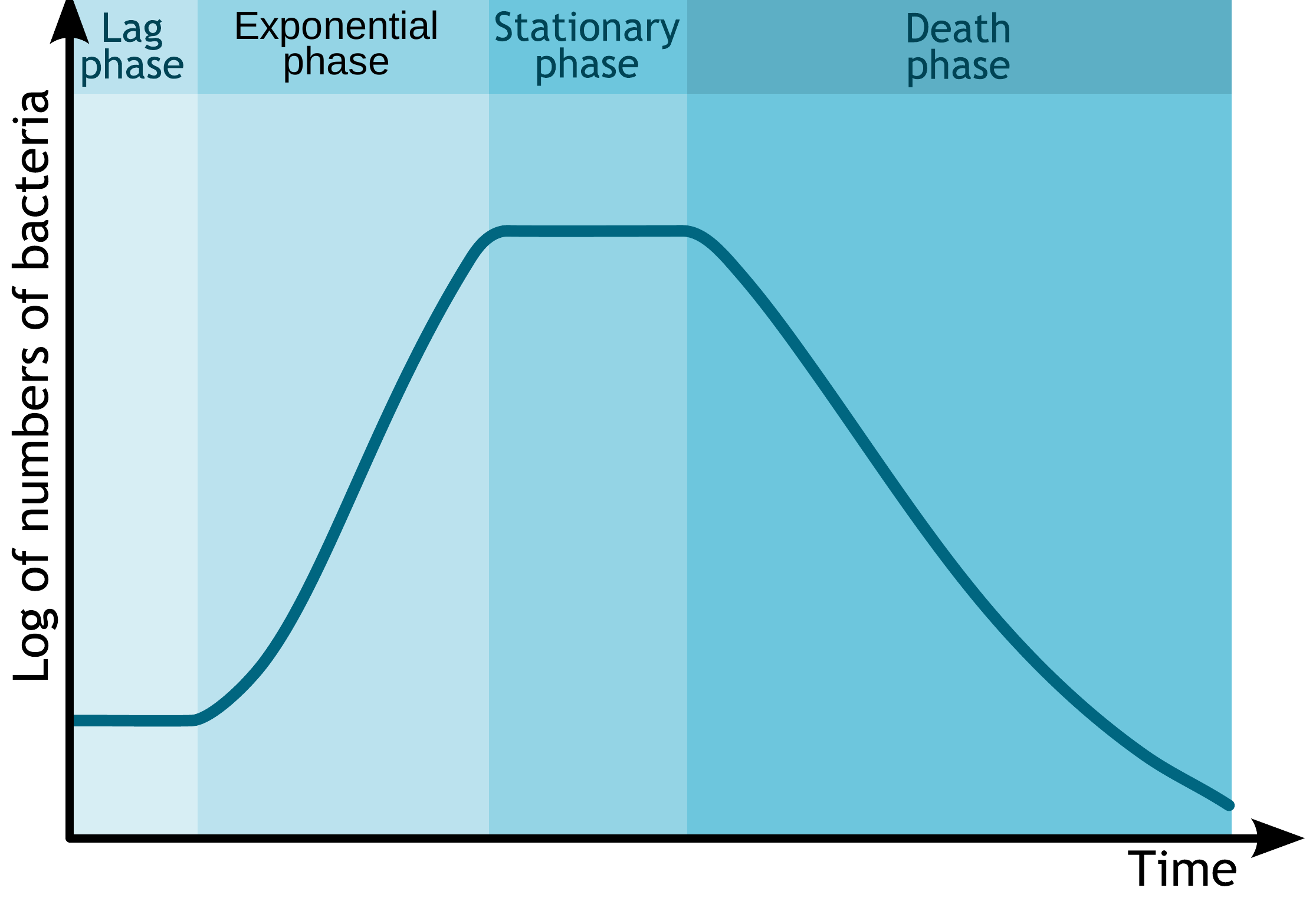}
  \caption{Typical bacterial growth in a suitable environment. \cite{Img_growth}}
  \label{fig:typical_growth}
\end{figure}

Depending on the environment, the metabolism of microorganism like E.Coli can have various macroscopical ``states'', the most studied of which is the growth state. When the medium is suitable for growth a population of E.Coli will, after a lag phase, first grow exponentially, then have a stationary state and finally die when all the resources are used, see Figure \ref{fig:typical_growth}. During the growth state, the cell is using the nutrients to produce all the elements necessary for the creation of new cells (amino acids, nucleotides, fatty acids,...). After $n$ such division, we can write:
\begin{equation}
N=N_02^n,
\end{equation}
where $n$ is the ``generation'' (number of division in the population of cells), $N_0$ is the initial number of cells and $N$ is the number of cells at population $n$. Using this equation we can immediately write:
\begin{equation}
n=\Delta \log(N)/\log(2),  
\end{equation}
where $\Delta\log(N)=\log(N)-\log(N_0)$. Thus after a growing time $\Delta t$, we can define the \textit{generation time}, $g=\Delta t/n$, as the average time it takes the cell to divide. Another equivalent way of measuring growth is by finding the slope at which cells are growing in the exponential phase, obtaining the \textit{growth rate}, $\lambda$ (hence $\lambda=log(2)/g$). These quantities are characterizing properties of the microorganism, for E.Coli the typical generation time is of 15-20 minutes.

In principle one could think of modeling metabolism using a comprehensive theoretical approach. This approach should then take into account all the elements involved in metabolism and in particular it should be possible to model reactions. This is generally done by considering it as thermodynamical processes, aided by enzymes, in which a free energy barrier is overcomed. Then assuming a Michaelis-Menten kinetic in which a kinetic constant is associated to a particular reaction it is possible to write equations that estimate the flux through a particular reaction. Nevertheless this constants may depend on many factors as temperature, enzymes or concentrations, making very difficult (if not impossible) to estimate it reliably. Furthermore not always the kinetic constant of a reaction will be the same when measured isolated or inside the cell. In principle it should be possible to measure it directly inside the cell but this is still an open problem in the experimental analysis of metabolism. Indirect measurements are possible via the isotopic labeling of the glucose \cite{sauer2006metabolic} where the glucose given to the cell is labeled using the $^{13}C$ isotope. It is then possible, by measuring the final products of the cell, to reconstruct from which pathways the glucose has been processed. Nevertheless, as pointed out in \cite{antoniewicz2013_review}: ``in general, the quality of flux estimates depends on: firstly the structure of the metabolic network model, secondly substrate labeling, thirdly isotopic labeling measurements, fourthly number of experiments performed, and finally to a lesser extent, the flux values''. Adding to these considerations the fact that the number of reactions involved increase very rapidly with the size of the organisms it is straightforward to understand that a reliable kinetic modeling for the analysis of the metabolism is nowadays still impossible.

\begin{figure}
\center
\includegraphics[scale=0.3]{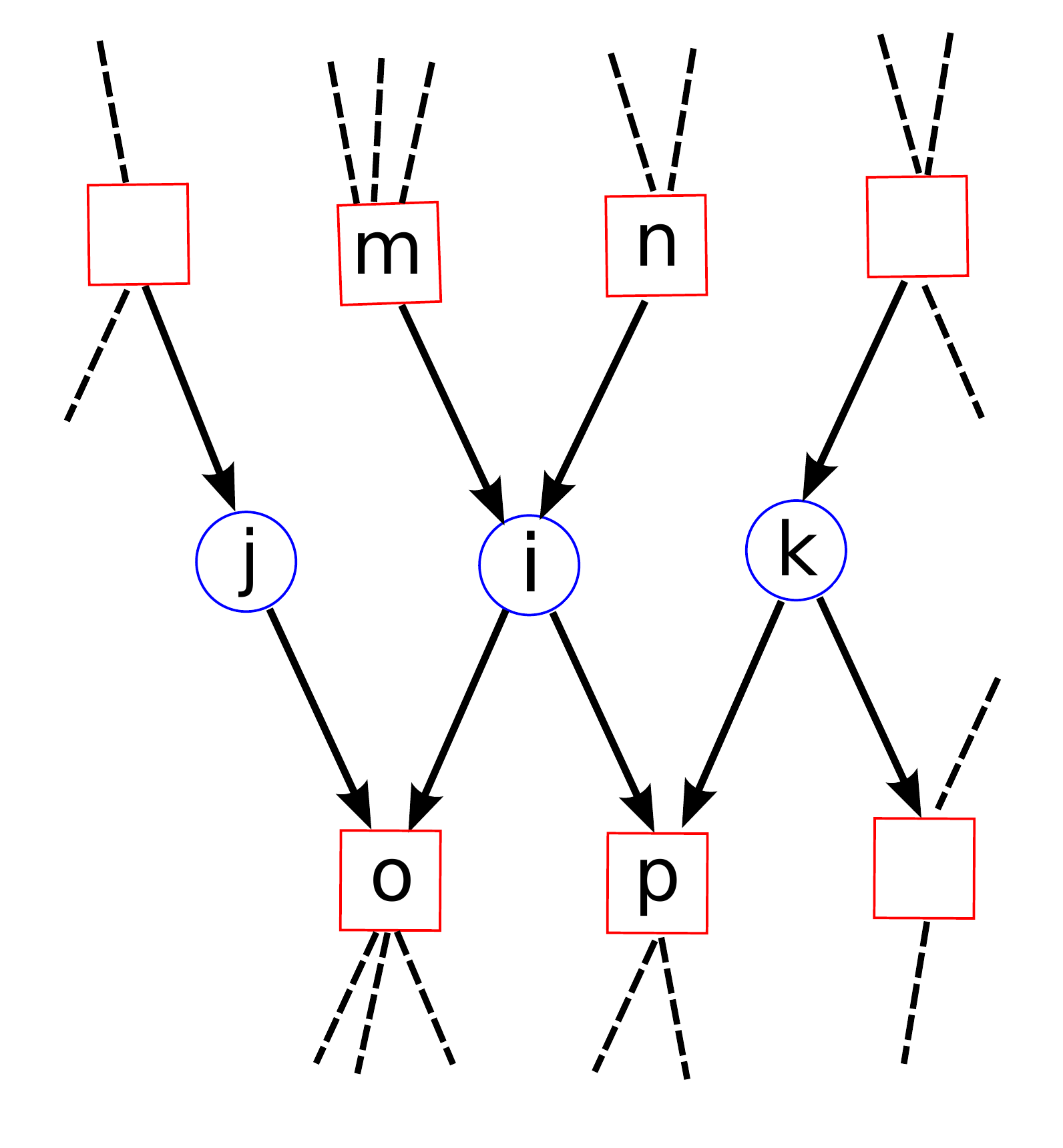}%0.25
\caption{Sketch of a bipartite network representing a metabolic network, where the squares are the metabolites and the circles the reactions.}
\label{fig:bip_sketch}
\end{figure}

On the other hand, in recent years a new holistic approach has been introduced in biological and biomedical research, the concept of \textit{Systems Biology}. In this approach it is considered that it is possible to understand a biological system without the need of a complete description of every single element of the system. In particular for metabolism, a common systems biology approach to overcome the theoretical limitations of a kinetic analysis has been to model metabolism as a bipartite network of reactions and metabolites or \textit{a metabolic network} (see Figure \ref{fig:bip_sketch}). In this network, each reaction uses a set of {\it reactants} to produce a set of {\it products} and each metabolite is the reactant (or product) of one or more reactions. The topology of the network is encoded in the {\it stoichiometric matrix} of the system, $\Xi$, with entries $\xi_i^m > 0$ if reaction $i$ produces metabolite $m$, $\xi_i^m < 0$ if reaction $i$ consumes metabolite $m$, and $\xi_i^m = 0$ if there is no link. The value of the stoichiometric matrix is the stoichiometric coefficient with which the metabolite is entering in the specific reaction. In general reactions are divided in three classes. {\it Nutrient in-takes} are defined as having $\xi_i^m=\delta_{m,n}$: these processes serve the purpose of supplying an individual metabolite ($n$ here) to the network. {\it Out-takes} are defined as having $\xi_i^m=-\delta_{m,o}$, and remove an individual metabolite ($o$ here) from the network. Finally, {\it core reactions} are the reactions responsible of intracellular processes. 

Using the stoichiometric matrix it is possible to write the dynamical mass balance conditions for the concentrations of the metabolites:
\begin{equation}
\label{eq:dyn_mass_balance}
\frac{d\bvec{x}}{dt}=\bvec{\Xi}\bvec{v},  
\end{equation}
where $\bvec{x}$ is the M dimensional vector of the concentrations and $\bvec{v}$ is the flux vector.

In metabolic modeling it is often useful to represent the capacity of the cell to grow with a ``biomass'' reaction. This reaction has in input all the necessary elements to make one gram of another cell. The stoichiometry of this reaction is experimentally obtained by considering the average weight of metabolites in a cell. Thus it is possible to infer the amount of each metabolite needed for the production of a new cell (see Section \ref{sec:real_net} for the biomass reaction of E.Coli).

In the following sections we will present two established approaches that have been used to study the metabolic network from a systems biology point of view: Flux Balance Analysis and Von Neumann approach.

\section{Flux Balance Analysis}
\label{bio:FBA}

Flux Balance Analysis (FBA) is a computational method developed by Palsson \cite{Palsson_FBA} to find possible biological states of a metabolic network. In FBA it is assumed that the concentrations of the metabolites are in a steady state (constant), or in a homeostatic condition :
\begin{equation}
\label{eq:omeostatic_cond}
\bvec{\Xi}\bvec{v}=0.  
\end{equation}
Note that this is a strong assumption but can be valid in some situations. Namely the steady state approximation can be considered generally valid because of the fast equilibration time of metabolite concentrations (seconds) with respect to the time scale of the genetic regulation (minutes). The convenience of this approach is that it is enough to find the solution of $M$ linear equations in $N$ variables instead of having to solve $M$ {\it differential} equations. 

The vectors of fluxes solving equation (\ref{eq:omeostatic_cond}) satisfy the Kirchoff law: the sum of the fluxes of the reactions producing a metabolite equals the sum of the fluxes consuming this metabolites. Hence from a chemical point of view these equations are stating that all the intermediate metabolites produced by the network are also completely consumed.

In a typical metabolic network, $N>M$, and in this case the dimension of the solution space is an hyperplane of size $N-rank(\Xi)$. In principle all the solution inside this space should be sampled, but even by Monte-Carlo methods it becomes unaffordable as the dimension of the spaces exceed a few tens \cite{Wiback_2004}.

There has been some attempts to understand mathematically and computationally how this solution space is organized. The first and most straightforward attempt is to search for the basis of the solution space that are on the edges of the cones of the solutions, thus these vectors can be written as
\begin{equation}
\label{eq:2}
\bvec{v}=\sum_i \beta_i \bvec{p}_i,
\end{equation}
where the vectors $\bvec{p}$ are called the extreme pathways \cite{Pals_extreme,Pals_extremeII} of the system. Clearly the decomposition in extreme pathway is not unique. In \cite{Pals_extremeII}, Singular Value Decomposition is used on the extreme pathways matrix to find the vectors that better describe the system. Another interesting attempt is the elementary mode analysis \cite{Schuster_1994,Schuster_extreme}. These elementary modes are minimal modes that satisfy basic thermodynamical constraints, in which the network can function. Contrary to the extreme pathways, the division in modes is unique and in \cite{Schuster_extreme} it is shown that some of these modes correspond to known biological pathways in the network of E.Coli. Still growing in scale the number of elementary modes grows very quickly and it becomes difficult to find it all. It thus seems that finding the extreme pathways is not simpler than sampling the configuration space.

It is thus clear that also if the solution space has been greatly simplified by assuming (\ref{eq:omeostatic_cond}), still a more direct and effective method for selecting the relevant biological solutions is needed. This has been done in FBA by imposing the maximization of an objective function, normally represented as a linear combination of fluxes, $\bvec{\alpha}\bvec{v}$, where $\bvec{\alpha}$ is a $N$ dimensional vector of coefficients. With this further assumption, the problem become a standard linear optimization problem or:
\begin{equation}
\label{eq:FBA_problem}
\max_{\bvec{\nu}_{min}\leq \bvec{\nu} \leq \bvec{\nu}_{max}}\left(\bvec{\alpha}.\bvec{v}\right)\quad\text{subject to}\quad\bvec{\Xi}\bvec{v}=0,
\end{equation}
where $\bvec{\nu}_{min}$ and $\bvec{\nu}_{max}$ are are the lower and upper bound to the fluxes. These bounds can be chosen to model a certain environment (e.g. to represent the presence or not of a certain metabolite in the environment), to define known physiological limitations (e.g. the ATP maintenance flux has a specific value) or to define the direction in which reactions function (e.g. if $\nu^i_{max}=0$ and $\nu^i_{min}< 0$, reaction $i$ has to function reversibly).

\begin{figure}[h]
\center
\includegraphics[scale=1]{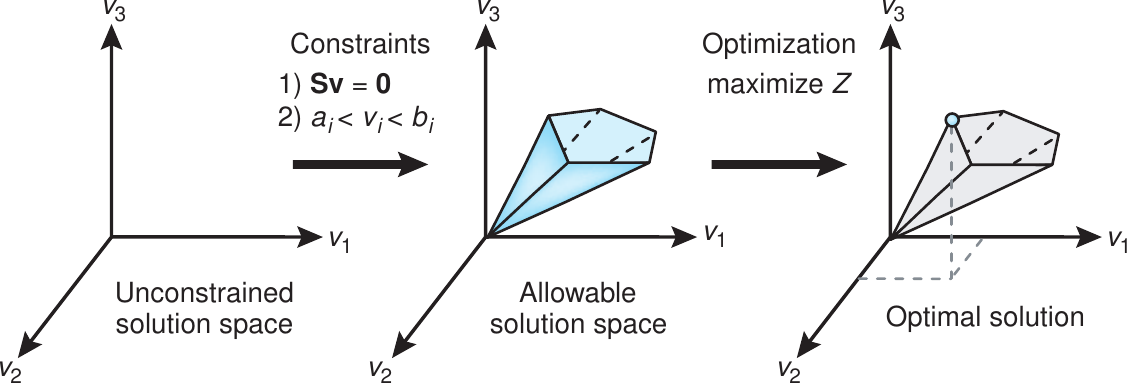}
\caption{Schematization of the procedure used in FBA. Taken from \cite{Orth_2010}.}
\label{fig:FBA_procedure}
\end{figure}

Under this kind of assumptions FBA amounts to maximize a linear function under a set of linear constraints. Hence each constraint is a plane in the variables space and the objective function is a line. It is then intuitive to understand that in order to maximize this function two cases are possible: the line and the surface defined by the constraints meet in one point or the line and the surface are parallel. In the first case there is only one solution while in the latter infinite. Furthermore in a real biological case it is highly unlikely that the line and the surface are parallel, thus in general in flux balance analysis only one solution is found \cite{Palsson_Theo}. A schematic representation of the procedure can be seen in Figure \ref{fig:FBA_procedure}.

Clearly the main biological assumption in FBA is contained in $\bvec{\alpha}$ and depending on the problem there can be many different choices for this function \cite{Schuetz_2007}. The most common and that has given the more interesting results is the biomass function. In this case the objective function is the biomass reaction and the problem is to find the configuration of the system that give the maximal biomass yield.

\begin{figure}[h]
\center
\includegraphics[scale=0.85]{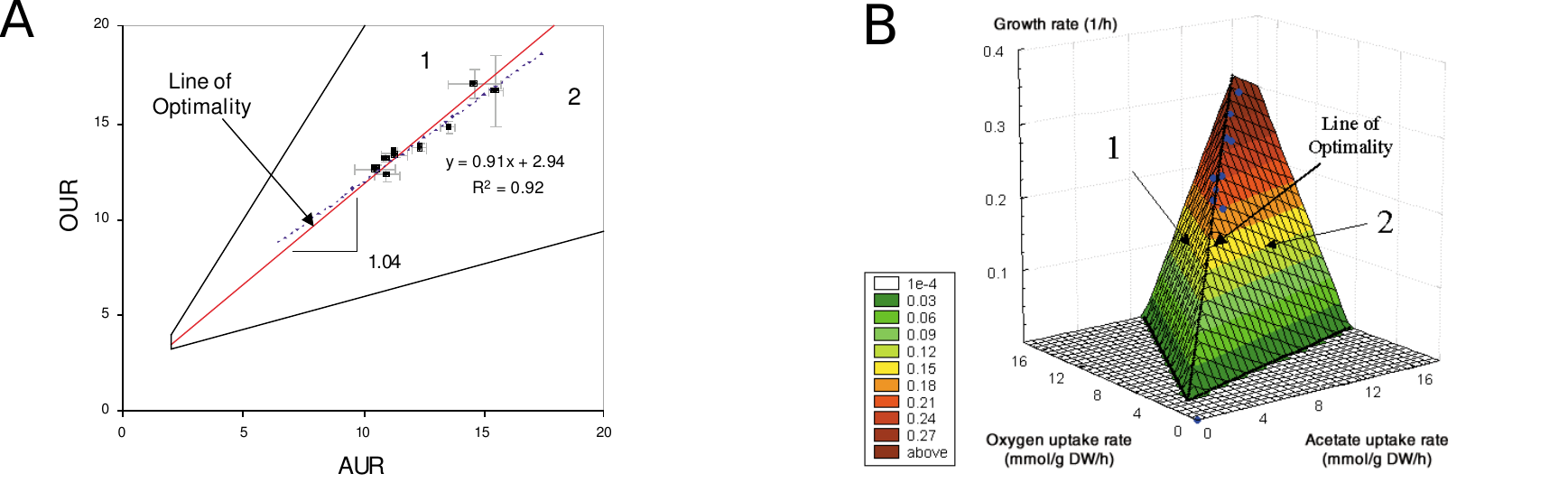}
\caption{Predictions and experimental results obtained in article \cite{Edwards}. (A) acetate uptake rate (AUR, mmol/g DW/h) versus oxygen uptake rate (OUR, mmol/g DW/h). The red line is the LO and the points are the experimental measurement. (B) 3D representation of the same results.}
\label{fig:FBA_acetate_growth}
\end{figure}

In the last 10 years, FBA has become in computational biology a very important tool to study and understand the metabolic networks. In \cite{Varma,Edwards} the authors managed to correctly reproduce experimental measurement using the metabolic FBA model \cite{Edwards_2000}. 
In particular, in \cite{Edwards} authors studied the dependence of the growth rate of E.Coli on the uptake of oxygen and the uptake of acetate both experimentally and computationally. This experiment was possible as uptakes are simple to measure because it is sufficient to measure the remaining metabolite in the environment at the end of growth. In Figure \ref{fig:FBA_acetate_growth} the results of this article are shown. In this Figure, there is the plot of the growth rate versus the uptakes studied ( the \textit{phenotype phase plane}) and of the \textit{line of optimality} (LO) that is the line of maximal growth. These results have also been reproduced using Succinate instead of acetate. It is thus clear from this article that the experimental results and the line of optimality are consistent, showing that the growth rates determined experimentally and the ones predicted by FBA are consistent.

\begin{figure}[h]
\center
\includegraphics[scale=1]{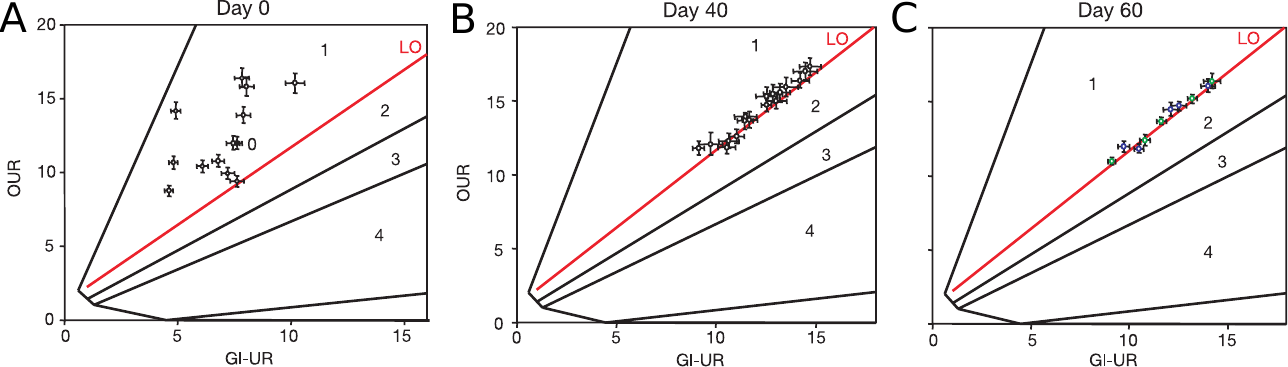}
\caption{Results obtained in article \cite{Ibarra_2002}. (A-C) glycerol uptake rate (Gl-UR) versus oxygen uptake rate (OUR) in a colony of E.Coli starting from day 0 to day 60. LO is the line of optimality and the dots and crosses are the experimental results.}
\label{fig:FBA_adaptive_evolution}
\end{figure}

In \cite{Ibarra_2002} the authors used the same methods to show that after adaptive evolution the cells converge to the growth values expected by the FBA analysis, thus substantiating the idea that this solution is the optimum of the system. This is presented in Figure \ref{fig:FBA_adaptive_evolution} where it is clear that the colony of bacteria undergoes an adaptive evolution in response to the environment they are in, finally converging (see C) to the line of optimality predicted by FBA.

\begin{figure}[h]
\center
\includegraphics[scale=0.7]{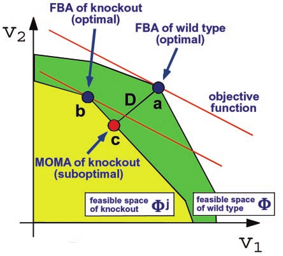}
\caption{Schematic of the MOMA method developed in \cite{Segre_MOMA}.}
\label{fig:MOMA_schematics}
\end{figure}

\begin{figure}[h]
\center
\includegraphics[scale=0.5]{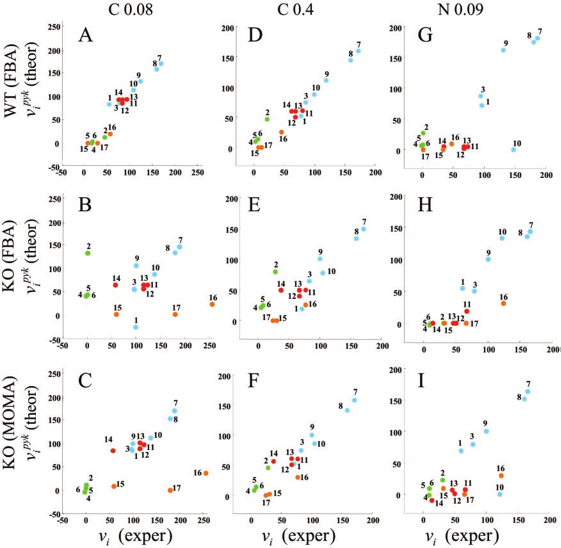}
\caption{Results of article \cite{Segre_MOMA}. The column correspond to different environment conditions. In each graph the measured fluxes of reactions from the central carbon metabolism are plotted versus the computed fluxes. The first row represents the FBA results for the Wild Type E.Coli (no mutation) while the second and third row are respectively the FBA and MOMA results for the mutated E.Coli.}
\label{fig:MOMA_results}
\end{figure}

In \cite{Segre_MOMA} the authors extended FBA to deal with the mutations and knock outs of genes, developing a quadratic programming method named minimization of metabolic adjustment (MOMA). In this method, if a mutation occurs, the system will search for a point in the solution space that has a minimal distance from the original (non-mutated) optimal solution. Thus defining $\bvec{w}$ as the flux vector obtained in the wild type organism and $\bvec{x}$ the new flux vector after the mutation, it is possible to define a euclidean distance between the two:
\begin{equation}
\label{eq:MOMA_distance}
D(\bvec{w},\bvec{x})=\sqrt{ \sum_{i=1}^N(w_i-x_i)^2}.
\end{equation}
Then the problem can be stated as: which is the vector $\bvec{x}$, solution of FBA, such that the euclidean distance between $\bvec{x}$ and $\bvec{w}$ is minimal? This is clearly a quadratic problem and is more difficult than FBA to solve but it is still possible. In Figure \ref{fig:MOMA_schematics} it is shown that using this procedure solutions that FBA could not find are sampled. Using this method, the authors compared the predictions of MOMA with experimental measurements of fluxes of the central carbon metabolism of E.Coli under a Pyruvate Kinase (pyk) knockout. The measurement of such fluxes has been done by combining NMR techniques and $^{13}C$ labeling \cite{Emmerling_2002} (see Section \ref{sec:metabolism}). The results are presented in Figure \ref{fig:MOMA_results} where it is clear that MOMA outsmarts FBA in the prediction of fluxes for the mutated organism (second and third row). Nevertheless it is interesting to note for both the wild type and the mutant, none of the methods presented until now is capable of reproducing the results in the environment on the right (G, H and I).

\begin{figure}
\center
\includegraphics[scale=0.8]{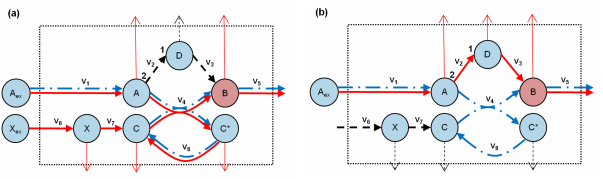}
\caption{In this sketch the differences between MD-FBA and FBA are highlighted. Results from FBA are represented by the blue arrows while results of MD-FBA are in red. The figure illustrates growth on two media: (a) growth on a medium in which both A and X are present; (b) growth on a medium including only metabolite A. FBA predicts the same growth rate, while MD-FBA predicts a different growth rates. Thus using MD-FBA it is possible to find alternative less efficient routes in order to account for dilution.}
\label{fig:md_fba}
\end{figure}

Other approaches have been dedicated to overcome some of the unrealistic assumptions made in FBA. A notable example is \cite{benyamini2009flux} where authors present the dilution-FBA (MD-FBA) method to overcome the limitations related to the steady state assumption. In this case, the metabolite dilution in growth is considered by changing the linear problem posed by FBA to a problem of the type:
\begin{align}
\max_{\bvec{v},\bvec{d}}\left(\bvec{\alpha}.\bvec{v}\right)\quad\text{subject to}\quad\bvec{\Xi}\bvec{v}-\bvec{d}=0,  
\end{align}
where $\bvec{d}$ is a ``dilution'' vector for which $d_j > 0$ if $v_j$ (taken from FBA) is non zero. Using this method it is possible to account for behaviours that in FBA are absent as shown in Figure \ref{fig:md_fba}.

After establishing the predictive power of FBA, this model was used to predict the capabilities and properties of real networks for various organisms. The robustness of the network to the deletion of chosen reactions was analyzed in \cite{Edwards_2008} using a method called Flux Variability Analysis. In this method the solutions to the FBA problem are sampled changing the bounds on the reactions. Many works have been dedicated in the study of alternative classes of objective functions in order to describe the behaviour of the organism in other environment or physiological situations (for a comprehensive list of these works: \cite{Kauffman_2003}).

Another interesting extension of FBA is Energy Balance Analysis, \cite{Beard_2002}, that is an approach to impose the thermodynamical constraint on the solutions of FBA by making FBA results more physically realistic. This approach has been developed because in standard FBA, it is possible to obtain solutions in which there are \textit{unfeasible cycles} or cycles of reactions that are unfeasible from a thermodynamical point of view \cite{beard2004thermodynamic}. Interestingly, in FBA solutions another type of cycle is present, the \textit{futile cycles}, or cycles of reactions in which two metabolic pathways run simultaneously in opposite directions, having no effect other than to dissipate energy. Nevertheless, it is now believed that these cycles make part of a regulatory mechanism inside metabolism \cite{Qian_2006}.

Even though FBA has been so successful, it is still difficult to imagine that all the complexity of a real metabolic network could be completely described by equations so simple as the one in FBA. But from the results, it is reasonable to imagine that during evolution the network optimized itself in some way. This is shown in \cite{Nishikawa_2008} where authors explain that the growth state naturally recruit fewer reactions than typical state and in \cite{lee2012optimal} where they show that this is mainly due to the presence of reversible reactions in the network. On the other hand, it is also reasonable to imagine that during evolution the network became robust to changes in the environment. Hence it is not possible that given the environment and the conditions on the reactions, the network can have only one way of arranging the fluxes in order to grow. This was experimentally observed in \cite{Balaban_2010} where the phenomenon of ``persistence'' is observed in a population of E.Coli. In this experiment they observed that in a population of E.Coli provided with the optimal environment for growth, not every element of the population was growing. Furthermore if the population is presented with an antibiotic, the bacteria that where not growing survived while the other died. This is clearly indicating that given the same environment and condition, an organism can choose different behaviours. Furthermore it is indicating that each behaviour has his evolutionary advantages. This is clearly a situation that FBA cannot picture.

This problem has been addressed in the FBA community , \cite{Mahadevan_2003}, where the authors search for the upper and lower bound on the fluxes of the solutions of FBA. They thus observe that the variability can be high and that many possible solutions with the same objective function value are possible. Hence hinting that it is possible that many solutions with different values of the fluxes exist but all giving the same ``macroscopical'' value of the growth rate.  In \cite{reed2004genome} they analyzed the correlation between reactions in different alternate optimal or suboptimal solutions showing that only a small subset of reaction have variable fluxes across optima, hence showing that optima is robust.

Another interesting open problem in FBA is the harsh dependence on the objective function used to compute the solution. As we can see in Figure \ref{fig:MOMA_results} while for the first two environments the results of FBA (and MOMA) are consistent with data, in the third environment the data and the predictions are not consistent. Hence in this case another function to maximize should be found or the parameters should be changed to find a correct solution.
\section{Von Neumann}
\label{sec:VN}
%\comment{Ha senso che dai risultati di \cite{DeMartino_genes}, dei metaboliti siano prodotti pi\`u di quanto sono consumati?} Si, saranno metaboliti che si accumulano nella rete....
The Von Neumann problem was first formulated to describe expansion in productive systems \cite{VonNeumann} as an autocatalytic process. The objective of this analysis was to determine the maximal production rate sustainable by a system composed of $M$ products connected to $N$ technologies, where the products are produced or consumed by the technologies. The analogy with the metabolic networks is striking and the biological states of a metabolic network based on this method were studied on random networks (\cite{DeMartino_theo}) and on real networks (\cite{DeMartino_genes,DeMartino_prodCap}). In this section we want to present this method, its results and its limitations.

The stoichiometric matrix of a metabolic network can always be written as:
\begin{equation}
\label{eq:stoich_input_output}
\bvec{\Xi}=\bvec{A}-\bvec{B},  
\end{equation}
where $\bvec{A}=\{a_i^m \geq 0\}$ are the coefficients of the products and $\bvec{B}=\{b_i^m \geq 0\}$ are the coefficients of the reactants. Consider then the metabolic network of $M$ metabolites and $N$ chemical reactions, with fluxes that evolve in discrete time steps $t=0,1,...$. Then define $S_i(t)$ as the flux of reaction $i$ at time $t$ and assume that the total input and output of metabolite $m$ at time $t$ is:
\begin{eqnarray}
I^{m}(t)=\sum_i S_i(t)b^{m}_i, \label{eq:VN_input_relation}\\
O^{m}(t)=\sum_i S_i(t)a^{m}_i, \label{eq:VN_output_relation}
\end{eqnarray}
For a theoretical derivation we can assume that the system is autocatalytic, thus the input of a certain metabolite at time step $t$ must come from the output at time step $t-1$ and the system has to satisfy:
\begin{equation}
C^{m}(t) \equiv O^{m}(t)-I^{m}(t+1) \geq 0 \quad \forall m=1,..,M,
\label{constr_VN_raw_form}
\end{equation}
or in words: the input at time $t+1$ must be less or equal than the output at time $t$.

Thus the Von Neumann problem (VN) can be stated as: \textit{find the maximal production rate $\rho^*$ sustainable by the system assuming a dynamical rule, $I^{m}(t+1)=\rho I^{m}(t)$ with constant production rate $\rho > 0$}.

Using (\ref{eq:VN_input_relation}) and the dynamical rule, we have that $S_i(t)=s_i\rho^t$ with constant $s_i \geq 0$. It is then possible to rewrite the linear constraint (\ref{constr_VN_raw_form}) as:
\begin{equation}
\label{eq:VN_constraint}
c^{m} \equiv \sum_{i=1}^{N} (a_i^m - \rho b_i^m )s_i \geq 0 \quad \forall m=1,...,M.
\end{equation}
Interestingly this equation is equal to (\ref{eq:omeostatic_cond}) if $\rho=1$ and $c^m=0$ for every $m$. Hence VN can be seen as a generalization of FBA, because only a subset of the possible solutions of VN satisfy the mass balance conditions. This assumption is reasonable in real systems where metabolites can be produced in excess because they can be used for other non metabolic cellular processes. Furthermore in VN no a priori decided objective function is needed in order to sample the fluxes. Thus in a nutshell, VN aims at recovering the production capability of a real metabolic system.

\begin{figure}[h]
\center
\includegraphics[scale=1]{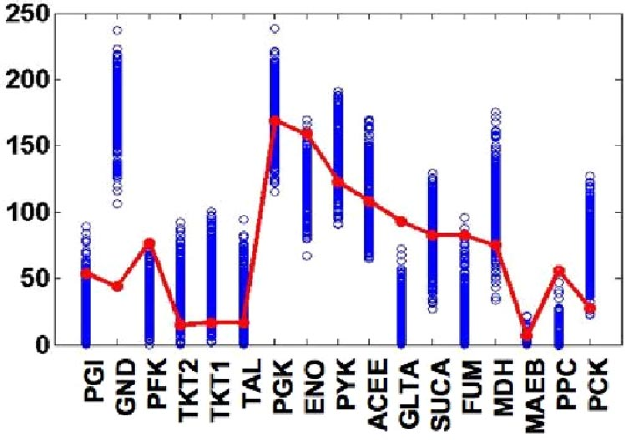}
\caption{Comparison of the value of the fluxes between theoretical values obtain by VN and experimental points (in red). (\cite{DeMartino_genes})}
\label{fig:VN_fluxes_comparison}
\end{figure}

In general if $\rho^* > 1$, the system is considered in expansion, as more elements are produced than consumed; if $\rho^* < 1$ the system is in contraction and finally if $\rho^*=1$ the system is stationary and the fluxes are constant in time.

The main underlying assumption in this model is that the network's physical state $\bvec{s}^*=\{s_i^*\}$ corresponds to one of those for which the production rate attains its maximum possible value $\rho^*$ compatible with constraints (\ref{eq:VN_constraint}). Hence, similarly to FBA, the main assumption of VN is to consider that the metabolic network has evolved into an ``optimal'' state in which the system undergoes the maximal possible expansion.

\begin{figure}[h]
\center
\includegraphics[scale=1]{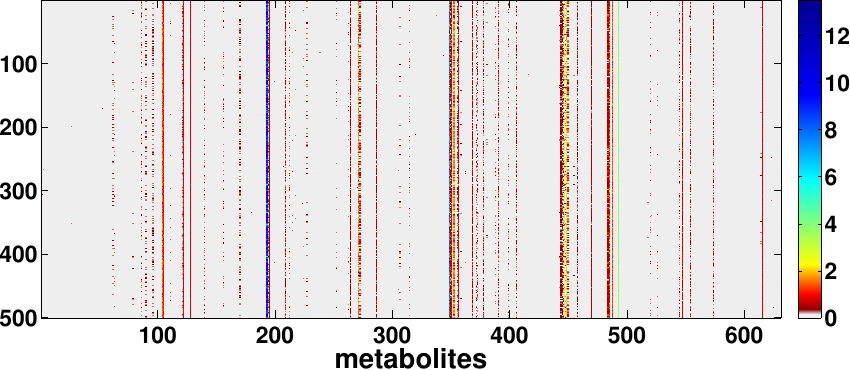}
\caption{Values of $c^m$ for different metabolites in 500 different solutions (\cite{DeMartino_genes}).}
\label{fig:VN_cm_distribution}
\end{figure}

This model has been fully characterized in random networks, \cite{DeMartino_theo}, where a typical phase transition occurs varying the ratio $N/M$. In this case the system crosses over from a contracting phase ($\rho^* < 1$) to an expanding phase ($\rho^*>1$), passing through a stationary regime at $\rho^*=1$. Computationally the problem is solved efficiently by a MinOver algorithm \cite{Krauth_1999} that, as proved rigorously in \cite{DeMartino_theo}, converge to a solution at fixed $\rho$ if at least one solution exists. Moreover, when multiple solutions occur, the algorithm provides a uniform sampling of the solution space.

\begin{figure}[h]
\center
\includegraphics[scale=1]{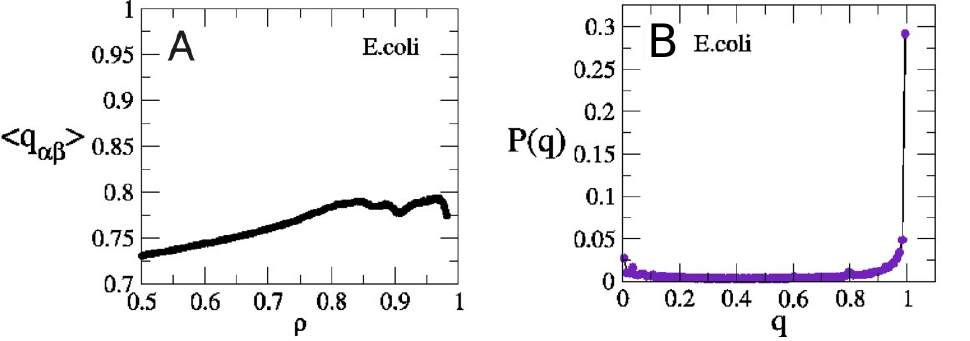}
\caption{(A) Mean overlap in 500 solutions of VN versus the value of $\rho$. (B) Distribution of the $q^{(i)}_{\alpha\beta}$ overlap. (\cite{DeMartino_genes}).}
\label{fig:VN_overlap}
\end{figure}

In \cite{DeMartino_genes} this method was applied to the metabolic network of E.Coli as the metabolism of a real network can still be modeled as an autocatalytic system if uptake reactions are included in the stoichiometric matrix. Though contrary to FBA, this uptake reactions are only from the outside to the inside of the cell, because in VN it is possible to have an accumulation of metabolites inside the cell.

In \cite{DeMartino_genes} it was observed that the fluxes resulting from the simulations using VN are consistent with the experimental ones (Figure \ref{fig:VN_fluxes_comparison}) thus showing that with this approach it is possible to reproduce the behaviour of real networks. Furthermore it was observed that the maximal production rate of the network is $\rho^*=1$, hence the network is in a stationary phase. However not all the $c^m$ are equal to $0$ (Figure \ref{fig:VN_cm_distribution}), thus the VN solutions differ from the ones of FBA.

To characterize solution space of VN it is possible to define an overlap between the fluxes in solution $\alpha$ and in solution $\beta$:
\begin{equation}
\label{eq:VN_overlap}
q_{\alpha \beta}=\frac{2}{N}\sum_{i=1}^N \frac{s_{i\alpha}s_{i\beta}}{s_{i\alpha}^2+s_{i\beta}^2}.
\end{equation}
Thus $q_{\alpha \beta}=1$ if $\bvec{s}_{\alpha}=\bvec{s}_{\beta}$ , whereas $q_{\alpha\beta}\neq 1$ if the value of the flux has changed for some (or all) the reactions passing from solution $\alpha$ to solution $\beta$. In Figure \ref{fig:VN_overlap} it is shown that for the metabolic network of E.Coli the mean of overlap over the solutions at $\rho^*$ is different from $1$ and that the distribution is peaked in $1$ but many different values of $q_{\alpha\beta}$ are possible. Thus this is showing that in the real network there are many possible solutions to the VN problem. Furthermore from Figure \ref{fig:VN_overlap} it is possible to say that only $30\%$ of the reactions are fixed to a given value (e.g. their state is frozen) in all solutions while the rest is free to fluctuate.

The stoichiometry of the real metabolic network presents groups of metabolites that form a \textit{conserved metabolite pools}. In the following we will see that taking into account these pools, it is possible to explain why in the metabolic network of E.Coli $\rho^*=1$. A conserved pool is a group of metabolites, $g$, for which:
\begin{equation}
\label{eq:VN_conserved_pools_first}
\sum_{m \in g}(a^m_i-b^m_i)=0 \quad \forall i.
\end{equation}
We can also characterize this property defining $z^m_g$ as:
\begin{eqnarray}
z^m_g=
\begin{cases}
1 \quad \text{if } m \in g \\ \\
0 \quad \text{otherwise}.
\end{cases}
\end{eqnarray}
Hence for each metabolite pool $g$, it is possible to write (\ref{eq:VN_conserved_pools_first}) as:
\begin{equation}
\label{eq:VN_conserved_pools}
\bvec{z}_g \bvec{\Xi}=0.
\end{equation}
Now we can rewrite equation (\ref{eq:VN_constraint}) as:
\begin{equation}
\label{eq:VN_constraint_pools}
\sum_{i=1}^N s_i\sum_{m=1}^Mz^m_g(a^m_i-b^m_i)+(1-\rho)\sum_{i=1}^Ns_i\sum_{m=1}^M z^m_g b^m_i \geq 0.
\end{equation}
The first term of the equation is $0$ due to the definition (\ref{eq:VN_conserved_pools}) whereas the second term has only two solutions: $\rho^* \leq 1$ and $\bvec{s}_g$ not constrained or $\rho^* > 1$ and $\bvec{s}_g=0$, where we referred to $\bvec{s}_g$ as the vector of fluxes of the reactions implied in the pool. The solution in which the fluxes are $0$ can be discarded for physical reasons if the pool is connected to the network. Hence the presence of metabolite pools imply that $\rho^*=1$ in real metabolic networks. This results has been generalized in \cite{DeMartino_2009} where conserved pools are studied rigorously in simplified reaction networks.

In conclusion VN approach is a way to show that some reactions tend ``naturally'' to satisfy mass balance whereas a consistent subset of reactions is not. All this without imposing an a priori objective function, thus overcoming the problem of the choice of the function to maximize. With this approach it is possible to explain experimental data. Furthermore the existence of conserved metabolite pools was shown to be a very important property in the growth of the metabolite network.

What is still not possible to do with this method is predict the growth rate of the organisms. An attempt has been done in \cite{DeMartino_2012} where it was observed that adding a biomass reaction VN naturally tend to switch it off, because the configuration of fluxes without this reaction was already a solution. Nevertheless it is possible to have values different from $0$ constraining the reaction with bounds, observing that the maximal $\rho$ sustainable by the system decrease as the bound increase. This clearly points out that in order to sustain the growth the system undergoes a stress. VN has also been compared with Network Expansion (next section), \cite{DeMartino_prodCap}, studying the metabolites that the network is producing in the VN solutions. Using principal component analysis it is possible to show that the principal mode is associated with the biomass production while the secondary mode contains the metabolites necessary for the survival of the cell. Also further generalizations have been considered, as the introduction of reversible processes in \cite{DeMartino_2010_bis}.

Being VN a relatively recent method, still many applications have to be made in order to understand its limitations or advantages compared to FBA.

\section{Network Expansion}
\label{sec:NE}
We will finally present a last method used to study the metabolic networks, Network Expansion (NE). This method, first introduced in \cite{Ebenhoh_2004} and then detailed in \cite{Ebenhoh}, is a method to find how the information about the environment affect the metabolic network. In essence, NE procedure can be idealized as follows: start from an assignment of the state of a seed of metabolites $U=\{1,..,M_{nutr}\}$ (in Figure \ref{steps_NE}, ON is black and OFF is white), propagate the information given by this condition (checking that all constraints are satisfied) until no further update is possible and  check the final configuration. It is simple to understand that in real networks the propagation will stop after a small number of steps unless the availability of additional compounds is invoked. Indeed, when complemented with assumptions on the presence of important `currency' metabolites, like $H_2O$, with this method it is possible to switch ON a consistent fraction of metabolites. Though, since not all the compounds can be synthesized from arbitrary seeds, the expansion process will in general not lead to a network containing all of the initial seed. The set of metabolites present at convergence is called the scope of the seed, $\Sigma(U)$.

NE has proved a very powerful method to analyze the relationship between structural and functional properties in real networks \cite{Ebenhoh,Ebenhoh_2005}. In particular in \cite{Ebenhoh_2005}, NE was applied to the complete set of reactions present in the KEGG database. Then the scope of all the compounds in this network was analyzed, using as initial seed this compound (plus water). The results show some functional properties of the network, such that some scope are more likely than other and that some compounds bring to the same scope. Furthermore it was analyzed the robustness of the scope size to deletion of reactions, finding the essential compounds inside the network. Finally the synthetic capabilities of various organisms was analyzed. Thus showing that with NE it is possible to analyze evolutionary differences between the organisms.

\begin{figure}
\center
\includegraphics[scale=0.3]{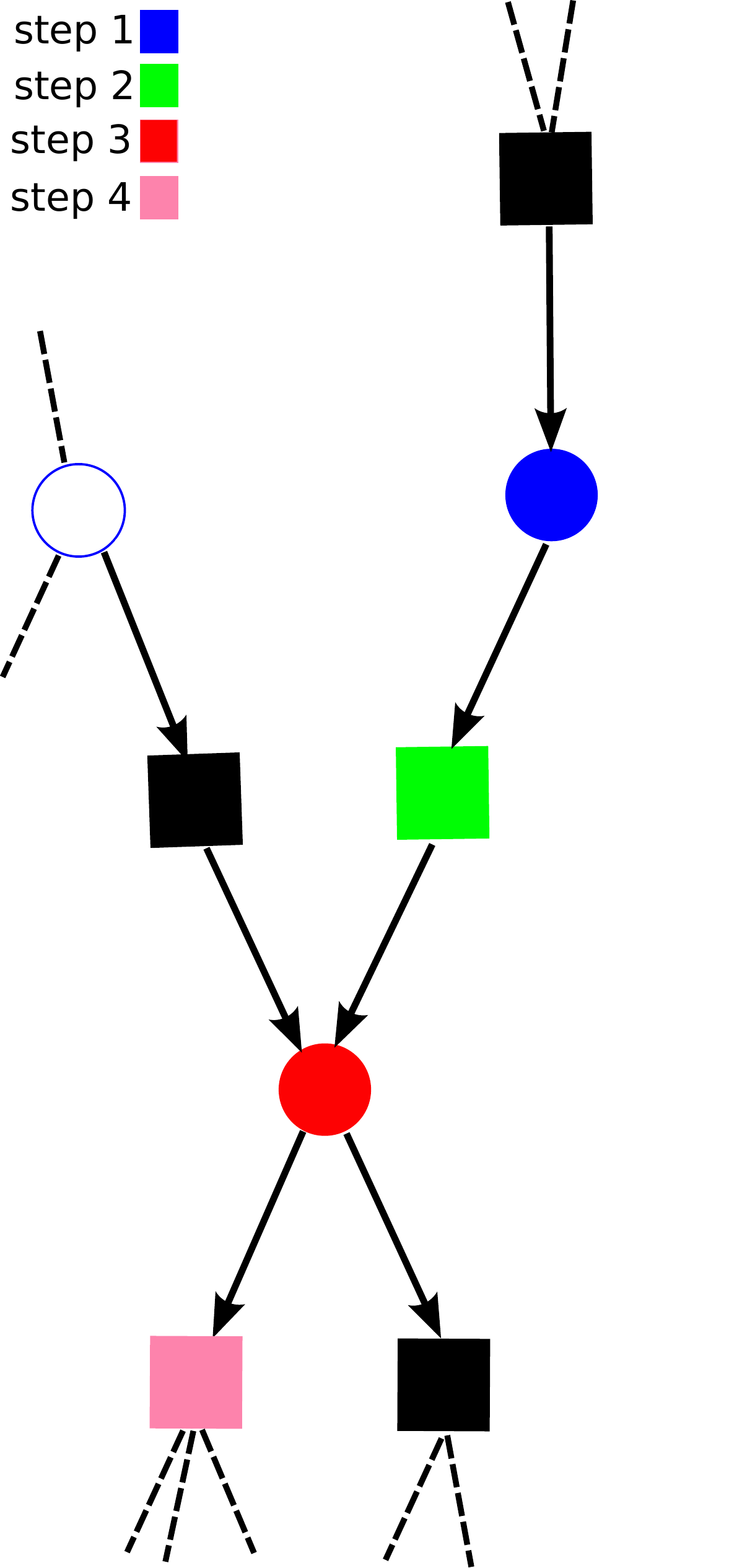}%0.25
\caption{Sketch of four steps of NE procedure in an idealized metabolic network. The metabolites contained in the seed are in black.}
\label{steps_NE}
\end{figure}

In \cite{Kruse_2008}, a metabolite $m$ is considered as \textit{producible} given a seed $U \subset \{1,..,M_{nutr}\}$ if there exist a flux configuration $\bvec{v}$ such that:
\begin{equation}
\label{eq:NE_producible}
[\bvec{\Xi} \bvec{v}]_m  > 0 \quad \text{and} \quad [\bvec{\Xi}\bvec{v}]_n \geq 0 \quad \text{for} \; n \notin U,
\end{equation}
or using equation (\ref{eq:stoich_input_output}) as:
\begin{equation}
\sum_{i=1}^N a_i^m v_i  > \sum_{i=1}^N b_i^m v_i \quad \text{and} \quad \sum_{i=1}^N a_i^n v_i \geq \sum_{i=1}^N b_i^n v_i \quad \text{for} \; n \notin U.
\end{equation}
Thus for a metabolite to be producible it has to be produced while consuming only the metabolites of the seed. The other metabolites, $n$, can be also produced as a ``by-product'' but cannot be consumed. The set of all the producible metabolites from a seed $U$ is called $P(U)$.

Hence if the fluxes inside the cell are stationary, the producible metabolites can increase their concentration while consuming only the nutrients. However if the fluxes are non-stationary this observation is not true anymore. For sure the concentration of all the metabolites that are not contained in $P(U)$ will eventually decrease to $0$ if the cell is growing. It is though possible to define the set of \textit{sustainable} metabolites ($S(U)$) which are the ones that the cell will continue to produce also if growing. This set is obtained by starting from the set of producible metabolites and keeping only metabolites for which all reactions producing it uses only metabolites of $P(U)$. If the resulting set is smaller, we repeat the procedure until convergence. Hence at convergence, metabolites contained in $S(U)$ will satisfy equation (\ref{eq:NE_producible}) with the additional constraint that all the reactions that produce it use only metabolites contained in $S(U)$. As NE is an expanding procedure and $S(U)$ is obtained by contracting $P(U)$ we can safely say that:
\begin{equation}
\label{eq:NE_ensemble_relations}
\Sigma(U) \subset S(U) \subset P(U)  
\end{equation}
Using these definition, in \cite{Kruse_2008} the scope and the set of sustainable metabolites has been sampled in the metabolic network of E.Coli and Methanosarcina barkeri using one compound as seed (plus water). This sampling is computationally intensive as for some calculations, several hundred of linear programming problems have to be solved. It is though observed that in general the scope of the metabolites is equal to the set of sustainable metabolites for many compounds. Furthermore it was observed that adding some cofactors in the NE procedure greatly amplifies this similarity.

NE can also be used to solve the problem: which seed are needed in order to obtain a set of target metabolites? This problem was treated in \cite{Handorf_2008} where a computational procedure to study this type of ``inverse'' problem is defined. First a set of target metabolites, that identify a ``universal'' set of necessary metabolites is identified. Then the minimal seed necessary to synthesize this organisms is identified by sampling the space of seeds using NE to expand the seed information. Finally this method has been applied to various organisms present in the KEGG database, showing that the results were consistent with the biological knowledge on various organisms.

%%% Local Variables: 
%%% mode: latex
%%% TeX-master: "thesis"
%%% End: 

\chapter{Constraint Satisfaction Problems for metabolic networks}
\label{chap:CSP_for_FBA_VN}
\section{The boolean version of the problem}
\label{sec:boolean_problem}
In order to define a Constraint Satisfaction Problem (CSP, for a definition see Section \ref{sec:CSP}) embodying realistic operational constraints, we focus on the characterization of the NESS induced by non-zero in- and out-fluxes of nutrients and sinks, respectively \cite{Beard_2008,ksch}, following two different (but related) schemes already presented in Chapter \ref{chap:biological_back}. In Flux-Balance-Analysis (FBA) it is assumed that fluxes in NESS ensure mass balance at each metabolite node in the network \cite{Kauffman_2003}, see Section \ref{bio:FBA} for further detail. (We do not consider here the optimization schemes that are typically coupled to such constraints in biological implementations of FBA \cite{Palsson_book,Feist}.) If we denote by $J_i$ the flux of reaction $i$ (with $J_i\geq 0$ for an irreversible reaction), this amounts to solving the system
\begin{equation}\label{mbe}
\sum_{i=1}^N \sigma_i^mJ_i=0~~,\qquad \forall m\in\{1,\ldots,M\}~~,
\end{equation} 
where $\sigma_i^m$ is the stoichiometric coefficient of metabolite $m$ in reaction $i$ (such that $\text{sgn}(\sigma_i^m)=\xi_i^m$). One easily understands that the above conditions are equivalent to Kirchhoff's node laws for the flow of matter through metabolite nodes and describe NESS with constant (time-independent) levels for each metabolite. A soft version of this model \cite{DeMartino_theo,dmm} assumes instead that intracellular concentrations may be allowed to increase linearly over time at constant rate, e.g. because some metabolites have to be available for processes outside of metabolism strictly defined. This simply leads to replacing (\ref{mbe}) with
\begin{equation}\label{vn}
\sum_{i=1}^N \sigma_i^mJ_i\geq 0~~,\qquad \forall m\in\{1,\ldots,M\}~~.
\end{equation} 
More formally, the above conditions can be seen to derive from Von Neumann's optimal growth scenario \cite{Gale_1989} and provide a useful means of characterizing a reaction network's production capabilities \cite{Imielinski_2006,DeMartino_genes,DeMartino_prodCap}, see Section \ref{sec:VN} for further detail. 

The general problem posed by (\ref{mbe}) and (\ref{vn}) consists, given the matrix $\widehat{\sigma}=\{\sigma_i^m\}$, in retrieving the flux vectors $\mathbf{J}=\{J_i\}$ satisfying the $M$ linear conditions. An interesting feature that is observed in the solutions of the above models is that a sizeable fraction of reactions carries a null flux in each solution \cite{DeMartino_genes,Nishikawa_2008}. This suggests that, to a first approximation, if one is interested in capturing certain aspects of NESS within a coarse-grained description it might suffice to just distinguish, for each reaction, the inactive state from the active one. 
%\com{In this approach it should be possible to study properties of the network that are not dependent on the specific value of fluxes. This includes: if the network is organized in dynamical modules and if there is a function-specific backbone of the network, informations on dynamical pathways, study of the possible changes in the organization of the network changing environmental conditions,...} 

We shall then introduce, for each reaction, a variable $\nu_i\in\{0,1\}$ (inactive/active). Similarly, we shall link to every metabolite a variable $\mu_m\in\{0,1\}$ that characterizes whether that particular chemical species is available ($\mu_m=1$) or not ($\mu_m=0$) to enzymes that process it. Our next task is to devise Boolean CSPs that embed the basic features underlied by (\ref{mbe}) and (\ref{vn}), respectively.

Starting from (\ref{mbe}), it is simple to understand that a minimal necessary requirement that is encoded in the mass-balance conditions is that, for each metabolite which is produced by an active reaction, there must be at least one active reaction consuming it, and vice-versa. This means that, for each compound $m$, all assignments of $\nu_i$'s are acceptable except those for which all active reactions either produce or consume it. We can therefore define the number of active reactions producing and consuming chemical species $m$ as
\begin{equation}
x_m\equiv \sum_{i\in\partial m_{\inn}} \nu_i\;\quad ~~~~~\text{and}~~~~~
y_m\equiv \sum_{i\in\partial m_{\outt}} \nu_i~~,
\end{equation}
and, in turn, introduce an indicator function $\Gamma_m\equiv \Gamma_m(\mu_m,\{\nu_i\})$ for every $m$ as
\begin{equation}
\label{constraint_FBA}
\Gamma_m = \delta_{\mu_m,0}\delta_{x_m,0} \delta_{y_m,0} + \delta_{\mu_m,1}(1-\delta_{x_m,0}) (1-\delta_{y_m,0})~~.
\end{equation}
Given a configuration $\{\nu_i\}$, metabolite $m$ will be said to be SAT when $\Gamma_m=1$, i.e. when either no reaction in which it is involved is active ($x_m=0$ and $y_m=0$) and the metabolite is unavailable ($\mu_m=0$), or when the metabolite is available ($\mu_m=1$) and at least one reaction produces it ($x_m>0$) and at least one reaction consumes it ($y_m>0$). Similarly, we define a reaction to be SAT when the indicator function $\Delta_i\equiv \Delta_i(\nu_i,\{\mu_m\})$, given by
\begin{equation}
\Delta_i =\delta_{\nu_i,0}+\delta_{\nu_i,1}\prod_{m \in \partial i}\mu_m\;,
\label{constraint_rea_tmp}
\end{equation}
with $\partial i=\partial i_\inn \cup \partial i_\outt$, equals 1. That is, $i$ can be active only if all its neighbouring metabolites  (including both substrates and products) are available. We note that (\ref{constraint_rea_tmp}) can actually be re-cast as
\begin{equation}
\Delta_i =\delta_{\nu_i,0}+\delta_{\nu_i,1}\prod_{m \in \partial i_{\inn}}\mu_m\;,
\label{constraint_rea}
\end{equation}
according to which $i$ can be active only if all of its inputs are available: it is indeed clear that if a reaction is active but one (say) of its products is unavailable, then the constraint imposed on the metabolite will either be violated or force that metabolite to become available. Notice that $\Delta_i=1$ does not imply that $i$ is active when all of its substrates are available. 

The CSP corresponding to (\ref{mbe}) can then be formulated as follows: {\it find a non-trivial assignment of $\nu_i$'s ($\nu_i$'s not all zero) such that all reactions and all metabolites are SAT, i.e. $\Gamma_m=1~\forall m$ and $\Delta_i=1~\forall i$, with $\Gamma_m$ and $\Delta_i$ given by (\ref{constraint_FBA}) and (\ref{constraint_rea}), respectively}. We shall call this CSP Hard Mass Balance, or Hard-MB for brevity.

In order to get a Boolean representation of (\ref{vn}), we note that the main difference between this case and that of (\ref{mbe}) is that, because of the soft constraint, it is no longer necessary that production fluxes are balanced by consumption fluxes. Therefore, while constraint (\ref{constraint_rea}) remains valid, (\ref{constraint_FBA}) has to be replaced by
\begin{equation}\label{gammam}
\Gamma_m = \delta_{\mu_m,0}\delta_{x_m,0} \delta_{y_m,0} + \delta_{\mu_m,1}(1-\delta_{x_m,0})\;.
\end{equation}
In other terms, metabolite $m$ can be available as soon as at least one reaction producing it is active ($x_m>0$). It is convenient to re-write $\Gamma_m$ for this case as
\begin{gather}
%\Delta_i =\delta_{\nu_i,0}+\delta_{\nu_i,1}\prod_{m \in \partial i_{in}}\mu_m\;, \label{constraint_rea_VN} \\
\Gamma_m = \delta_{\mu_m,0}\delta_{x_m,0} + \delta_{\mu_m,1}(1-\delta_{x_m,0})\;, 
\label{constraint_VN}
\end{gather}
so that the constraint at each metabolite node only includes incoming degrees of freedom, making the directionality inherent in the corresponding CSP explicit. It is straightforward to see that  (\ref{gammam}) or (\ref{constraint_VN}), together with (\ref{constraint_rea}), which retains validity, return the same configurations.  

The CSP corresponding to (\ref{vn}) is then the following: {\it find a non-trivial assignment of $\nu_i$'s such that all reactions and all metabolites are SAT, i.e. $\Gamma_m=1~\forall m$ and $\Delta_i=1~\forall i$, with $\Gamma_m$ and $\Delta_i$ given by (\ref{constraint_VN}) and (\ref{constraint_rea}), respectively}. We shall call this CSP Soft Mass Balance, or Soft-MB for brevity.

Note that the constraints behind the two problems can be written compactly as
\begin{gather}
\Gamma_m = \delta_{\mu_m,0}\delta_{x_m,0} (\delta_{y_m,0})^\alpha + \delta_{\mu_m,1}(1-\delta_{x_m,0}) (1-\delta_{y_m,0})^{\alpha}\\
\Delta_i =\delta_{\nu_i,0}+\delta_{\nu_i,1}\prod_{m \in \partial i_{\inn}}\mu_m
\end{gather}
where $\alpha=1$ for Hard-MB and $\alpha=0$ for Soft-MB. 

We shall be interested in solutions obtained upon fixing the probability that a nutrient is available, which we denote below as $\rho_{\inn}$ ( see Appendix \ref{inputs_and_outputs} ).

\section{Real metabolic network}
\label{sec:real_net}
In this thesis we will focus on Escherichia Coli (E.Coli), K-12 MG1655 strain. This bacterium was first discovered in 1885 by a German pediatrician, Theodor Escherich (from which the name) and is commonly present in the intestine of warm blooded organisms. A pathogenic strain of E.Coli (O157:H7) can cause food poisoning in humans but most of the strains are harmless. A special strain (K-12 MG1655) is generally used in laboratories, constituting a typical benchmark in the theoretical study of metabolism.

\begin{figure}[h]
\center
\includegraphics[scale=0.45]{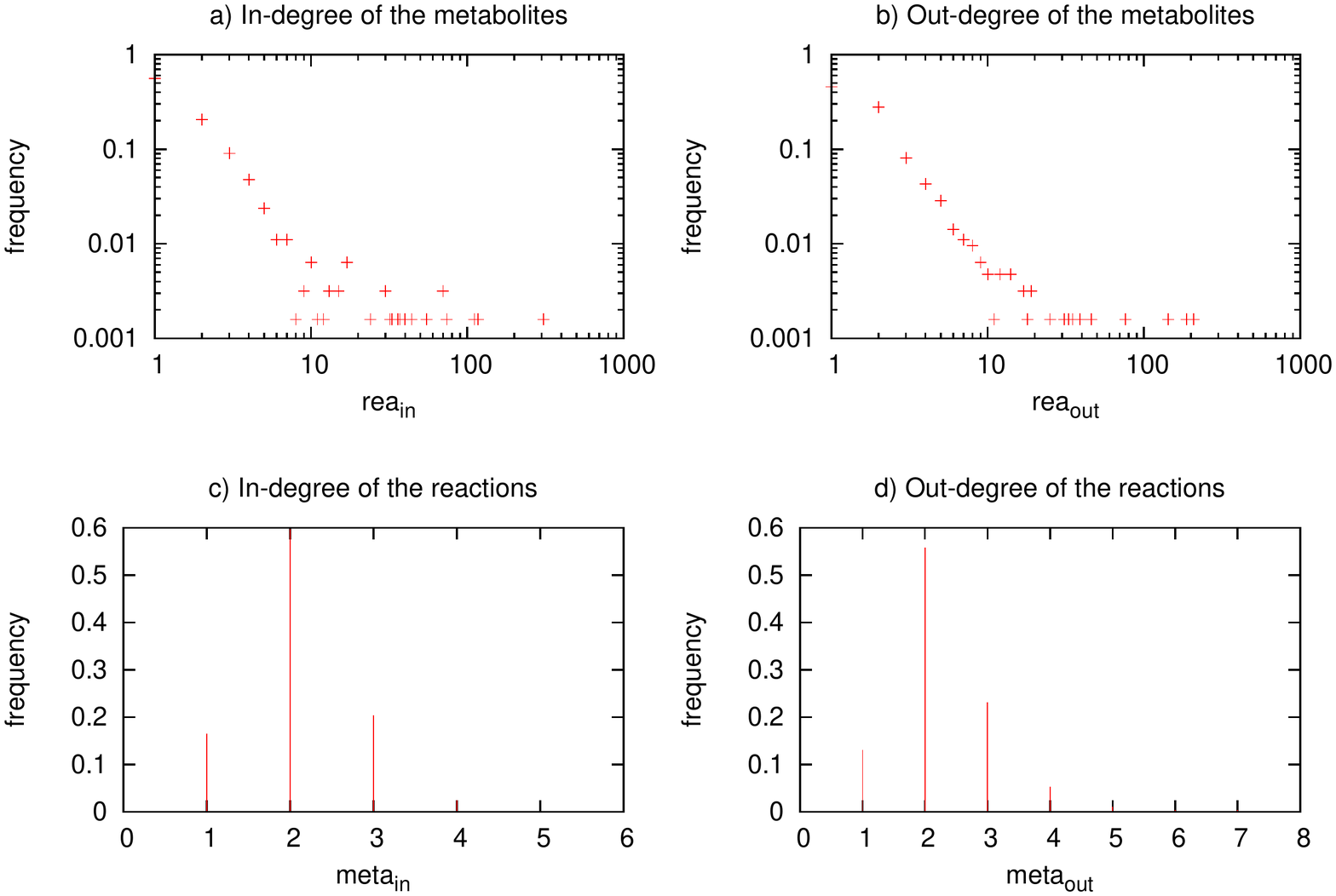}
\caption{In- and out-degrees of E.Coli for metabolites and reactions.}
\label{fig:deg_Ecoli}
\end{figure}

E.Coli is generally considered a model organism as it is inexpensive to grow in laboratory due to its little generation rate and to the fact that the K-12 strain is easily cultivated. Furthermore its metabolism is complete but enough simple to be studied by theoretical methods. From DNA sequencing \cite{blattner1997sequencing}, experimental evidences and modeling predictions \cite{keseler2011ecocyc} it has been possible over the years to reconstruct the stoichiometric matrix that describe the metabolism of E.Coli to a very good precision.  The minimal environment for the optimal growth of E.Coli is called the \verb+M-9+ minimal medium \cite{sambrook1989molecular} : $Na_2HPO_47H_2O$, $KH_2PO_4$, $NaCl$, $NH_4Cl$, $CaCl_2$, $MgSO_4$, $MgSO_47H_2O$, thiamine, carbon source (e.g. glucose) and $H_2O$. In the study of the metabolic network of E.Coli, this medium is equivalent to considering that 7 metabolites are present: $CO_2$, $K$,  $NH_4$ , $Pi$, $O_2$, $SO_4$ and $glc$. 

\begin{longtabu} to \linewidth {|X[l]|X[c]|X[l]|X[c]|}
\hline
Reactant & Coeff. & Reactant & Coeff. \\
\hline
L-Alanine              & $ 0.488     $  &  L-Glutammate  & $0.25      $ \\ 
L-Asparagine           & $ 0.229     $  &  Water         & $45.561 $ \\ 
L-Aspartate            & $ 0.299     $  &  ATP           & $45.732   $ \\ 
AMP                    & $ 0.001     $  &  L-Glutammine  & $0.25      $ \\ 
L-valine               & $ 0.402     $  &  FAD           & $10^{-5}      $ \\ 
NADH                   & $ 3 \; 10^{-5}  $  &  NAD           & $0.00215   $ \\ 
Acetile-CoA            & $ 5 \; 10^{-5}  $  &  UDP glucose   & $0.003     $ \\ 
NADP                   & $ 1.3 \; 10^{-4}$  &  NADPH         & $4 \; 10^{-4}  $ \\ 
Coenzime A             & $ 6 \; 10^{-6}  $  &  Succinil-CoA  & $3 \; 10^{-6}  $ \\ 
Putrescine             & $ 0.035     $  &  L-Metionine   & $0.146     $ \\ 
L-Arginine             & $ 0.218     $  &  Spermidine    & $0.007     $ \\ 
L-Proline              & $ 0.21      $  &  L-Serine      & $0.205     $ \\ 
CTP                    & $ 0.126     $  &  dTTP          & $0.0247    $ \\ 
L-Cisteine             & $ 0.087     $  &  GTP           & $0.203     $ \\ 
Glicine                & $ 0.582     $  &  L-Tirosine    & $0.131     $ \\ 
5-Methyltetrahydrofolate & $ 0.05      $  &  L-Treonine     & $0.241     $ \\ 
Glycogen              & $ 0.154     $  &  L-Istidine    & $0.09      $ \\ 
dCTP                   & $ 0.0254    $  &  dGTP          & $0.0354    $ \\ 
dATP                   & $ 0.247     $  &  L-Lisine      & $0.326     $ \\ 
L-Fenilalanine         & $ 0.176     $  &  L-Leucine     & $0.482     $ \\ 
L-Isoleucine           & $ 0.276     $  &  L-Triptofane  & $0.054     $ \\ 
ADP           & $ -45.560     $  &  Pi  & $-45.563$ \\ 
H           & $ -45.560     $  &  Ppi  & $-0.7302$ \\ 
\hline
\caption{\protect\rule{0cm}{1cm} Table of the values of the biomass reaction as used in this thesis, taken from \cite{reed2003expanded}. If the coefficient is negative, it means that this metabolite is produced by the reaction while it is consumed otherwise.}
\label{biomass_rea_components}
\end{longtabu}

The topology of the network is very different between reactions and metabolites. As we can see from Figure (\ref{fig:deg_Ecoli}), the degree of metabolites have a typical scale free distribution as opposed to the degree distribution of reactions that is peaked around two in-/out-metabolites. This is because some metabolites, (like water or h) contribute in many reactions while in average a reaction is using 2 metabolites to produces 2 products.

In order to write a network that is possible to treat theoretically but that is realistic, it is important to choose how to treat the externals. Conventionally in metabolic analysis, there are external metabolites that are brought inside the cell by {\it uptakes} and some metabolites can be expelled from the cell by means of {\it outtakes}. Generally uptakes and outtakes are represented by the same reversible reaction. It is commonly agreed that the metabolic network of E.Coli has 127 uptakes.

As already introduced in Section \ref{bio:FBA}, another very important reaction that has to be added to the metabolic network is the biomass reaction from which it is possible to simulate growth. This reaction has the form as in Table \ref{biomass_rea_components}.

\section{Random Reaction Networks (RRNs)}
\label{sec:RRN}
We define a random reaction network (RRN) to be a bipartite random graph with two types of nodes, representing respectively chemical species (or \textit{metabolites}) and enzymes (or \textit{reactions}). We shall denote by $N$ and $M$, respectively, the number of reactions and that of metabolites. Both $N$ and $M$ will be taken to be large, i.e. $N,M\gg 1$. For the sake of simplicity, we shall assume here that each reaction has a well defined operational direction, so that  the bipartite graph is directed. Its topology will be encoded in an adjacency matrix $\widehat{\xi}$, with entries $\xi_i^m=1$ if reaction $i$ produces metabolite $m$, $\xi_i^m=-1$ if reaction $i$ consumes metabolite $m$, and $\xi_i^m=0$ otherwise. We furthermore define $\partial m_{\inn}$ (resp. $\partial m_{\outt}$) as the set of reactions producing (resp. consuming) $m$; likewise, for each reaction $i$, $\partial i_{\inn}$ (resp.\ $\partial i_{\outt}$) will denote the set of its substrates (resp.\ products).

The topology of the RRN is specified by the probability distributions of the degrees of the two node types. For metabolite nodes  we shall assume that the in- and out-degrees $\ell_{\inn}\equiv|\partial m_{\inn}|$ and $\ell_{\outt}\equiv|\partial m_{\outt}|$ are independent random variables, both distributed according to a Poissonian with parameter $\lambda$, i.e.
\begin{equation}
\mathit{D}_M(\ell) = e^{-\lambda}\frac{\lambda^{\ell}}{\ell !}\;.
\label{degree_M}
\end{equation}
Metabolites having $(\ell_{\inn},\ell_{\outt})=(0,0)$ are disconnected from the network and will be ignored in what follows. We shall generically assume that $\lambda\geq \lambda_p=1$ (the percolation threshold), ensuring the existence of a `giant' connected subgraph. Metabolites with $(\ell_{\inn},\ell_{\outt})=(0,\ell\ge1)$ represent the substrates that the reaction network derives from the environment (the `nutrients'), whereas metabolites with $(\ell_{\inn},\ell_{\outt})=(\ell\ge1,0)$ will be considered to be the final products or sinks (e.g. excreted compounds or molecules that are employed in processes other than chemical reactions) of the network. The fraction of such `leaves' (nutrients or sinks) is given by $e^{-\lambda}(1-e^{-\lambda})\simeq e^{-\lambda}$ for large enough $\lambda$. The dependence of the results on this external variables is one of the most interesting problem both in real and in random networks and this problem will be addressed in the following sections. 

\begin{figure}
\center
\includegraphics[scale=0.35]{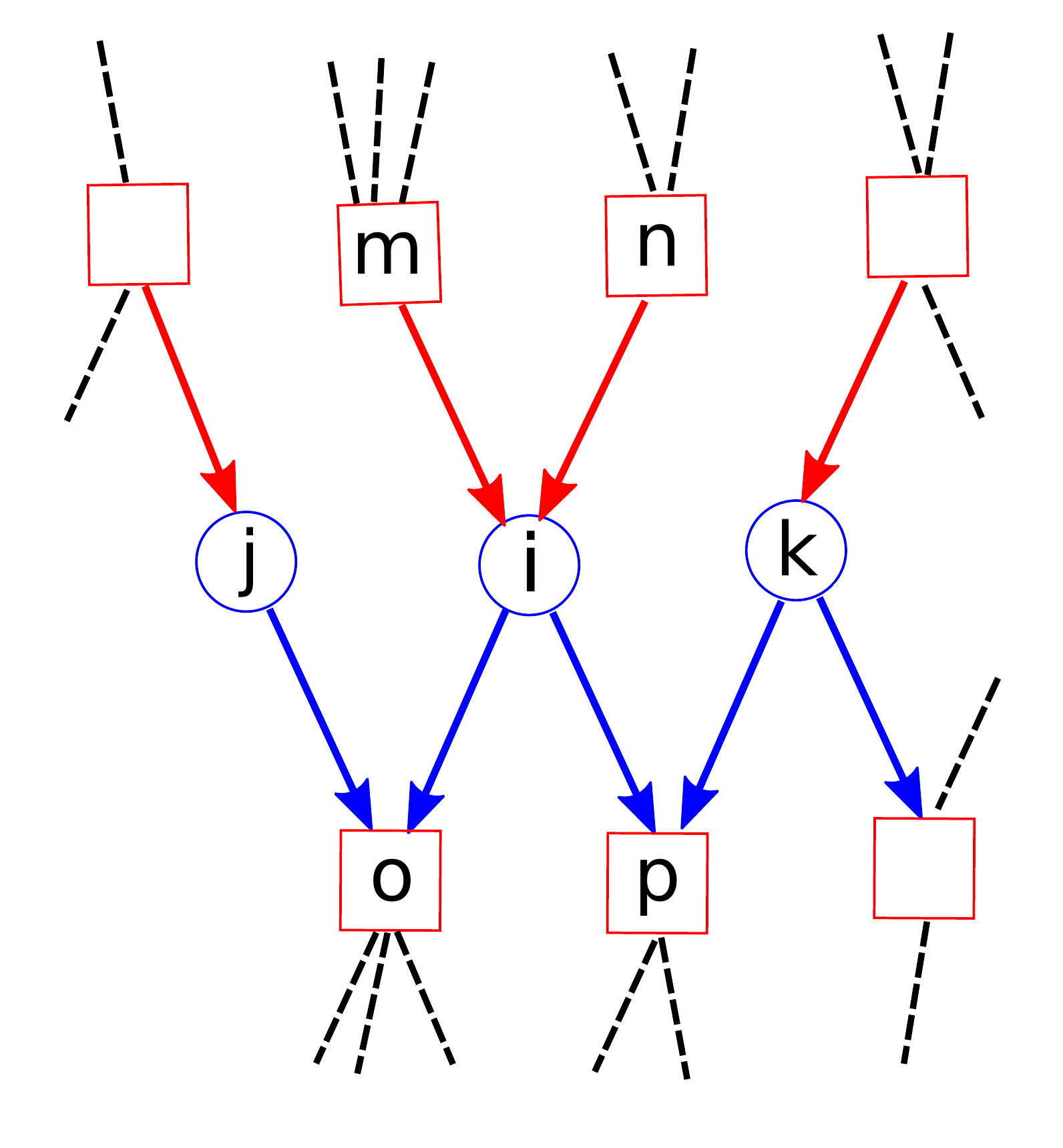}%0.25
\caption{Sketch of a random reaction network of the type discussed in the text. $m,n,o$ and $p$ denote metabolites (squares), $i,j$ and $k$ are instead reactions (circles).  Red (resp. blue) links carry substrate-like (resp. product-like) couplings with $\xi_i^m=-1$ (resp. $\xi_i^p=1$).   \label{bip_sketch}}
\end{figure}

For the reaction nodes, the quantities $|\partial i_{\inn}|\equiv d_\inn $ and $|\partial i_{\outt}|\equiv d_{\outt}$ will be assumed to be independent random variables, both distributed according to
\begin{equation}
\mathit{D}_R(d)= q\delta_{d,2}+(1-q)\delta_{d,1}\;,
\label{degree_R}
\end{equation} 
where $0\leq q\leq 1$ is a parameter. In other words, reactions can be of four different types according to their in- and out-degrees ($(d_{\inn},d_{\outt}) \in \{(1,1),(1,2),(2,1),(2,2)\}$) and $q$ weights the relative number of bi-component reactions (as inputs, outputs or both). The only structural control parameters that we shall use in the following are the mean degrees of metabolites ($\lambda$) and of reactions ($q$). A sketch of the network is given in Fig. \ref{bip_sketch}.
In order to construct the random network, it is important to notice that the parameters of the system and $N$ and $M$ are related by
\begin{equation}
\label{rel_N_M_q_lam}
(1+q)N=\lambda M,
\end{equation}
or $N=\lambda M/(1+q)$.

\begin{figure}
\center
\includegraphics[scale=0.35]{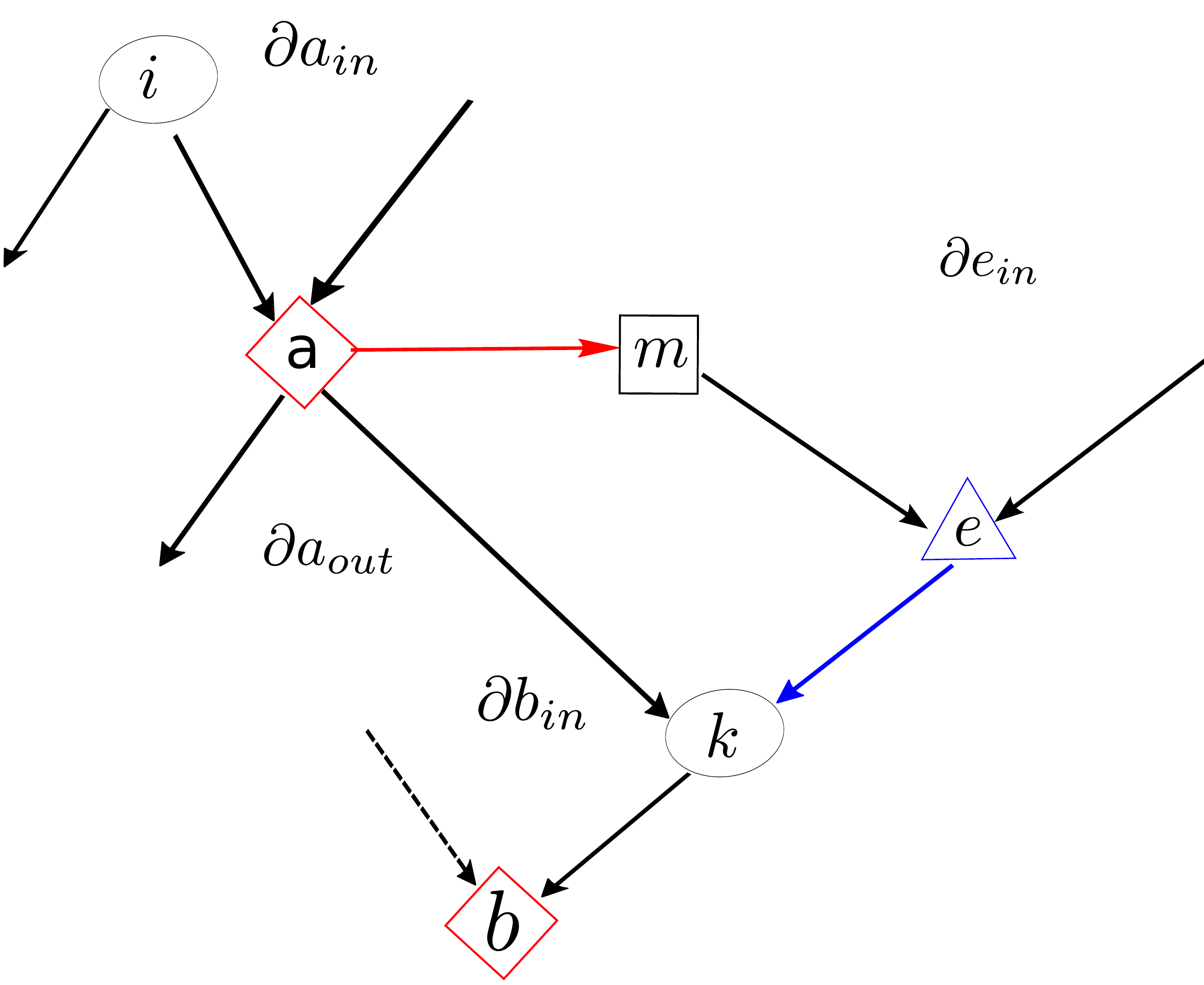}%0.25
\caption{Sketch of the factor graph representation of a RRN. Here squares represent metabolite variables, circles reaction variables, hexagons metabolite-constraints and triangles reaction-constraints.\label{im:factor_graph_sketch}}
\end{figure}

In order to compute the equations for the system it is useful to refer to the factor graph corresponding to the problem under consideration. A factor graph is a representation of the bipartite network with two classes of nodes, one representing variables (reaction activities and metabolite availabilities, in our case) and the other representing constraints (the so-called factor nodes). (Fig. \ref{im:factor_graph_sketch} displays a portion of the factor graph together with the messages that are exchanged between nodes in the message-passing procedure constructed below.)

In the following we will use letters $a,b,..$ for the metabolite constraints and $e,f,..$ for the reaction constraints. Furthermore we introduce the condensed notations: if $a$ is the constraint of metabolite $m$, then $\partial a^R=\partial a \backslash m$ is the set of reactions involving metabolite $m$; if $e$ is the constraint of reaction $i$, then $\partial e^M=\partial e \backslash i$ is the set of metabolites involved in reaction $i$. Moreover by dividing in two groups reactions producing and consuming a given metabolite, we call $\partial a^R_i$ the set of reactions in the same group as $i$, excluding $i$, and $\partial a^R_{\neg i}$ the opposite group.

Using this notation it is possible to rewrite the constraints (\ref{constraint_rea}) and (\ref{constraint_VN}) introduced in Section \ref{sec:boolean_problem} for the Soft-MB case in a form taking into consideration the factor graph representation, that will be useful in the computation of the equations of the system:
\begin{gather}
 \Gamma_a(\mu_m,\nu_{\neigRea}) =\delta_{\mu_m,0}\prod_{\neigIn}(1-\nu_j)+\delta_{\mu_m,1}(1-\prod_{\neigIn}(1-\nu_j))\;, 
\nonumber\\ \label{constraint_VN_final}\\ \nonumber
 \Delta_e(\mu_{\neigMeta},\nu_i) =\delta_{\nu_i,0}+\delta_{\nu_i,1}\prod_{n \in \partial e^M}\mu_n\;,
\end{gather}
where we have substituted $\delta_{x_m,0}=\prod\limits_{j \in \partial a^R_{in}}(1-\nu_j)$, with $a^R_{in}$ being the set of reactions producing the metabolite whose constraint is $a$.

In the same way, for the Hard-MB case, the constraints in the new notation can be written as:
\begin{gather}
  \Gamma_a(\mu_m,\nu_{\neigRea}) =\delta_{\mu_m,0}\prod_{\neigIn}(1-\nu_j)\prod_{\neigOut}(1-\nu_j)+\delta_{\mu_m,1}(1-\prod_{\neigIn}(1-\nu_j))(1-\prod_{\neigOut}(1-\nu_j))\;, \nonumber\\ \label{constraint_FBA_final} \\ \nonumber
  \Delta_e(\mu_{\neigMeta},\nu_i) =\delta_{\nu_i,0}+\delta_{\nu_i,1}\prod_{n \in \partial e^M}\mu_n\;.
\end{gather}

%%% Local Variables: 
%%% mode: latex
%%% TeX-master: "thesis"
%%% End: 

\chapter{Statistical Mechanics methods}
\label{sec:stat_mech}
\section{Constraint Satisfaction Problem}
\label{sec:CSP}

\begin{figure}
  \center
  \includegraphics[scale=0.35]{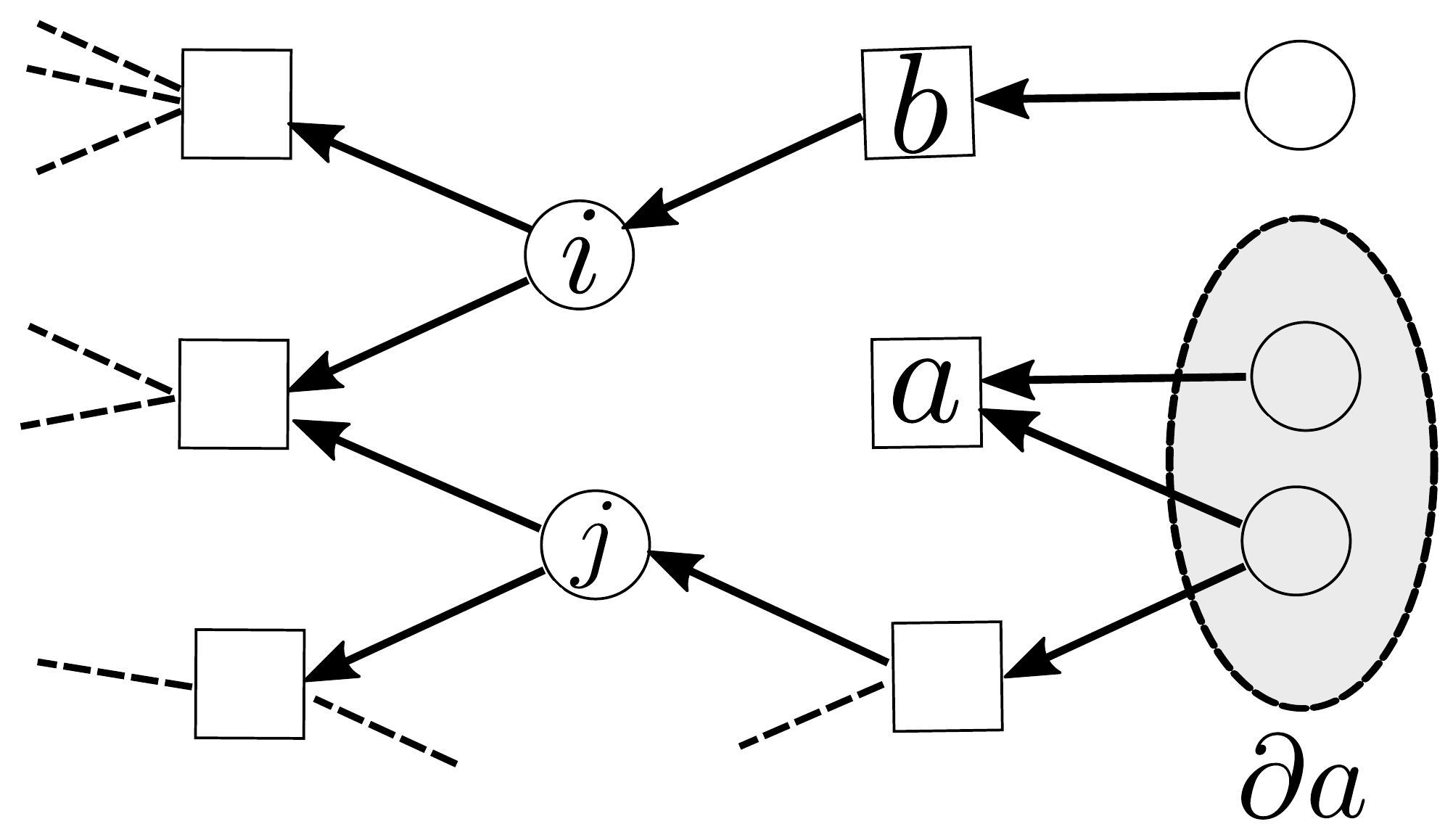}
  \caption{Sketch of the factor graph representation of a CSP.}
  \label{fig:sketch_fg_generic}
\end{figure}

A constraint satisfaction problem (CSP) is a class of problems in which a set of variables have to satisfy constraints. In Statistical Physics and in Computer Science such problems are considered as a set of $N$ variables with states $(\sigma_1,...,\sigma_N)$ and a set of $M$ constraints $\psi_a(\sigma_{\partial a})$, where $\partial a$ represent the variables involved in constraint $a$ and where $\psi_a(\sigma_{\partial a})=1$ if $\sigma_{\partial a}$ satisfy the constraint, 0 otherwise. It is possible to define also a ``soft'' version of the CSP where the constraints can be violated. In this case a weigth of the form $\psi^{\beta}_a(\sigma_{\partial a})=e^{-\beta(1-\psi_a(\sigma_{\partial a}))}$ can be defined. See Figure \ref{fig:sketch_fg_generic} for a factor graph representation of a generic CSP. 

In a CSP it is generally convenient to write the Hamiltonian of the system as $H=\sum_{a}\left(1-\psi_a(\sigma_{\partial a})\right)$ and the partition function as:
\begin{equation}
Z=\sum_{\{\sigma\}}\prod_a \psi^{\beta}_a(\sigma_{\partial a})=\sum_{\{\sigma\}}\prod_a e^{-\beta (1-\psi_a(\sigma_{\partial a}))},  
\end{equation}
that is equal to the number of solution at $\beta\rightarrow\infty$.

When confronted with the study of a CSP the common approaches are: decision (understanding if one solution is possible), sampling (manage to sample the solutions according to a given distribution) or optimization (search for optimal configurations). In optimization the function that is generally optimized is the Hamiltonian defined above, thus finding a solution to an optimization problem in CSPs is equivalent to find the configurations for which all constraints are satisfied (for $\beta\rightarrow \infty$) or the number of satisfied constraints is maximal (for finite $\beta$).

In the theory of computational complexity \cite{cook1971complexity,moore2011nature}, CSPs are classified depending on the resources needed to study these problems. Problems for which the worst case can be solved in polynomial time are in $P$, while problems for which the answer can be only \textit{verified} in polynomial time are called $NP$ (nondeterministic polynomial time). Then a problem is considered NP-complete if, taken a problem $p \in$ NP, any other NP problem can be transformed into $p$ in polynomial time. A long lasting problem is to understand if NP (or NP-complete) problems can be solved in polynomial time \cite{cook2006p}. This is very important because as we saw, if a polynomial algorithm is able to find a solution of an NP-complete problem then all NP problems could be solved polynomially. Generally many problems can be P, NP or NP-complete depending on the value of the parameters of the system. For example two typical problems that are NP-complete only for $k$ or $q \geq 3$ are:
\begin{itemize}
\item k-satisfiability (k-SAT). In a generic satisfiability problem, variables take boolean values ($0$ or $1$) and constraints are formed as boolean expressions written using AND, OR and NOT (to negate a variable). A $k$-satisfiability is the same problem in which all constraints involve exactly $k$ variables.
\item q-coloring (q-COL). In this problem each variable can take one of $q$ possible colors and the constraint assume that two neighbouring vertices cannot take the same color.
\end{itemize}

\section{Random graphs}
\label{sec:random_graphs}
A graph is a collection of {\it nodes} that are connected by {\it links} with its topology encoded in the adjacency matrix, $A$, such that each element $A_{ij}$ is $1$ if there is a link between node $i$ and node $j$ and $0$ otherwise. The graphs has no geometrical representation as the nodes are not spatially organized. Then a random graph is defined as a network whose topology is generated by a stochastic process. The concept of random graph has first been introduced by P. Erd\"os and A. R\'eyni \cite{erdHos1959random} and independently by Gilbert \cite{gilbert1959random}.

For simplicity in the following we will restrict ourselves to the case of random graphs with a fixed number of vertices $N$. Depending on the process chosen for the assignment of the links, each graph $G$ has a probability $p(G)$ of being generated at the end of the process and these probabilities define an {\it ensemble} of random graphs. We will use in the following the notation $\mathbf{G}_{N,M}$ for the ensemble of random graphs with $N$ vertices and $M$ links. 

The mean of a generic quantity $\alpha(G)$ on this ensemble is given by $\sum p(G)\alpha(G)$ where the sum is intended as on all graphs with $N$ vertices. The number of graphs with $N$ vertices and $M$ links can be computed by considering that the number of possible links between $N$ vertices is $n_L=N(N-1)/2$ and between these it is possible to choose $M$ links in ${n_L \choose M} = \frac{n_L!}{M!(n_L-M)!}$ ways.

It is possible to construct a random graph in $\mathbf{G}_{N,M}$ by starting with a network with $N$ vertices and no links and assigning each link randomly. The only rules to follow in this case are: each link must connect distinct vertices (no self-links) and any pair of vertices can be connected at most once (no repeated couples). It is important to notice that in the thermodynamic limit ($N\rightarrow \infty$) the ensemble $\mathbf{G}_{N,M}$ is equivalent to the ensemble $\mathbf{G}_{N,p}$ where each link has a probability $p=2M/N(N-1)$ of being present. We will thus in the following consider that $N$ is finite but $N\gg1$ (thus graphs in $\mathbf{G}_{N,p}$) in order to compute the  properties of random graphs as calculations are easier.

The probability that a given vertex has a degree $d=k$ is given by:
\begin{equation}
  \label{eq:rg_degree_prob}
  P(d=k)= {N-1 \choose k} p^k (1-p)^{N-1-k},
\end{equation}
where ${N-1 \choose k}$ is the probability of choosing $k$ neighbouring vertices in the remaining $N-1$, $p^k$ is the probability that $k$ vertices exist and $(1-p)^{N-1-k}$ is the one that the remaining $N-1-k$ vertices are not present. An interesting particular case of the $G_{N,p}$ ensemble is the case of random graphs with fixed mean degree $\cc$. In this case we have $M=\cc N/2$, $p=\cc/(N-1)$ and
\begin{equation}
  \label{eq:rg_degree_prob_fixed_mean}
  P(d=k)=\lim_{N\rightarrow \infty}\frac{(N-1)!}{k!(N-1-k)!} \left(\frac{\cc}{N-1}\right)^k \left(1-\frac{\cc}{N-1}\right)^{N-1-k}=\frac{\cc^k}{k!}e^{-\cc},
\end{equation}
that is a Poissonian distribution. This type of distribution has the characteristic that the probability of having a vertex with degree $d \gg \cc$ decreases exponentially.

In random graphs typical loops have size $\log(N)$ and this is simply showed by choosing a vertex $i$ on a random graph with fixed connectivity $k$. Then it is possible to define a distance $d$ from this vertex as: the neighbours of $i$ are at $d=1$, the neighbours of the neighbours of $i$ are at $d=2$ and so on. So the probability that a loop has size $d$ is equal to the probability that at distance $d$, $i$ is extracted again or:
\begin{equation}
  \label{eq:3}
  p_{L}(d)=\frac{k^d}{N},
\end{equation}
where $d>0$ as no self links exist. Thus the size of a typical loop in a random graph is equal the distance at which $p_{L}(d)=\mathit{O}(1)$ i.e. for which $d_{L}=\log(N)$.

The structure of random graphs has been thoroughly studied \cite{gilbert1959random,erdHos1959random,bollobas2001random}: for $\cc < 1$ the typical graph is a forest, i.e. a union of trees, then as $\cc$ grows some trees merge together but the typical graphs is acyclic, finally at $\cc=1$ the first cycles are formed and for $\cc > 1$  a phase transition occurs in which a sizeable fraction of vertices is in the same connected component, hence called {\it giant component}.

\begin{figure}
\center
\includegraphics[scale=0.35]{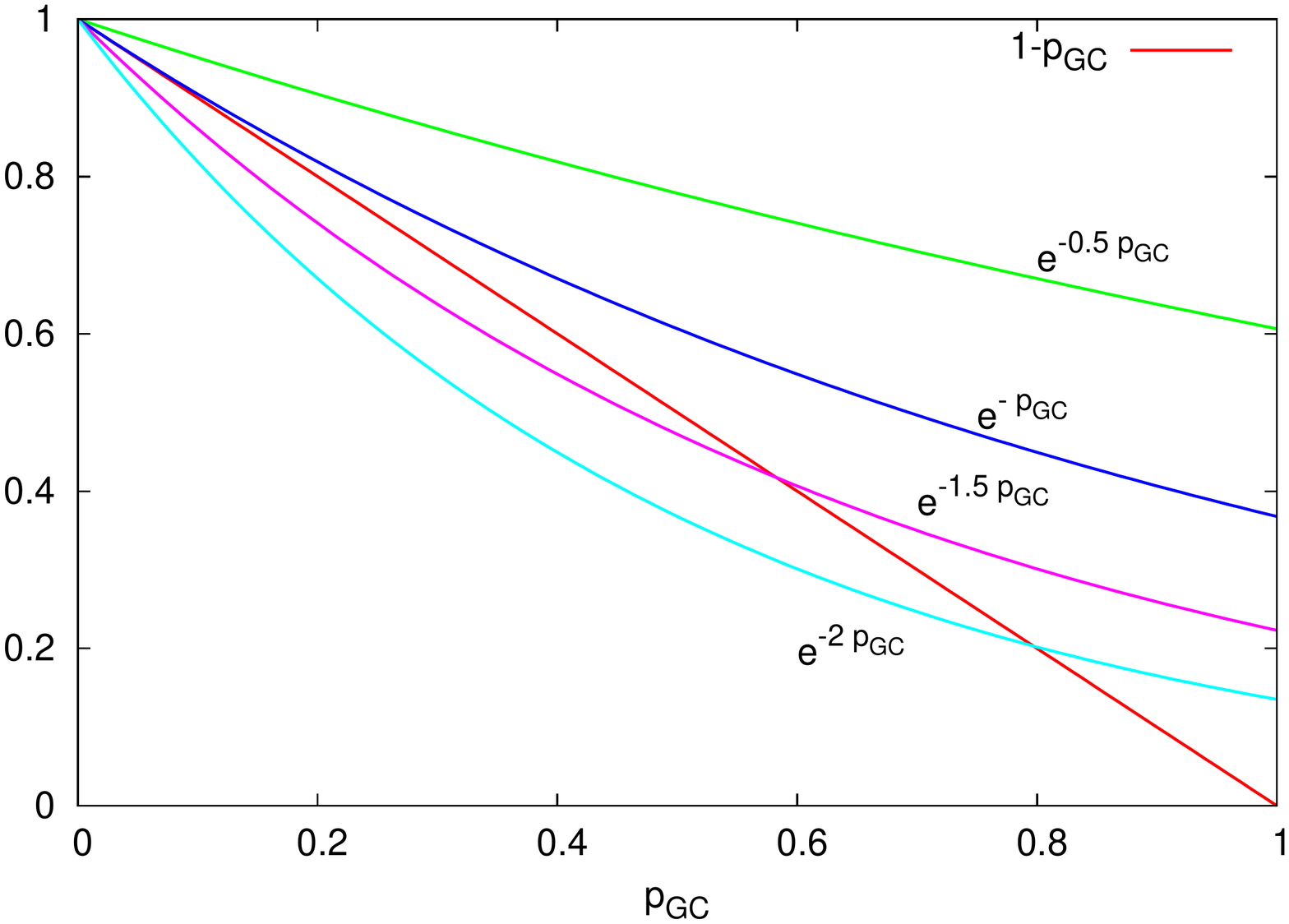}%0.25
\caption{Plot of the r.h.s. and l.h.s. of equation (\ref{eq:rg_proba_m}) for various values of $\cc$.   \label{fig:trascendent_sol}}
\end{figure}

It is possible to derive formally all of these structural characteristics by studying the probability that a vertex belong to the giant component, $p_{GC}$. This will be done in the following trying to make an intuitive derivation more than a rigorous one. It is possible to write:
\begin{equation*}
  1-p_{GC}=\sum_{k=0}^{\infty}e^{-\cc}\frac{\cc^k}{k!}(1-p_{GC})^k,
\end{equation*}
because $1-p_{GC}$ is the probability that the new vertex is {\bf not} in the giant component, thus this is equal to the probability that this vertex has $k$ neighbours ($e^{-\cc}\frac{\cc^k}{k!}$) and that neither of these neighbours is in the giant component ($(1-p_{GC})^k$). From this equation we obtain:
\begin{equation}
  \label{eq:rg_proba_m}
  1-p_{GC}=e^{-\cc p_{GC}}.
\end{equation}
This equation admit always the solution $p_{GC}=0$ but for $\cc>1$ there is also a solution for $p_{GC}>0$ that suggest the presence of a giant component. This equation has no analytical solution but it is simple to find a numerical solution by searching the interesection points between $1-p_{GC}$ and $e^{-\cc p_{GC}}$ as shown in Figure \ref{fig:trascendent_sol}. From this graph it is clear that for $\cc<1$ no solution to this equation exist while for $\cc>1$ there is a giant component. Hence random graphs with fixed mean degree undergo a phase transition in the critical point $\cc_p=1$.

As random graphs are used to analyze more complicated systems, the topology of the graph can become substantially different from the one we presented. It is possible to construct a bipartite random graph, where there are two type of vertices and edges connect only vertices of a different type. Another type of graph generally used in CSPs is the factor graph, that in its simplest form is a bipartite graph where one vertex is a variable (with indexes $i,j,..$) and the other vertex is a function node (constraint, with indexes $a,b,...$) (for a detailed description see Section \ref{sec:RRN}). A random factor graph is a factor graph in which the links between variables and constraints are assigned at random. It is though important to notice that the properties presented in this section hold also on these other types of random graphs.

\section{Cavity Method}
\label{sec:cavity_met}

The cavity method is a method by which it is possible to estimate the free energy and the magnetization of a system \cite{Parisi_cavity,mezard1988spin}. Although introduced to study the Sherrington Kirkpatrick model \cite{kirkpatrick1978infinite}, this method has provided theoretical and practical tools to analyze many other problems. It is in principle equivalent to the replica method developed in the study of spin glasses \cite{kirkpatrick1978infinite,mezard1988spin} but it gives a more intuitive picture of which is the correlation structure in the system.

In the following we will consider that the graph under study is a generic random factor graph. In general, for the following assumptions to be true, it is important that the graph is locally tree-like. Indeed this is always true in a random network where typical loops are of size $\log(N)$ (see Section \ref{sec:random_graphs}). In the cavity method we imagine of removing a node, computing all the interesting quantities (``cavity'' quantities) and then put the node back in. It is thus possible to use the ``cavity'' information to recover the complete information about the system.

Another way of seeing this problem is by considering that each variable send to its neighbours a message that represent the ``belief'' that the variable has about the state of its neighbours. This is the typical picture in Computer Science and is known as Belief Propagation (BP) \cite{Pearl_BP}. In \cite{kabashima1998belief,yedidia2005constructing} it has been shown the equivalence between the two approaches. Therefore in the following we will use equivalently cavity equations or BP equations.

%In \ref{subsec:bethe_approx} we will present a brief argument that gives the idea of what is the relationship between the Bethe approximation and the cavity method (or BP), this will then be useful to understand what are the limitation of the cavity approach as explained in \ref{subsec:pure_states}.

%In \ref{subsec:algoritmic} we will briefly present how the algorithm to find cavity solutions is functioning. Finally in \ref{subsec:1d_ising} and \ref{subsec:q-col} we will review two problems for which it is possible to compute the cavity equations. Indeed \ref{subsec:q-col} is interesting because of its structure in the solution space from which it is possible to explain why the approach is limited in some region of the parameters.
The cavity method can be represented formally considering that each variable send to its neighbours a ``message'', $m_{i\rightarrow j}(\sigma_i)$, that is the marginal probability law (cavity marginal) of $\sigma_i$ in a graph where the link $<i,j>$ is not present. In a generic factor graph two types of messages are possible $m_{i\rightarrow a}$ (from variable to constraint) and $m_{a \rightarrow i}$ (from constraint to variable). Nevertheless in a first part we will consider only one type of message and then we will generalize for generic factor graphs. In this case, $\psi_a(\sigma_{\partial a})=\psi_{ij}(\sigma_i,\sigma_j)$ (see Section \ref{sec:CSP}) and it is possible to write (for a full derivation see Section \ref{subsec:bethe_approx}):
\begin{equation}
  \label{eq:generic_cavity_eq}
m_{i\rightarrow j}(\sigma_i)=\frac{1}{z_{i\rightarrow j}}\prod_{k \in \partial i \setminus j} \left( \sum_{\sigma_k}m_{k\rightarrow i}(\sigma_k)\psi_{ik}(\sigma_i,\sigma_k) \right)\;,
\end{equation}
where:
\begin{equation}
  \label{eq:1}
 z_{i\rightarrow j}=\sum_{\sigma_i}\prod_{k \in \partial i \setminus j} \left( \sum_{\sigma_k}m_{k\rightarrow i}(\sigma_k)\psi_{ik}(\sigma_i,\sigma_k) \right).
\end{equation}
Clearly all variables with in or out degree $0$ (the leaves of the graph) have $m_{i\rightarrow j}=const$.

From the cavity marginal it is straightforward to recover the marginal of the variable by adding back the removed link, obtaining
\begin{equation}
  \label{eq:4}
  m_i(\sigma_i)=\frac{1}{z_i} \prod_{k \in \partial i}\left( \sum_{\sigma_k} m_{k\rightarrow i} \psi_{ik}(\sigma_i,\sigma_k)\right)\;,
\end{equation}
where $z_i=\sum\limits_{\sigma_i} \prod\limits_{k \in \partial i}\left(\sum\limits_{\sigma_k} m_{k\rightarrow i} \psi_{ik}(\sigma_i,\sigma_k)\right)$. Furthermore the magnetization of the variables can be recovered as
\begin{equation}
  \label{eq:generic_cavity_marginal}
  \avg{\sigma_i}=\sum_{\sigma_i}m_i(\sigma_i)\sigma_i.
\end{equation}
It is also possible to compute the free energy of the system by understanding that the complete partition function of the system can be written as:
\begin{equation}
  \label{eq:partition_function_generic_cavity}
  Z=z_i\prod_{j\in \partial i}z_{j \rightarrow i}\prod_{k \in \partial j \setminus i}z_{k\rightarrow j}...=z_i \prod_{j\in \partial i}\frac{z_j}{z_{ij}}\prod_{k\in \partial j \setminus i}\frac{z_k}{z_{jk}}...=\frac{\prod_i z_i}{\prod\limits_{<i,j>}z_{ij}},
\end{equation}
where we have defined:
\begin{equation}
  \label{eq:6}
  z_{ij}=\sum_{\sigma_i,\sigma_j}m_{j\rightarrow i}(\sigma_j)m_{i\rightarrow j}(\sigma_i)\psi_{ij}(\sigma_i,\sigma_j).
\end{equation}

Hence the free energy can be written as:
\begin{equation}
  \label{eq:free_energy_generic_cavity}
  F=-T\log(Z)=\sum_i f_i -\sum_{<i,j>}f_{ij},  
\end{equation}
where $f_i=-T \log(z_i)$ and $f_{ij}=-T \log(z_{ij})$. Thus by this derivation it is clear that the free energy of a generic graph can be divided in two contributions: $f_i$ coming from the variables and $f_{ij}$ coming from the links. 

\begin{figure}[h]
  \center
  \includegraphics[scale=1]{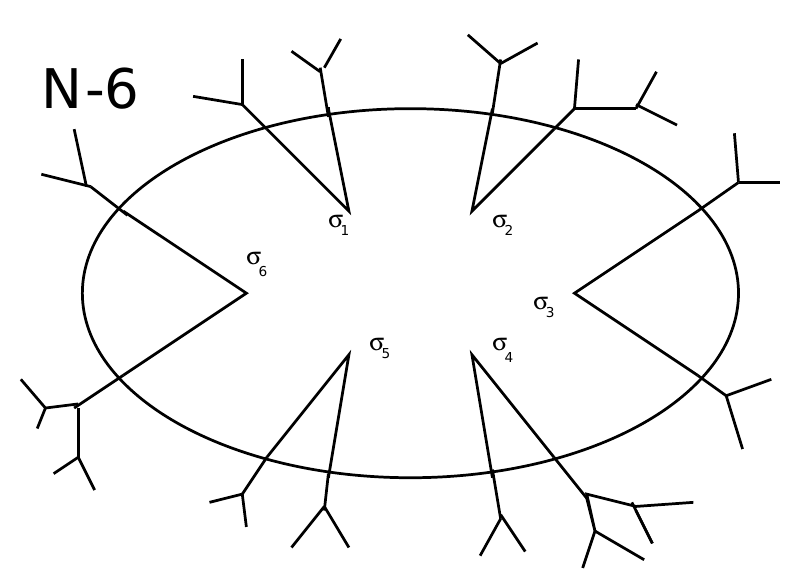}
  \caption{Representation of a $\mathcal{G}_{N,6}$ cavity graph on a lattice with $k=2$. Hence $q=6$ spins have $2$ neighbours and all other $N-6$ remaining spins have $3$ neighbours (taken from \cite{Parisi_cavity}).}
  \label{fig:cavity_graph}
\end{figure}

It is instructive to compute the free energy at zero temperature on a Bethe lattice with fixed $k+1$ connectivity \cite{Parisi_cavity} as in this case the computation is simpler because the relevant variable is the internal energy $U=\lim\limits_{N\rightarrow \infty}E_N/N$. In this case one can define a ``cavity'' graph $\mathcal{G}_{N,q}$ of $q$ randomly chosen spins that have only $k$ neighbours while the remaining $N-q$ spins all have $k+1$ neighbours (see Figure \ref{fig:cavity_graph}). Furthermore the $q$ cavity spins are fixed to values $\sigma_1,..,\sigma_q$. 

The complete energy of the system correspond to the $\mathcal{G}_{N,0}$ graph but we will see that it is possible to compute it by understanding the properties of the $\mathcal{G}_{N,q}$ graphs. On these graphs it is possible to:
\begin{itemize}
\item (iteration) add a new cavity spin $\sigma_0$ thus changing a $\mathcal{G}_{N,q}$ graph to a $\mathcal{G}_{N+1,q-k+1}$ graph ($\delta N=1$,$\delta q= -k+1$)
\item (link addition) add a link between two randomly chosen cavity spins and then optimizing its spin value; in this case we are transforming a $\mathcal{G}_{N,q}$ graph to a $\mathcal{G}_{N,q-2}$ graph ($\delta N = 0$ and $\delta q=-2$).
\item (site addition) add a new spin $\sigma_0$, connect it to $k+1$ cavity spins and then optimizing the values of the $k+2$ spins; in this case we are transforming a $\mathcal{G}_{N,q}$ graph to a $\mathcal{G}_{N+1,q-k-1}$ graph ($\delta N = 1$ and $\delta q=-k-1$).
\end{itemize}

Thus if we start with a $\mathcal{G}_{N,2(k+1)}$ cavity graph, we can obtain a $\mathcal{G}_{N,0}$ graph by doing $k+1$ link addition. Furthermore, starting from the same graph and doing $2$ site addition it is possible to obtain a $\mathcal{G}_{N+2,0}$ cavity graph. Therefore the variation of the energy going from $N$ to $N+2$ spins is related to the average energy shift $\Delta E^{S}$ for a site addition and $\Delta E^{L}$ for link addition as:
\begin{equation}
  \label{eq:cavity_link_site}
  E_{N+2}-E_{N}=2\Delta E^{S}-(k+1)\Delta E^{L}.
\end{equation}
Finally, using the fact that the total energy is asymptotically linear in N, the energy density for the ground state is:
\begin{equation}
  \label{eq:cavity_ground_state_T0}
  U=\lim\limits_{N\rightarrow \infty}E_N/N=\frac{E_{N+2}-E_{N}}{2}=\Delta E^{S}-\frac{k+1}{2}\Delta E^{L}.
\end{equation}
\section{A simple application: 1D Ising Model}
\label{sec:1d_ising}

\begin{figure}
  \center
  \includegraphics[scale=0.2]{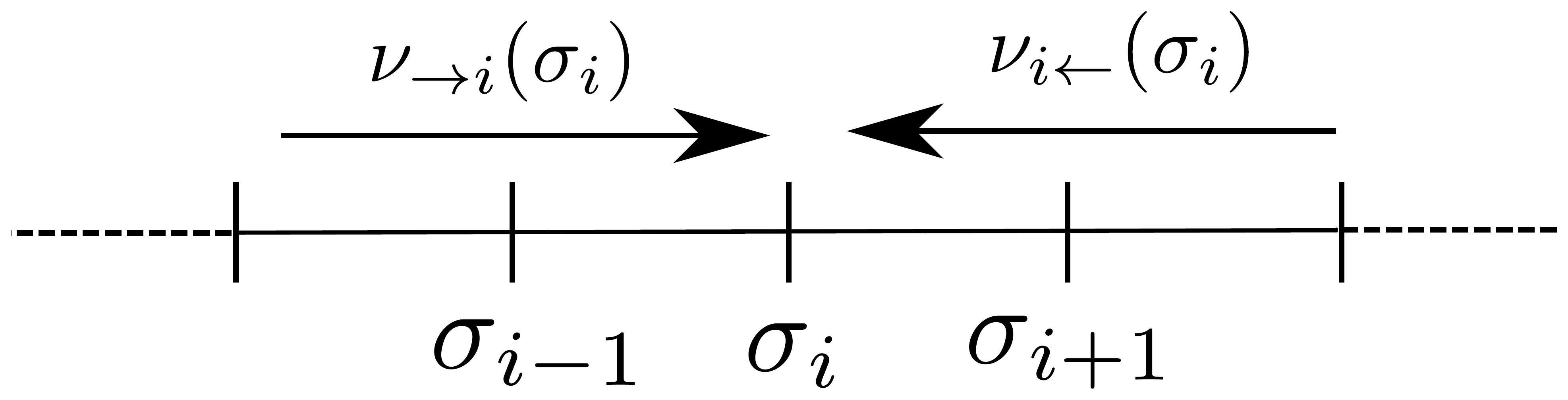}
  \caption{Sketch of the 1-d Ising model under consideration}
  \label{fig:ising1d}
\end{figure}
We would like to present now a straightforward application of the cavity method to the one-dimensional Ising model. This example should clarify how it is possible to apply the cavity method to practical cases. Consider a ferromagnetic Ising model on a line with $N$ spin variables, $\underline{\sigma}=(\sigma_1,...,\sigma_N)$, such that $\sigma_i\in\{+1,-1\}$, see Figure \ref{fig:ising1d}. In this case, we can write:
\begin{equation}
  \label{eq:ising_chain_prob}
  p_{\beta}(\underline{\sigma})=\frac{1}{Z}e^{-\beta E(\underline{\sigma})} \;,\; E(\underline{\sigma})=-\sum_{i=1}^{N-1}\sigma_i\sigma_{i+1}-h\sum_{i=1}^N \sigma_i.
\end{equation}

In a model of this kind, the marginal of a variable $i$ is given by:
\begin{equation*}
p_i(\sigma_i)=\sum_{\sigma_k, k \neq i} e^{-\beta E(\underline{\sigma})}/Z,
\end{equation*}
where $Z=\sum_{\underline{\sigma}}e^{-\beta E(\underline{\sigma})}$. This marginal is easily rewritten as
\begin{equation*}
p_{i}(\sigma_i)=\frac{Z_{\rightarrow i}Z_{i \leftarrow}}{Z}\;\nu_{\rightarrow i}(\sigma_i)~e^{-\beta h \sigma_i}~\nu_{i \leftarrow}(\sigma_j),
\end{equation*}
where 
\begin{eqnarray}
\nonumber
 & \nu_{\rightarrow i}(\sigma_i)&=\sum\limits_{\sigma_1,..,\sigma_{i-1}}\exp\left( \beta \sum\limits_{k=1}^{i-1} \sigma_{k}\sigma_{k+1} + \beta h \sum\limits_{k=1}^{i-1}\sigma_k \right)/Z_{\rightarrow i},\\\label{eq:ising_chain_cavity_marg} \\ \nonumber
& \nu_{i \leftarrow}(\sigma_i)&=\sum\limits_{\sigma_{i+1},..,\sigma_N}\exp\left( \beta \sum\limits_{k=i+1}^N \sigma_{k-1}\sigma_{k} + \beta h \sum\limits_{k=i+1}^N\sigma_k \right)/Z_{i \leftarrow},
\end{eqnarray}
and $Z_{\rightarrow i}=\nu_{\rightarrow i}(-1) + \nu_{\rightarrow i}(+1)$ (and the same for $_{i \leftarrow}$).

It is thus possible to interpret $\nu_{\rightarrow i}$ as the distribution of $\sigma_i$ in a graphical model in which all factor nodes adjacent to $i$ except the one on its left are removed. It is now possible to write some recursive equations for the messages as
\begin{eqnarray}
\nonumber 
  \nu_{\rightarrow i+1}(\sigma)&=\sum_{\sigma'}\nu_{\rightarrow i}(\sigma') e^{\beta \sigma' \sigma + \beta h \sigma'}/Z_{\rightarrow i},\\   \label{eq:ising_chain_cavity_eq}\\ \nonumber   
  \nu_{i+1 \leftarrow}(\sigma)&=\sum_{\sigma'}\nu_{i \leftarrow}(\sigma') e^{\beta \sigma' \sigma + \beta h \sigma'}/Z_{i \leftarrow},
\end{eqnarray}
The boundaries have in this case $\nu_{\rightarrow 1}=\nu_{i \leftarrow}=1/2$. Thus starting from the boundary and evolving messages using equation (\ref{eq:ising_chain_cavity_eq}) until convergence it is possible in linear time to compute the marginal $p(\sigma_i)$. As all messages are distributions over binary variables, it is possible to parametrize it with a single real number
\begin{equation*}
  u_{\rightarrow i}=\frac{1}{2\beta}\log\frac{\nu_{\rightarrow i}(+1)}{\nu_{\rightarrow i}(-1)},
\end{equation*}
that can be seen as an local magnetic field as $\nu_{\rightarrow i}=e^{\beta u_{\rightarrow i}\sigma}$.

As we have seen the one dimensional Ising model is simple to treat but contains most of the concept used in the cavity method. We saw that it is possible to write the marginal in function of cavity marginals $\nu_{\rightarrow i}$ and $\nu_{i \leftarrow}$ and that these quantities satisfy some recursive equations (\ref{eq:ising_chain_cavity_eq}). From these equations it is thus possible to recover the marginal of each variable in the system. This approach can clearly be further refined, as it is possible to compute the free energy and the correlations inside the system, but this is probably out of the scope of this thesis.
\section{Bethe approximation and the cavity approach}
\label{subsec:bethe_approx}

This paragraph is based on \cite{Yedidia} where classical methods in Statistical Mechanics are presented from an interesting point of view for our purpose. Indeed in this paragraph we will show that the solutions to the cavity (BP) equations describe the system in the Bethe approximation \cite{bethe1935statistical,peierls1936statistical,onsager1936electric}. Thus using this result it will be possible in Section \ref{subsec:pure_states} to justify why and in which system the cavity method can fail.

We want to find the probability distribution $P(\underline{\sigma})$ of a system of $N$ elements with states $\underline{\sigma}=(\sigma_1,...,\sigma_N)$ ($\sigma_i=\{-1,1\}$). In this case, it is possible to write the Gibbs free energy as
\begin{equation}
G(P(\underline{\sigma}))= U -T S,
\end{equation}
where the internal energy, $U$, and the entropy, $S$ can be computed as
\begin{eqnarray}
&U &= \sum_{\underline{\sigma}}P(\underline{\sigma})E(\underline{\sigma}), \\
&S &= -\sum_{\underline{\sigma}} P(\underline{\sigma})\ln P(\underline{\sigma}).
\end{eqnarray}
For a physical system in equilibrium, the Gibbs free energy is minimal. Hence using a Lagrangian in which the constraint on the probability $P(\underline{\sigma})$ is enforced via a lagrangian multiplier $\lambda$,
\begin{equation}
\mathit{L}= G(P(\underline{\sigma}))-\lambda \left(\sum_{\underline{\sigma}}P(\underline{\sigma})-1\right),
\end{equation}
it is possible to find the $P(\underline{\sigma})$ for which $G$ is minimum. This is
\begin{equation}
P(\underline{\sigma})=\frac{\exp(-E(\underline{\sigma})/T)}{Z},
\end{equation}
or the Boltzmann distribution as expected for a generic physical system. Furthermore substituting the value of $P(\underline{\sigma})$ in $G(P)$ we see that at equilibrium the Gibbs Free energy is equal to the Helmotz free energy, $F\equiv -T\ln Z$. 

This is a trivial statement in Statistical Mechanics, but the procedure is useful: if the form of the energy and the correlation structure of the system are known (or assumed), then it is possible to compute the equations the probability has to satisfy. Hence with this procedure it is possible to compute the Gibbs free energy of a system using a reasonable approximation (the belief) to joint probability between the variables. Then minimizing the free energy it is possible to find the equations that the belief has to satisfy in this approximation.

%it is possible to state a problem as: find the approximate joint  probability (the belief) of the variables that best describe the system then compute the Gibbs free energy with this probability. Thus minimizing the free energy it is possible to find the equations that the belief has to satisfy in this approximation.

For example for an Ising model:
\begin{eqnarray}
E(\underline{\sigma})=-\sum_{(i,j)}J_{ij}\sigma_i \sigma_j-\sum_i h_i\sigma_i,
\end{eqnarray}
it is possible to find the equations that the system has to satisfy in \textit{Mean Field Approximation} \cite{weiss1907hypothese,curie1894symetrie} in which the variables of the system are independent, or:
\begin{equation}
  P(\underline{\sigma})=\prod_i b_i(\sigma_i).
\end{equation}
Given that the $\sigma_i$ are bimodal, the only way to write the $b_i$'s is:
\begin{equation}
b_i(\sigma_i)=\frac{1+\mu_i\sigma_i}{2}.
\end{equation}
Therefore:
\begin{equation}
G=-\sum_{(ij)} J_{ij} \mu_i \mu_j -\sum_i h_i \mu_i  + T \sum_i\left[ \frac{1+\mu_i}{2} \ln \left(\frac{1+\mu_i}{2}\right) + \frac{1-\mu_i}{2} \ln \left(\frac{1-\mu_i}{2}\right) \right].
\end{equation}
Minimizing with respect to $\mu_i$ one obtain the self consistency equations:
\begin{equation}
\mu_i=\tanh\left( \frac{\sum_j J_{ij} \mu_j + h_i}{T}\right),
\end{equation}
that are the classical equations for Ising in Mean Field.

\begin{figure}
  \center
  \includegraphics[scale=0.3]{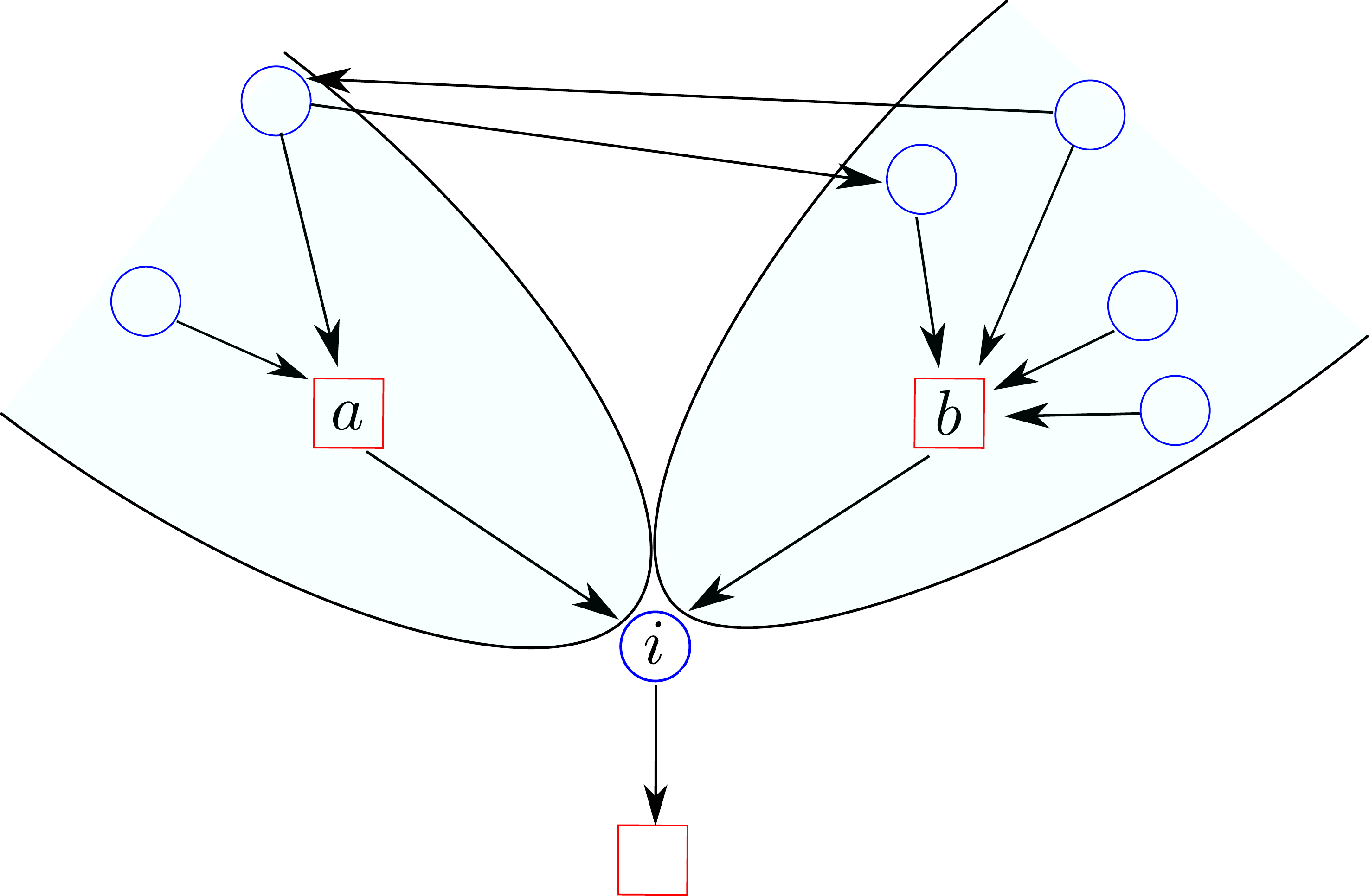}
  \caption{Sketch of the messages sent in part of a factor graph. In this situation, the Bethe Approximation consist in considering that the messages arriving from vertices $a$ and $b$ are not correlated although it is clear that this is not true.}
  \label{fig:bethe_approx}
\end{figure}

Finally it is possible to derive the equations in the {\it Bethe Approximation} \cite{bethe1935statistical,peierls1936statistical,onsager1936electric} in which only the first neighbours are correlated (see sketch in Figure \ref{fig:bethe_approx}), hence:
\begin{equation}
P(\underline{\sigma})=\prod_{(ij)}b_{ij}(\sigma_i,\sigma_j)\prod_i b_i(\sigma_i)^{1-q_i},
\end{equation}
this approximation is actually exact on a tree-like topology, where $q_i$ is the degree of node $i$.

Then again it is possible to write:
\begin{equation}
S=-\sum_{ij}\sum_{\{\sigma_i,\sigma_j\}}b_{ij}(\sigma_i)\ln(b_{ij}(\sigma_i,\sigma_j)) - \sum_i (1-q_i) \sum_{\{\sigma_i\}} b_i(\sigma_i) \ln(b_i(\sigma_i)),
\end{equation}
and
\begin{equation}
H=-\sum_{ij} \sum_{\{\sigma_i,\sigma_j\}}b_{ij}(\sigma_i,\sigma_j)(J_{ij}(\sigma_i,\sigma_j)+h_i(\sigma_i)+h_j(\sigma_j))-\sum_i (1-q_i) \sum_{\{\sigma_i\}} b_i(\sigma_i) h(\sigma_i),
\end{equation}
where the first term is the energy given by the links and the second term is the excess of energy consequence of the fact that each node is counted more than once. Having this it is straightforward to write the Gibbs Free Energy and using the lagrange multipliers, $\lambda_i$ and $\lambda_{ij}$, to enforce the conditions on the probabilities:
\begin{eqnarray}
&\sum\limits_{\{\sigma_i\}} b_i(\sigma_i)&=1, \nonumber\\ \nonumber\\ \label{eq:bethe_normalization_cond}
&\sum\limits_{\{\sigma_i,\sigma_j\}} b_{ij}(\sigma_i,\sigma_j)&=1,  \\ \nonumber\\\nonumber 
&\sum\limits_{\{\sigma_j\}} b_{ij}(\sigma_i,\sigma_j)&=b_i(\sigma_i),
\end{eqnarray}
the equations that the beliefs should satisfy are obtained or
\begin{eqnarray}
& b_i(\sigma_i)&=\frac{1}{Z_i}\exp\left[ -\frac{E_i(\sigma_i)}{T}+\frac{\sum_j \lambda_{ij}(\sigma_i)}{T(q_i-1)}\right], \nonumber \\   \label{eq:bethe_belief_equations} \\ \nonumber
& b_{ij}(\sigma_i,\sigma_j)&=\frac{1}{Z_{ij}}\exp\left[- \frac{E_{ij}(\sigma_i,\sigma_j)}{T}+\frac{\lambda_{ij}(\sigma_i)}{T}+\frac{\lambda_{ij}(\sigma_j)}{T}\right],
\end{eqnarray}
where $Z_i$ and $Z_{ij}$ are constants that enforce the normalization conditions and where $E_i(\sigma_i)=-h_i(\sigma_i)$ and $E_{ij}(\sigma_i,\sigma_j)=-J_{ij}(\sigma_i,\sigma_j)-h_i(\sigma_i)-h_j(\sigma_j)$.   Finally by using the normalization condition (3rd of equations (\ref{eq:bethe_normalization_cond})) it is possible to write a self consistent equation for the lagrangian multipliers $\lambda_{ij}$:
\begin{equation}
  \label{eq:bethe_lagrangian_mult}
  \frac{1}{Z_i}\exp\left[ -\frac{E_i(\sigma_i)}{T}+\frac{\sum_j \lambda_{ij}(\sigma_i)}{T(q_i-1)}\right]= \sum_{\sigma_j}\frac{1}{Z_{ij}}\exp\left[- \frac{E_{ij}(\sigma_i,\sigma_j)}{T}+\frac{\lambda_{ij}(\sigma_i)}{T}+\frac{\lambda_{ij}(\sigma_j)}{T}\right].
\end{equation}
Using this equation it is possible to find the $\lambda_{ij}$ that minimize the Gibbs Free Energy thus solving the problem. In \cite{yedidia2000generalized} authors showed that
\begin{equation}
  \label{eq:bethe_lagrangian_BP_relation}
  \lambda_{ij}=T\ln \prod_{k\in \partial j \setminus i}m_{k\rightarrow j}(\sigma_j),
\end{equation}
where $m_{k\rightarrow j}$ is defined by:
\begin{equation}
  \label{eq:generic_cavity_eq}
m_{i\rightarrow j}(\sigma_i)=\frac{1}{z_{i\rightarrow j}}\prod_{k \in \partial i \setminus j} \left( \sum_{\sigma_k}m_{k\rightarrow i}(\sigma_k)\psi_{ik}(\sigma_i,\sigma_k) \right)\;,
\end{equation}
where:
\begin{equation}
  \label{eq:1}
 z_{i\rightarrow j}=\sum_{\sigma_i}\prod_{k \in \partial i \setminus j} \left( \sum_{\sigma_k}m_{k\rightarrow i}(\sigma_k)\psi_{ik}(\sigma_i,\sigma_k) \right).
\end{equation}
Thus we showed that solving the cavity (BP) equations is equivalent to search for the minimum of the Gibbs free energy in a system in which the Bethe approximation hold. This is a very important result as it can help to understand how and why these equations are failing in some systems.

\begin{figure}[h]
  \center
  \includegraphics[scale=0.3]{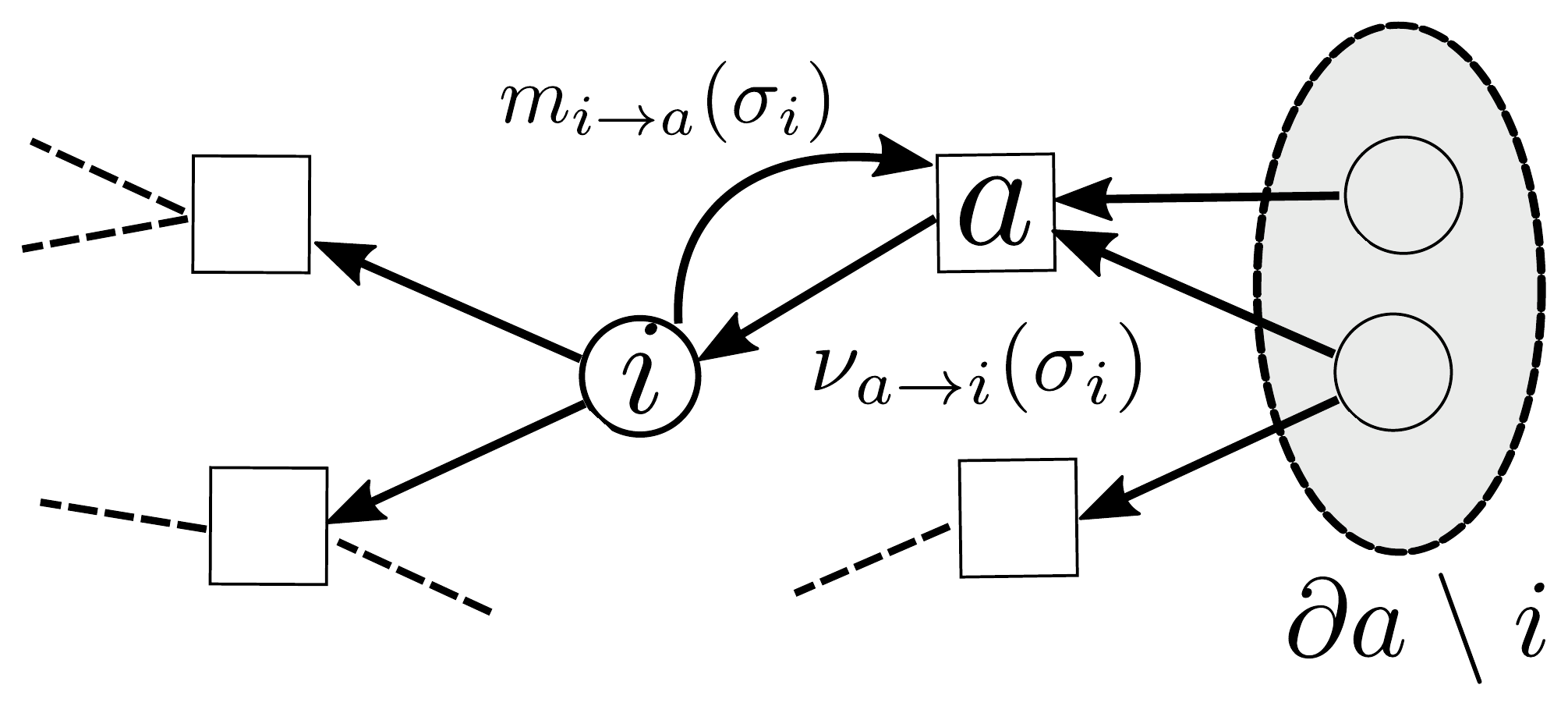}
  \caption{Sketch of the cavity messages in a factor graph}
  \label{fig:cavity_mess_fg}
\end{figure}

In cases in which it is not possible to make the simplification of removing the factor node (see Figure \ref{fig:cavity_mess_fg}), equation (\ref{eq:generic_cavity_eq}) is written as
\begin{eqnarray}
\nonumber 
& m_{i \rightarrow a}(\sigma_i)&=\prod_{b \in \partial i \setminus a} \nu_{b \rightarrow i}(\sigma_i)/Z_{i \rightarrow a},
\\  \label{eq:generic_cavity_eq_factor_graph} \\ \nonumber
&\nu_{a \rightarrow i}(\sigma_i)&=\sum_{\underline{\sigma}_{\partial a} \setminus i}\psi_a(\underline{\sigma}_{\partial a})\prod_{k \in \partial a \setminus i} m_{k \rightarrow a}(\sigma_k)/Z_{a \rightarrow i}.
\end{eqnarray}
As before it is possible to define the quantities $Z_i=\sum\limits_{\sigma_i}\prod\limits_{b \in \partial i} \nu_{b \rightarrow i}(\sigma_i)$, $Z_a=\sum\limits_{\underline{\sigma}_{\partial a}}\psi_a(\underline{\sigma}_{\partial a})\prod\limits_{k \in \partial a} m_{k \rightarrow a}(\sigma_k)$ and $Z_{ai}=\sum\limits_{\sigma_i}m_{i \rightarrow a}\nu_{a \rightarrow i}$.

Using these definitions, the free energy (\ref{eq:free_energy_generic_cavity}) can be written as
\begin{equation}
  \label{eq:generic_cavity_free_energy_factor_graph}
  F=-T\log(Z)=\sum_i f_i+ \sum_a f_a -\sum_{<i,a>}f_{ia},    
\end{equation}
where
\begin{eqnarray*}
  &f_i&=-T \log(Z_i) ,\\ 
  &f_a&=-T \log(Z_a), \\ 
  &f_{ia}&=-T \log(Z_{ia}).
\end{eqnarray*}
Hence still the free energy is divided between a contribution given by the nodes (factor node $f_a$ and variable node $f_i$) and a contribution given by the links ($f_{ia}$).

\section{Algorithmic point of view}
\label{sec:algoritmic}
From an algorithmic point of view, the cavity method can be solved by using a Belief Propagation (BP) algorithm. As already presented in Section \ref{sec:cavity_met} in the BP approach, cavity marginals are considered as the ``beliefs'' that a variable has about the state of its neighbours. Following this logic, it is possible to write an algorithm in which the beliefs satisfying the cavity equations (see Section \ref{sec:cavity_met}) are computed iteratively until convergence. Thus at convergence it is possible to obtain the marginal on any variable of the system, $\avg{m_i}$. As presented in Section \ref{subsec:bethe_approx}, the BP picture is equivalent to what in physics is called the Bethe Approximation where neighbours to variable $x$ are considered indipendent between each other when $x$ is removed. Hence this approximation is valid until no long range correlation is present, i.e. in systems with few short loops or not undergoing a phase transition (this will be discussed in detail in Section \ref{subsec:pure_states}). Nevertheless it has been observed that BP is able to converge also in systems in which these assumptions do not hold \cite{GBP_2D}. In these cases it has to be checked if the approximate results still are describing the properties of the system.

In many problems in which BP is applied, it is observed that at convergence of the algorithm a fraction of the variables has a higly polarized marginal, thus being ``frozen'' variables. The size of this fraction depend on which part of the phase space we are exploring (see Section \ref{sec:q-col} for a better understanding). This property has been exploited to develop a {\it Warning Propagation} (WP) algorithm \cite{Braunstein_2005_survey} in which each constraint sends to its variable a message $0$ if the variable has to be switched off, $1$ if the variable has to be on and $*$ if it cannot constrain the variable. This is equivalent of considering a cost function that counts the {\it number} of violated constraints. Using WP the information given by the variables frozen during BP iterations is propagated, thus using these two algorithms (WP+BP) together it is possible to improve the convergence of BP \cite{RicciT_Semerjian,ksatJSTAT}.

When it is important to recover the {\it configurations} of the system that satisfies the equations, it is possible to resort to a Decimation Procedure \cite{RicciT_Semerjian}. In this procedure first the BP marginals are computed, then the most biased marginals are set according to the marginal, thus the ``hard'' information is propagated by WP and finally the process is repeated until all variables are set or a contradiction is found.

If one is interested in studying the solutions of the cavity equations for typical graphs extracted from a given ensemble, it is possible to use a Population Dynamics Algorithm (POPDYN). In this algorithm the network is constructed (following the degree distribution of the ensemble) while the algorithm is running. Thus in this way, it is possible to study the equations for a population of variables in the ensemble under study. Hence at convergence the marginal of the variables averaged over the ensemble, $\overline{\avg{m}}$, is obtained, where $\overline{\bullet}$ denotes the average over the probability distribution of the degrees.

All the algorithm used, specified to the CSPs presented in Section \ref{sec:boolean_problem} are presented in Appendix \ref{chap:algorithms}.

\section{Failure of the cavity approach}
\label{subsec:pure_states}
As already explained in Section \ref{subsec:bethe_approx}, the cavity equations are derived under a Bethe approximation. Clearly there are situations in which this approximation is strongly wrong and in these cases the equations can in principle have no solutions and the algorithm could not converge. In this Section we would like to revise the possible causes for which the algorithm will not work.

As already presented in the latter Section, in the Bethe approximation it is assumed there is a unique state of the system or that if we remove variable $x$ from the system, its neighbours are indipendent between each other. This assumption do not hold in cases in which the topology of the graph has many short loops or when the phase space of the system can be divided in many ``pure'' states \cite{mezard1988spin}. In these cases the messages that neighbours are sending are not independent hence the BP equations are not exact. Nevertheless in many cases in which these things are not true, BP algorithm reaches a fixed point.

The problem of the division of the phase state in many pure states is well known in Statistical Physics as it has been treated in the study of spin glasses \cite{mezard1988spin}. A spin glass is a system of spin variables where the interaction terms are assigned at random. The cavity method has been developed to deal with these problems and it is understood that the cavity equations correspond to the Replica Symmetric (RS) phase of the system. In this phase only one pure state exist and thus the cavity equations are valid. However this approximation is not always holding because in these systems it is possible to have what is called a Replica Symmetry Breaking (RSB). In this situation, an exponential number of pure states is present in the system, hence bringing to the failure of the RS approach. Nevertheless using a step by step breaking of the symmetry of the replicas it is possible to find the equations also in this part of the phase space. Following this ideas, the authors of \cite{mezard2001bethe}, developed a cavity method at 1-RSB step with which it is possible to find solutions (in population dynamics, see Section \ref{sec:algoritmic}) also in the phase in which standard cavity equations do not hold.

Another approach proposed to overcome this problem is the one in \cite{yedidia2000generalized}. In this approach the authors search for a finer approximation of the Gibbs Free Energy using the Kikuchi approximation \cite{kikuchi1951theory}. In this approximation the system is divided in overlapping clusters and then recombined trying to consider all the contributions given by the various clusters. Then using the ideas presented in \ref{subsec:bethe_approx} it is possible to find some ``generalized'' belief propagation equations that overcome some of the problems presented above.

\section{A complicated example: q-Col}
\label{sec:q-col}

The q-coloring problem has already been introduced in \ref{sec:CSP}: in a nutshell, q-col is the problem of assigning a color to variables of a given graph using $q$ possible colors such that no two neighbouring variables have the same color. This problem has been widely studied in both Computer Science and Statistical Physics \cite{bollobas2001random,achlioptas2005rigorous,mulet2002coloring,van2002random}. In this section we will briefly introduce the results obtained in \cite{zdeborova2007phase} but we recommend the interested reader to read the article for a complete derivation. The problem for the purpose of this thesis is interesting because it presents the typical characteristics of complex systems like spin glasses still being a CSP.

Consider a graph with $N$ vertices that can have states $s_i=1,..,q$ and $M$ edges, then it is possible to define an Hamiltonian in the form of a Potts model \cite{wu1982potts}
\begin{equation}
  \label{eq:hamiltonian_qcol}
  H(\{s\})=\sum_{(i,j)}\delta(s_i,s_j).
\end{equation}
Hence this Hamiltonian give a positive contribution if neighbours have the same colors while no contribution is present when neighbours have different colours. From this Hamiltonian it is possible to derive a Gibbs measure over the configurations as
\begin{equation}
  \label{eq:gibbs_meas_qcol}
  p(\{s\})=\frac{1}{Z_0} e^{-\beta H(\{s\})}.
\end{equation}
\begin{figure}
  \center
  \includegraphics[scale=0.9]{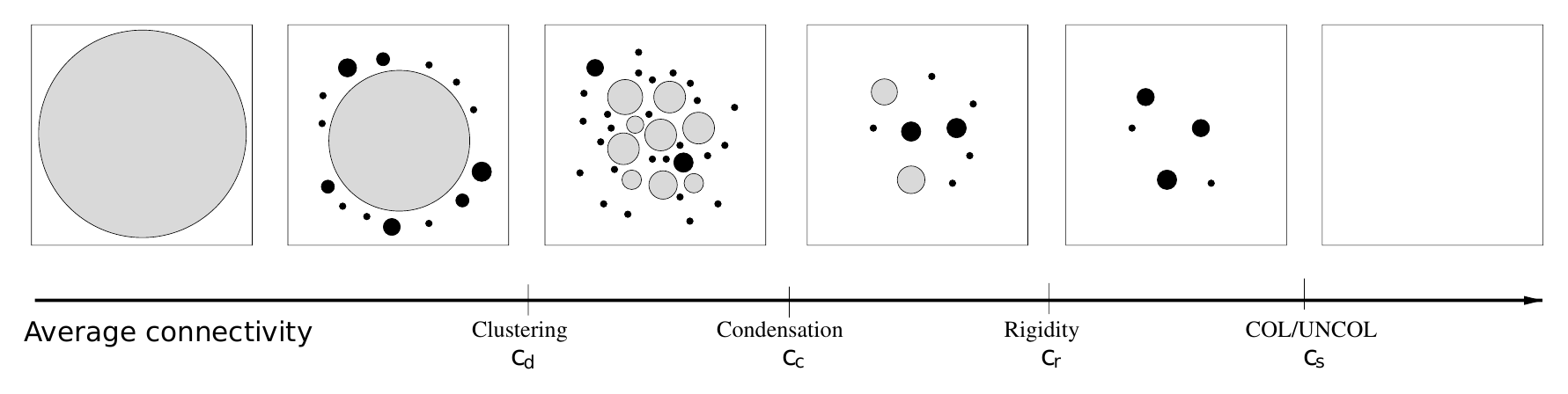}
  \caption{Sketch of the set of solutions in the q-col problem as the average connectivity, $c$, changes. Black clusters have all variables frozen while grey clusters no (taken from \cite{zdeborova2007phase}).}
  \label{fig:q-col}
\end{figure}

Using the cavity formalism presented in \ref{sec:cavity_met}, it is straightforward to derive the cavity equations:
\begin{equation}
  \label{eq:9}
\mess{m}{i}{j}_{s_i}=\frac{1}{\Z{i}{j}} \prod_{k\in i-j}\sum_{s_k} e^{-\beta \delta_{s_i,s_k}}\mess{m}{k}{i}_{s_i}= \frac{1}{\Z{i}{j}} \prod_{k \in i-j} \left(1-(1-e^{-\beta})\mess{m}{k}{i}_{s_i}\right).
\end{equation}
It is also possible to derive the free energy for this system by recovering the free energy contribution for the variables:
\begin{equation*}
  \Delta F^i=-T \log\left(\sum_s \prod_{k\in i}\left(1-(1-e^{-\beta})\mess{m}{k}{i}_{s_i} \right)\right),
\end{equation*}
and for the edges
\begin{equation*}
  \Delta F^{ij}=-T \log\left(1-(1-e^{-\beta})\sum_s \mess{m}{j}{i}_s \mess{m}{i}{j}_s\right),
\end{equation*}
thus obtaining:
\begin{equation*}
  f(\beta)=\frac{1}{N}\left(\sum_i \Delta F^i - \sum_{ij} \Delta F^{ij} \right).
\end{equation*}
As already discussed all this derivation is made under the Bethe approximation (or RS approximation). Nevertheless the q-col problem present a very complex behaviour when the average connectivity of the graph, $c$, varies. In Figure \ref{fig:q-col} the various phases in which the system undergoes are presented (taken from \cite{zdeborova2007phase}) where in black are represented the cluster of solutions with frozen variables (see Section \ref{sec:algoritmic}). From this Figure it is clear that for $c < c_d$ a large cluster of solutions exist and also if some cluster of solution appears, most of the solutions are still contained in the biggest one. This is the phase where BP converges without problem. Then in the region where $c_d < c < c_c$, many clusters of solutions are formed and also the giant cluster is divided in an exponential number of small one. In this region the BP algorithm starts to have problems, mainly due to the fact that messages coming from two neighbouring variables can be contradictory if in different clusters. From this point on the search for the solution become almost impossible (so far only the most probable unfrozen clusters have been sampled \cite{dall2008entropy}) and a condensed phase is formed where some cluster contain almost all of the solutions. When $c_r < c < c_s$, a rigidity transition is happening, where most of the variables become frozen. Finally for $c > c_s$ no coloring is possible.

\section{References}
For this Chapter we used many standard references. From \cite{Zamponi_note} the simplified description of the cavity method was taken, while from \cite{mezard2009information} we inspired ourselves for the example on the 1D Ising model. From \cite{Yedidia} we derived the relationship between BP and Bethe approximation and from \cite{krzkakala2008phase} we took the analysis on the q-Col. Throughout the Chapter we used \cite{mezard1988spin} as a base reference for spin glasses, cavity and replicas.

%%% Local Variables: 
%%% mode: latex
%%% TeX-master: "thesis"
%%% End: 

\chapter{Results on random reaction networks}
\label{sec:random_case}
\section{Cavity equations}
\label{cavity_derivation}
As already explained in Chapter \ref{sec:stat_mech}, CSPs as Soft-MB or Hard-MB, can be solved efficiently on random networks by the belief propagation algorithm \cite{Pearl_BP} or equivalently by the replica symmetric cavity method \cite{Parisi_cavity}. In this method, the marginal of a variable is computed by creating a ``cavity'' inside the system, removing a subpart of the network. Thus it is possible to obtain a ``cavity marginal'' and then reintroduce the variables removed. Finally the complete marginal of the variables follows directly from the cavity marginals.

From an algorithmic point of view, it is possible to search for the solutions to the cavity equations by an iterative procedure, in which ``messages'' are exchanged between variable and function nodes (see Section \ref{sec:algoritmic}). For the two CSPs defined in the latter Section, eight type of messages are required:  $\mess{\psi}{a}{m}_{\mu_m}$, $\mess{\psi}{m}{a}_{\mu_m}$, $\mess{\eta}{e}{m}_{\mu_m}$, $\mess{\eta}{m}{e}_{\mu_m}$, $\mess{\psi}{a}{i}_{\nu_i}$, $\mess{\psi}{i}{a}_{\nu_i}$, $\mess{\eta}{e}{i}_{\nu_i}$, $\mess{\eta}{i}{e}_{\nu_i}$. Each message represents the belief that a variable (function) has about its neighbouring function (variable) state. The messages can be divided in two classes: from function nodes to variable nodes and from variable nodes to function nodes. The first class of messages, e.g. $\mess{\psi}{a}{m}_{\mu_m}$, is the probability that metabolite $m$ is in state $\mu_m$ when there is only the function metabolite $a$. While the second class, e.g. $\mess{\psi}{m}{a}_{\mu_m}$, is the probability that metabolite $m$ is in state $\mu_m$ when the edge $(am)$ is not present. 

\begin{figure}
  \centering
  \includegraphics[width=8.5cm]{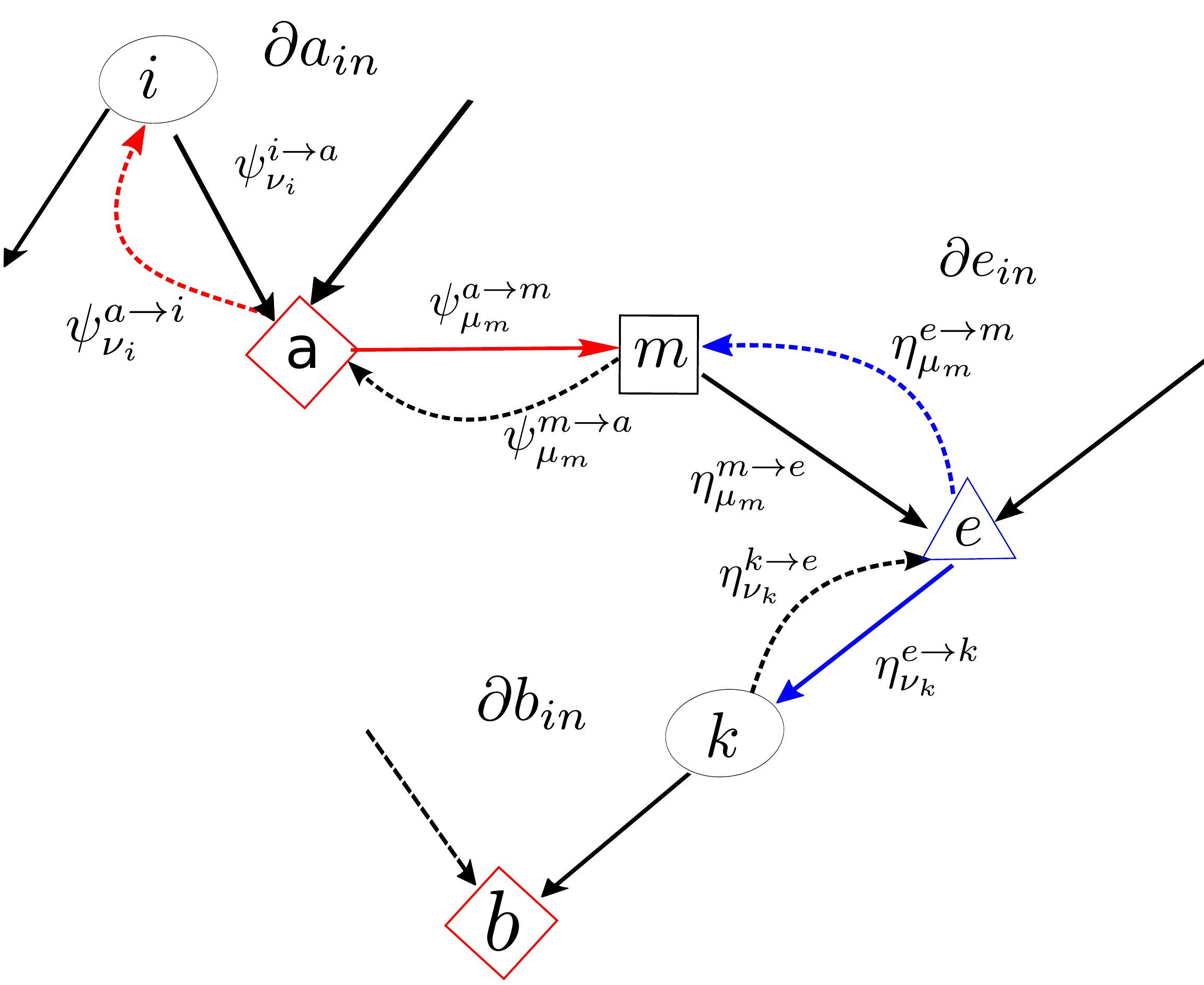}
  \includegraphics[width=8.5cm]{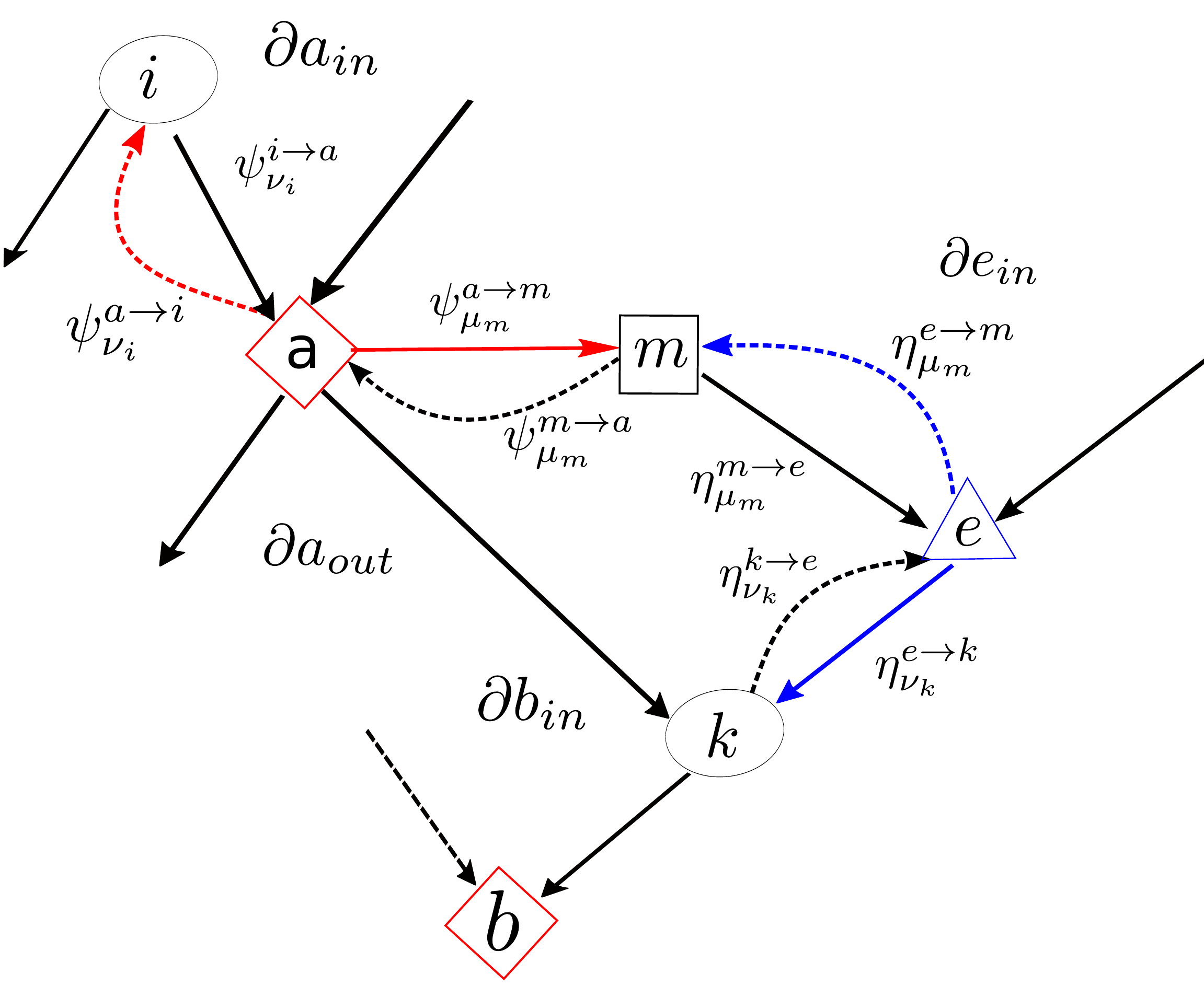}
  \caption{Summary of the cavity method messages for Soft-MB (left) and Hard-MB (right) constraints.   \label{schema_cavity}}
\end{figure}

The factor graph representation of our RRN is given in Figure \ref{schema_cavity} for Soft-MB and Hard-MB.

Introducing a parameter $\alpha$ to interpolate between Soft-MB ($\alpha=0$) and Hard-MB constraints ($\alpha=1$), the equations to be satisfied by the messages in the two CSPs under study can be written as follows:
\begin{align*}
 & \begin{cases}
    \mess{\psi}{m}{a}_{\mu_m}=\prod\limits_{f \in \neigRea[m]}\mess{\eta}{f}{m}_{\mu_m}/\Z{m}{a} \\ \\
    \mess{\psi}{a}{m}_{\mu_m}=\sum\limits_{\{\nu_j\}}\Gamma(\mu_m,\nu_{\neigRea})\prod\limits_{k\in \neigRea}\mess{\psi}{k}{a}_{\nu_k}/\Z{a}{m}
  \end{cases}
  \\ \nonumber \\
&  \begin{cases}
    \mess{\psi}{i}{a}_{\nu_i}=\mess{\eta}{e}{i}_{\nu_i}\left( \prod\limits_{b \in \neigMeta[i]_{in} \backslash a}\mess{\psi}{b}{i}_{\nu_i}\right)^\alpha \prod\limits_{b \in \neigMeta[i]_{out} \backslash a}\mess{\psi}{b}{i}_{\nu_i}/\Z{i}{a} \\ \\
    \mess{\psi}{a}{i}_{\nu_i}=\sum\limits_{\{\nu_j\},j\neq i}\sum\limits_{\mu_m}\Gamma(\mu_m,\nu_{\neigRea})\mess{\psi}{m}{a}_{\mu_m}\prod\limits_{k\in \neigRea \backslash i}\mess{\psi}{k}{a}_{\nu_k}/\Z{a}{i}
  \end{cases}
\end{align*}
\begin{align*}
&  \begin{cases}
    \mess{\eta}{i}{e}_{\nu_i}=\left(\prod\limits_{b \in \neigMeta[i]_{in}}\mess{\psi}{b}{i}_{\nu_i}\right)^\alpha\prod\limits_{b \in \neigMeta[i]_{out}}\mess{\psi}{b}{i}_{\nu_i}/\Z{i}{e}  \\ \\
    \mess{\eta}{e}{i}_{\nu_i}=\sum\limits_{\{\mu_n\}}e^{\theta\nu_i}\Delta(\nu_i,\mu_{\neigMeta})\prod\limits_{n \in \neigMeta}\mess{\eta}{n}{e}_{\mu_n}/\Z{e}{i}
  \end{cases} 
  \\ \nonumber \\
&  \begin{cases}
    \mess{\eta}{m}{e}_{\mu_m}=\mess{\psi}{a}{m}_{\mu_m}\prod\limits_{f \in \neigRea[m] \backslash e}\mess{\eta}{f}{m}_{\mu_m}/\Z{m}{e} \\ \\
    \mess{\eta}{e}{m}_{\mu_m}=\sum\limits_{\{\mu_n\},n\neq m}\sum\limits_{\nu_i}e^{\theta\nu_i}\Delta(\mu_{\neigMeta},\nu_i)\mess{\eta}{i}{e}_{\nu_i}\prod\limits_{n\in \neigMeta \backslash m}\mess{\eta}{n}{e}_{\mu_n}/\Z{e}{m}
  \end{cases} 
\end{align*}
The equations we have just written hold in Soft-MB or Hard-MB case, with the difference that in Soft-MB case reaction nodes are connected only to output metabolite functions, while in Hard-MB all metabolite functions are connected to reaction nodes. Another caution we have to take is that the reaction function node is connected only to the input metabolites [see equations (\ref{constraint_VN_final}) and (\ref{constraint_FBA_final})] regardless of the constraints used. Writing explicitly the constraints we can compute the cavity equations obtaining, for the metabolite constraints,
\begin{align*}
&  \begin{cases}
    \mess{\psi}{m}{a}_{\mu_m}=\prod\limits_{f \in \neigRea[m]}\mess{\eta}{f}{m}_{\mu_m}/\Z{m}{a} \\ \\
    \mess{\psi}{a}{m}_{\mu_m}=\left[\delta_{\mu_m,0}\prod\limits_{\neigIn}\mess{\psi}{j}{a}_0\left(\prod\limits_{\neigOut}\mess{\psi}{j}{a}_0\right)^{\alpha}+\delta_{\mu_m,1}\left(1-\prod\limits_{\neigIn}\mess{\psi}{j}{a}_0\right)\left(1-\prod\limits_{\neigOut}\mess{\psi}{j}{a}_0\right)^{\alpha}\right]/\Z{a}{m}
  \end{cases}
  \\ \nonumber \\
  &  \Z{a}{m}=\left(1-\prod_{\neigIn}\mess{\psi}{j}{a}_0\right)\left(1-\prod_{\neigOut}\mess{\psi}{j}{a}_0\right)^{\alpha}+\prod\limits_{\neigIn}\mess{\psi}{j}{a}_0\left(\prod\limits_{\neigOut}\mess{\psi}{j}{a}_0\right)^{\alpha}
  \\ \nonumber \\
&  \begin{cases}
    \mess{\psi}{i}{a}_{\nu_i}=\mess{\eta}{e}{i}_{\nu_i}\left(\prod\limits_{b \in \neigMeta[i]_{in} \backslash a}\mess{\psi}{b}{i}_{\nu_i}\right)^{\alpha}\prod\limits_{b \in \neigMeta[i]_{out} \backslash a}\mess{\psi}{b}{i}_{\nu_i}/\Z{i}{a} \\ \\
    \mess{\psi}{a}{i}_{\nu_i} = \left[\mess{\psi}{m}{a}_0(1-\nu_i)\prod\limits_{\neigGroupI}\mess{\psi}{j}{a}_0\left(\prod\limits_{\neigOppositeI}\mess{\psi}{j}{a}_0\right)^{\alpha}+\right.\\ \\
    \qquad\qquad\left. +\mess{\psi}{m}{a}_1\left(1-\prod\limits_{\neigOppositeI}\mess{\psi}{j}{a}_0\right)^{\alpha}\left((1-\prod\limits_{\neigGroupI}\mess{\psi}{j}{a}_0)+\nu_i\prod\limits_{\neigGroupI}\mess{\psi}{j}{a}_0\right)\right]/\Z{a}{i}
  \end{cases}
  \\ \nonumber \\
&\Z{a}{i}=\mess{\psi}{m}{a}_0\prod\limits_{\neigGroupI}\mess{\psi}{j}{a}_0\left(\prod\limits_{\neigOppositeI}\mess{\psi}{j}{a}_0\right)^{\alpha}+ \mess{\psi}{m}{a}_1 \left(1-\prod\limits_{\neigOppositeI}\mess{\psi}{j}{a}_0\right)^{\alpha}\left(2-\prod\limits_{\neigGroupI}\mess{\psi}{j}{a}_0\right)
\end{align*}
and, for the reaction constraints,
\begin{align*}
&  \begin{cases}
    \mess{\eta}{i}{e}_{\nu_i}=\left(\prod\limits_{b \in \neigMeta[i]_{in}}\mess{\psi}{b}{i}_{\nu_i}\right)^{\alpha}\prod\limits_{b \in \neigMeta[i]_{out}}\mess{\psi}{b}{i}_{\nu_i}/\Z{i}{e}  \\ \\
    \mess{\eta}{e}{i}_{\nu_i}=\left[\delta_{\nu_i,0}+e^{\theta}\delta_{\nu_i,1}\prod\limits_{n\in \neigMeta}\mess{\eta}{n}{e}_1\right]/\Z{e}{i} 
  \end{cases} 
  \\ \nonumber \\
& \Z{e}{i}=1+e^{\theta}\prod_{m \in \neigMeta}\mess{\eta}{m}{e}_1
  \\ \nonumber \\
&  \begin{cases}
    \mess{\eta}{m}{e}_{\mu_m}=\mess{\psi}{a}{m}_{\mu_m}\prod\limits_{f \in \neigRea[m] \backslash e}\mess{\eta}{f}{m}_{\mu_m}/\Z{m}{e} \\ \\
    \mess{\eta}{e}{m}_{\mu_m}=\left[\mess{\eta}{i}{e}_0+e^{\theta}\mess{\eta}{i}{e}_1\mu_m\prod\limits_{n\in \neigMeta \backslash m}\mess{\eta}{n}{e}_1\right]/\Z{e}{m}
  \end{cases}
  \\ \nonumber \\
& \Z{e}{m}=2\mess{\eta}{i}{e}_0+e^{\theta}\mess{\eta}{i}{e}_1\prod\limits_{n\in \neigMeta \backslash m}\mess{\eta}{n}{e}_1
\end{align*}

Using these equations, we can iterate until convergence the algorithms presented in Appendix \ref{sec:PopDyn_explanation}, finding solutions that satisfy the constraints and obtaining the cavity marginals. We can then compute the real marginals of the variable nodes as:
\begin{align}
& p(\mu_m)=\mess{\psi}{a}{m}_{\mu_m}\prod\limits_{f \in \neigRea[m]}\mess{\eta}{f}{m}_{\mu_m}/Z^m,
\nonumber \\ \label{proba_rel} \\ \nonumber
& p(\nu_i)=\mess{\eta}{e}{i}_{\nu_i}\left(\prod\limits_{b \in \neigMeta[i]_{in}}\mess{\psi}{b}{i}_{\nu_i}\right)^\alpha\prod\limits_{b \in \neigMeta[i]_{out}}\mess{\psi}{b}{i}_{\nu_i}/Z^i,
\end{align}
where:
\begin{align}
& Z^m=\sum_{\mu_m}\mess{\psi}{a}{m}_{\mu_m}\prod\limits_{f \in \neigRea[m]}\mess{\eta}{f}{m}_{\mu_m}.
\nonumber \\ \label{normalizations} \\ \nonumber
& Z^i=\sum_{\nu_i}\mess{\eta}{e}{i}_{\nu_i}\left(\prod\limits_{b \in \neigMeta[i]_{in}}\mess{\psi}{b}{i}_{\nu_i}\right)^\alpha\prod\limits_{b \in \neigMeta[i]_{out}}\mess{\psi}{b}{i}_{\nu_i},
\end{align}

The main assumption behind the cavity method is that the messages coming from two neighbouring nodes are independent. This clearly depends on the length of loops in the network: if the length of typical loops grows with the system size (as in RRN), then the above assumption can be valid, at least in the thermodynamic limit. As we can see clearly from Figure \ref{schema_cavity} short loops are not present in the Soft-MB, but they arise in the Hard-MB problem. If the assumption breaks down, then message passing algorithms may fail to converge. Though we will see in Section \ref{sec:BP_Decimation} that this is not the case here.

\section{Population dynamics}
\label{sec:pop_dyn}
In a nutshell, the setup presented in Chapter \ref{chap:CSP_for_FBA_VN} aims at retrieving Boolean patterns of activity of reactions (or of metabolite availabilities) induced, on network architectures defined by $q$ and $\lambda$, by the fact that a certain set of metabolites (nutrients) is available from the outset, which happens with probability $\rho_\inn$. Ideally, one would like to devise a method to sample configurations $(\boldsymbol{\nu}=\{\nu_i\},\boldsymbol{\mu}=\{\mu_m\})$ with a controlled probability given by
\begin{equation}
P\big(\boldsymbol{\nu},\boldsymbol{\mu}\big) \propto \prod_{m=1}^M \Gamma_m \prod_{i=1}^N \Delta_i e^{\theta \nu_i}\;,
\label{meas}
\end{equation}
which forbids states that don't satisfy all constraints. The `chemical potential' $\theta$ appearing above can be tuned externally in order to concentrate the measure around configurations with a different average fraction $N^{-1} \sum_i \avg{\nu_i}$ of active reactions, where angular brackets represent the average with respect to the measure (\ref{meas}). In order to find the configurations of reaction and metabolite variables that solve the above CSPs one may resort to statistical mechanics techniques. In particular, we have used the cavity method to derive the belief propagation (BP) equations (see Section \ref{sec:algoritmic} for details) and then employed a population dynamics algorithm (see Appendix \ref{sec:PopDyn_explanation} for details) in order to sample the corresponding solutions and, in turn, characterize the behaviour of the system in the typical case.

We shall concentrate here on the  scenario that emerges for different  $q$ and $\lambda$ upon varying two parameters, namely the chemical potential $\theta$ and the probability $\rho_{\inn}$ that nutrients are available. In specific, we have computed the average reaction activity and the average metabolite availability following two protocols:  first, by gradually reducing $\theta$ starting from a large, positive value, and, second, by doing the reverse. Averages obtained in these ways will be denoted, respectively, by $\overline{\avg{\cdots}}_+$ and $\overline{\avg{\cdots}}_-$. These averages (that we call magnetizations, using a statistical physics jargon) need not coincide, in which case the two quantities will display hysteresis when plotted against the chemical potential. Generally, the presence of hysteresis is a main characteristic of a discontinuous (first order) phase transition, while for continuous (second order) ones no hysteresis is observed, as also happens in cases where no phase transition takes place.

\subsection{Soft Mass-Balance}

\begin{figure}
  \includegraphics[width=17cm]{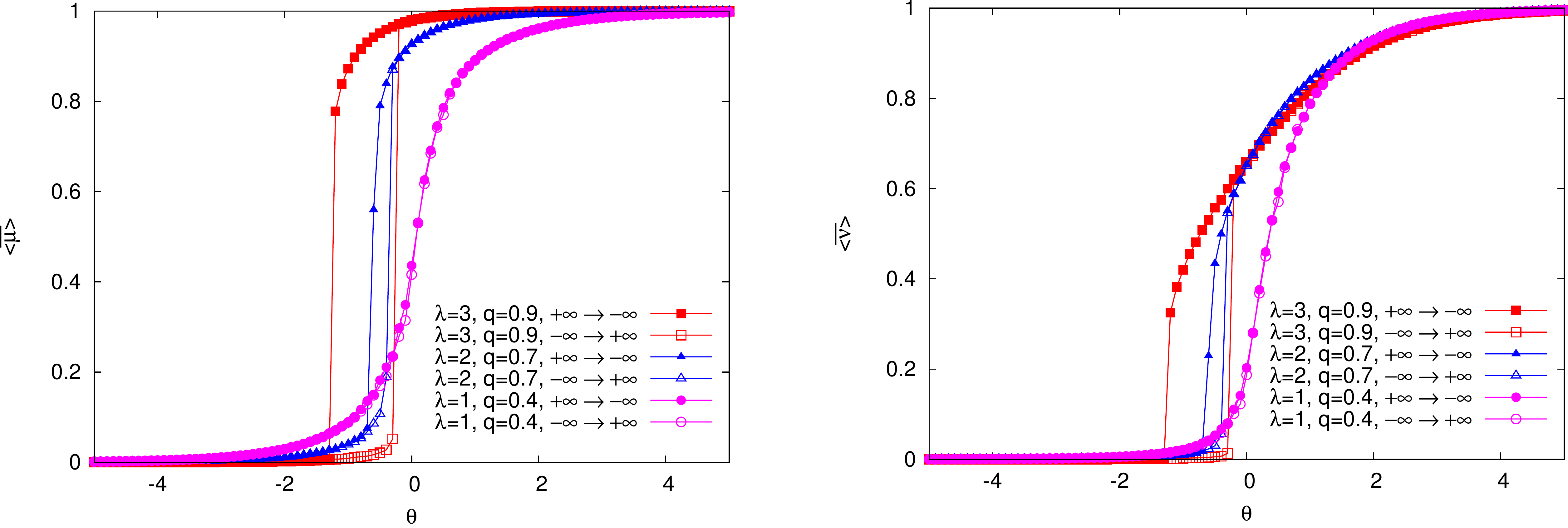}
  \caption{ \label{soft-MB_inTHETA}Soft-MB: behaviour of the average fraction of available metabolites, $\overline{\avg{\mu}}$ (left) and of  the average fraction of active reactions, $\overline{\avg{\nu}}$ (right) versus $\theta$ for different values of the parameters $\lambda$ and $q$ and fixed $\rho_{\inn}=0.5$.} 
\end{figure}

The average fractions of available compounds (metabolites) and active reactions obtained upon varying $\theta$ at fixed $\rho_{\inn}=0.5$ for Soft-MB is displayed in Fig. \ref{soft-MB_inTHETA}.  
One sees that, expectedly, larger values of $\theta$ lead, on average, to larger fractions of available metabolites and of active reactions. For large enough values of $\lambda$ and $q$, however, as the $\overline{\avg{\cdots}}_+$ and $\overline{\avg{\dots}}_-$ averages become steeper functions of $\theta$, the curves obtained by increasing and decreasing $\theta$ no longer coincide. Notice that, while for lower $\lambda$ and $q$ solutions can be found over a broad range of values of the magnetizations, when $\lambda$ and $q$ increase the average metabolite availability seems to concentrate in small ranges close to the extremes 0 and 1, distinguishing solutions with few available metabolites from solutions with a large fraction of available compounds. This type of picture is however not observed for reactions (we shall return to this point later on).

\begin{figure}
\centering
\includegraphics[width=17cm]{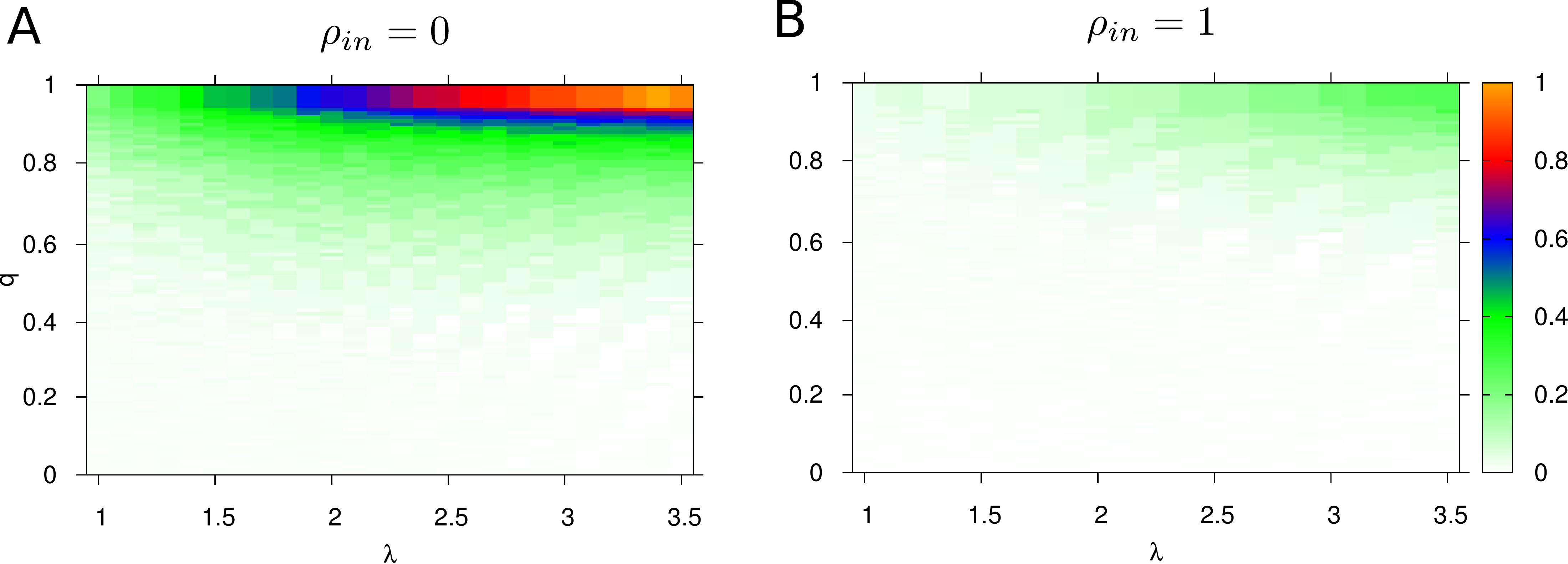}
\caption{Map of the values of $\Delta\mu$ (normalized by the same maximum: 4.852) for the Soft-MB problem in the $(\lambda,q)$ plane. The spacing in $q$ is equal to $0.01$, while it is $0.1$ in $\lambda$. A) $\rho_{\inn}=0$; B) $\rho_{\inn}=1$.}
  \label{phase_diagram_soft-MB}
\end{figure}

A simple way to quantify the onset of hysteresis is by measuring the quantity (we focus for simplicity on metabolites)
\begin{gather}
\Delta \mu =\int_{-\infty}^{+\infty} \left( \overline{\avg{\mu}}_+- \overline{\avg{\mu}}_- \right) d\theta ~~,
\end{gather}
which vanishes when $\overline{\avg{\mu}}_+=\overline{\avg{\mu}}_-$ and generically differs from 0 in presence of hysteresis. A map of the values of $\Delta\mu$ in the parameter space $(\lambda, q)$ is presented in Figure \ref{phase_diagram_soft-MB} for the limiting choices $\rho_\inn=1$ and $\rho_\inn=0$.

While hysteretic behaviour can be found practically all throughout the $(\lambda,q)$ plane, it becomes stronger at high enough $\lambda$ and $q$, where an abrupt jump in the magnetizations takes place. The presence of such a large hysteresis, and the coexistence of low and high magnetization solutions, signal a non trivial structure in the space of solutions to the CSP. Such a non-trivial structure appears also in many other well-known CSP, as the random k-XORSAT \cite{xorsat} and random k-SAT \cite{ksatScience,ksatJSTAT}, and is the origin of the onset of long range correlations, that have important consequences on the behavior of searching algorithms \cite{PNAS}.
Away from the hysteretic portion, $\overline{\avg{\mu}}_+$ and $\overline{\avg{\mu}}_-$ vary smoothly with $\theta$, allowing one to sample easily solutions with any magnetization not in the jump.

On the other hand, the overall structure of the solutions (in terms of $\Delta\mu$) appears to vary weakly with $\rho_\inn$. This strongly suggests that main observed effects (e.g. the jump and the hysteresis) are essentially due to topology of the network, rather than to the boundary conditions.
It is interesting to observe that the hysteretic region shrinks as $\rho_\inn$ increases, suggesting that, within the constraints imposed by Soft-MB, a larger repertoire of available nutrients stabilizes the output by allowing to achieve higher values of the magnetization for smaller values of $\theta$.

\subsection{Hard Mass-Balance}

\begin{figure}
\centering
\includegraphics[width=17cm]{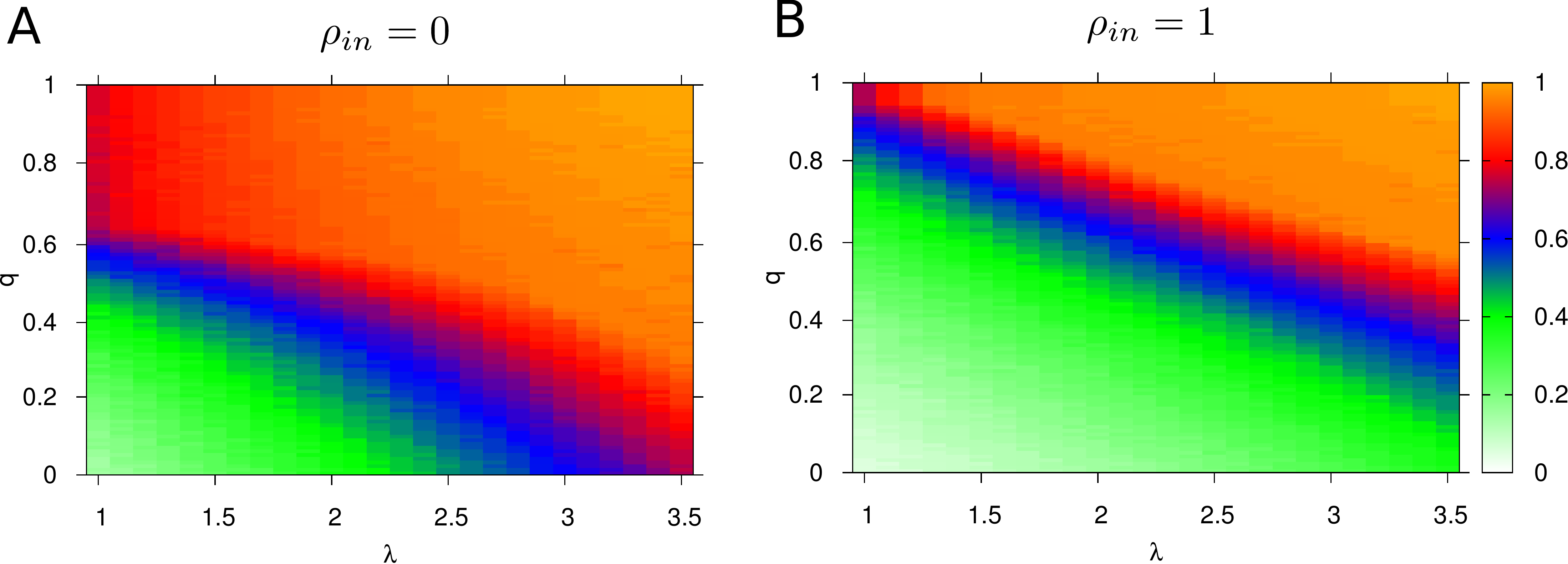}
\caption{\label{phase_diagram_HMB}Map of the values of $\Delta\mu$ (normalized by the same maximum: 12.643) for the Hard-MB problem in the $(\lambda,q)$ plane.  The spacing in $q$ is equal to $0.01$, while it is $0.1$ in $\lambda$. A) $\rho_{\inn}=0$; B) $\rho_{\inn}=1$.}
\end{figure}

\begin{figure}
\includegraphics[width=17cm]{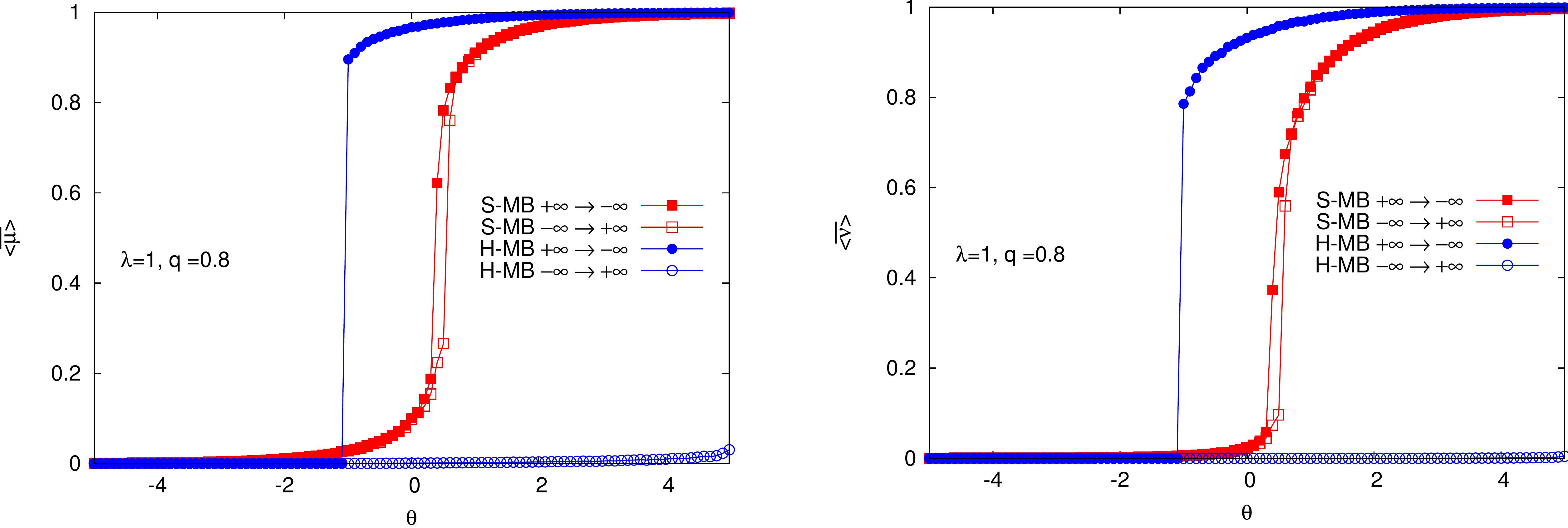}
\caption{Behaviour of $\overline{\avg{\mu}}_+$ and $\overline{\avg{\mu}}_-$ (left) and $\overline{\avg{\nu}}_+$ and $\overline{\avg{\nu}}_-$ (right) as functions of $\theta$ at $\lambda=1$, $q=0.8$ and $\rho_{\inn}=0.5$ for the Soft- and Hard-MB problems.}
\label{VN_FBA_inTHETA_l1}
\end{figure}

\begin{figure}
\includegraphics[width=17cm]{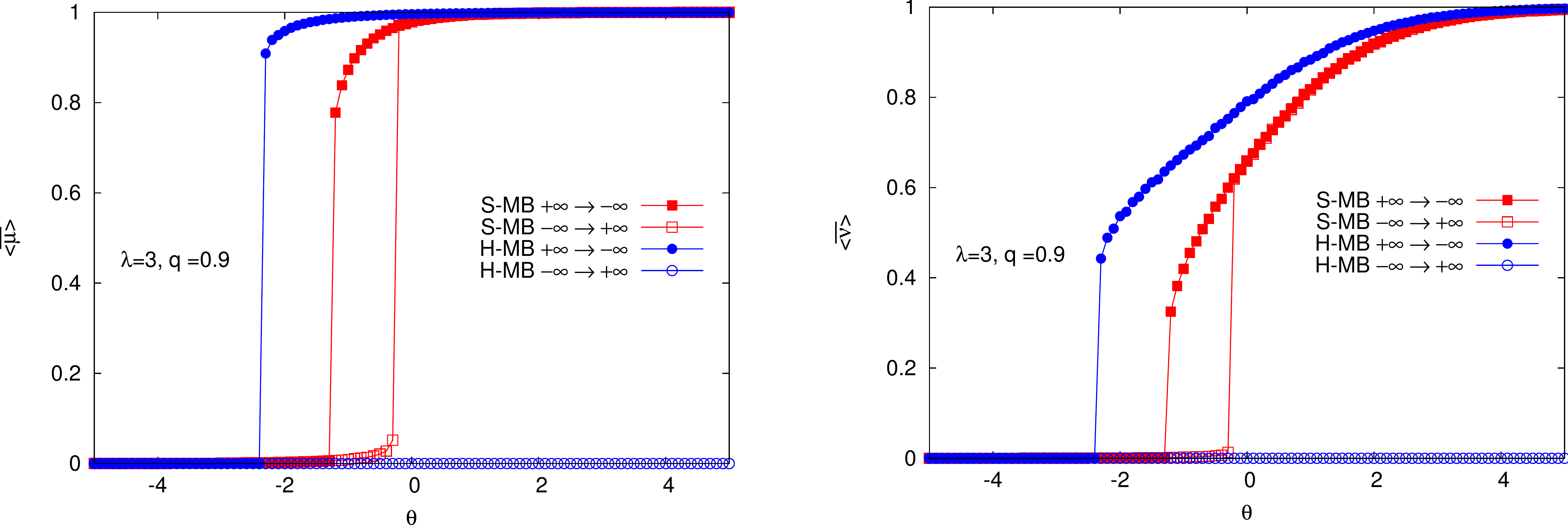}
\caption{Behaviour of $\overline{\avg{\mu}}_+$ and $\overline{\avg{\mu}}_-$ (left) and $\overline{\avg{\nu}}_+$ and $\overline{\avg{\nu}}_-$ (right) as functions of $\theta$  at $\lambda=3$, $q=0.9$ and $\rho_{\inn}=0.5$ for the Soft- and Hard-MB problems.}
\label{VN_FBA_inTHETA_l3}
\end{figure}

The $\Delta \mu$-map for the Hard-MB case is displayed in Figure \ref{phase_diagram_HMB}.
In contrast with the Soft-MB case, Hard-MB solutions display strong hysteresis for all choices of $\lambda$, $q$ and $\rho_\inn$. Furthermore, comparing the results at $\rho_{\inn}=0$ and $\rho_{\inn}=1$, it is clearly seen that, again, changing $\rho_\inn$ (i.e. increasing the number of available nutrients) has little influence on the overall structure of the phase space. Rather, its main effect is that of reducing the magnitude of hysteresis cycles. It is interesting to note that the maximum value of $\Delta \mu$ in Hard-MB is more than double than the one in Soft-MB.

The presence of strong hysteresis markedly distinguishes the solution space of the two CSPs. A comparison between the behaviour of the magnetization obtained in the Soft- and Hard-MB cases for selected parameter values is displayed in Figs \ref{VN_FBA_inTHETA_l1} and \ref{VN_FBA_inTHETA_l3}.
In first place, one sees that the limiting value of the average magnetization for $\theta\to\pm\infty$ in the Hard-MB problem is identical to that of the Soft-MB problem, suggesting that in specific cases the Hard-MB CSP may actually acquire a strong directional nature (like the Soft-MB case), despite the fact that in Hard-MB substrates and products are highly correlated between each other. Secondly, the increasing-$\theta$ protocol appears to be unable to identify active solutions in the Hard-MB case, suggesting that the Hard-MB constraints bias solutions towards activating a large fraction of metabolite nodes.
In the Hard-MB case, it seems that it is possible to start from the all-on configuration and gradually switch off the network, but it is very difficult to switch on part of the network starting from the all-off configuration: for this reason the all-off solution is very stable in the Hard-MB case.

\begin{figure}
\centering
\includegraphics[width=17cm]{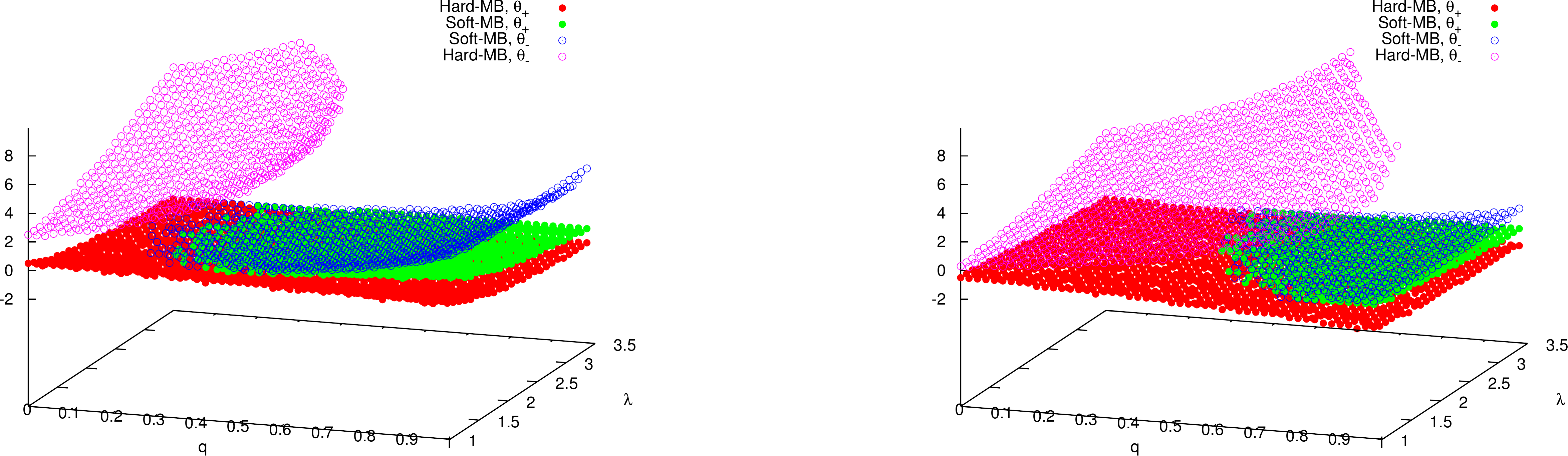}
\caption{Spinodal values for the existence of upper and lower branches of $\overline{\avg{\mu}}$ ($\theta_+$ and $\theta_-$, respectively) for Soft-MB and Hard-MB. Left panel: $\rho_{\inn}=0$; Right panel: $\rho_{\inn}=1$.}
  \label{spinodal_theta}
\end{figure}

Another way to visualize the hysteresis as a function of the parameters $\lambda$ and $q$ is by plotting the spinodal points $\theta_+$ and $\theta_-$, i.e. the endpoints of the upper and lower branches of $\overline{\avg{\mu}}$, respectively. This is shown in Fig. \ref{spinodal_theta} for both Soft-MB and Hard-MB.

\begin{figure}
\includegraphics[width=17cm]{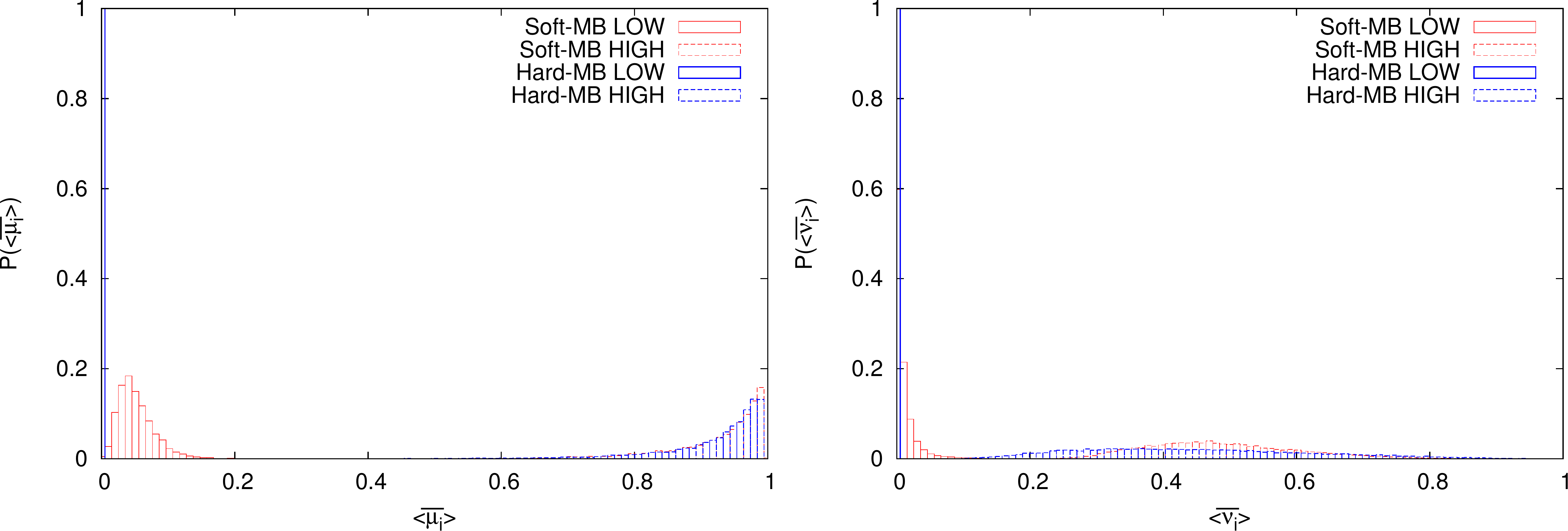}%0.27
\caption{\label{VN_FBA_histogram_mu}Histogram of the values of the average availability of metabolites (left) and reactions (right) for $\lambda=3$, $q=0.8$ and $\rho_{\inn}=0.5$. The value of $\theta$ has been chosen for both CSPs at the transition, so that both high and low values of $\overline{\avg{\mu}}$ and $\overline{\avg{\nu}}$ are possible. In specific, the $\theta$ values for HIGH solutions correspond to $(\overline{\avg{\mu}},\overline{\avg{\nu}}) \simeq (0.92,0.5)$ for both CSPs, while those for LOW solutions corresponds to $(\overline{\avg{\mu}},\overline{\avg{\nu}}) \simeq (0,0)$ for Hard-MB and $(\overline{\avg{\mu}},\overline{\avg{\nu}}) \simeq (0.06,0.014)$ for Soft-MB.}
\end{figure}

A more quantitative description of the stabililty of the null solution in the Hard-MB case is given in Fig. \ref{VN_FBA_histogram_mu}, where we display the  distribution of values of the magnetization for metabolites and reactions obtained for a value of $\theta$ at the transition.

From the distribution of metabolite availabilities one clearly sees that, generically, fluctuations are larger in Soft-MB than in Hard-MB, implying that, while Soft-MB sustains non-trivial solutions over a wide range of values of the magnetizations, Hard-MB only admits solutions with a large and tightly constrained value of the average metabolite availability. Interestingly, the overall structure of the distributions changes when one considers reactions, for which both Soft- and Hard-MB can lead a large variability (much larger, in turn, than what occurs for metabolites). This is consistent with our constraints, which do not impose to activate a reaction even when all of its neighbouring metabolites are available. Note that both for reactions and metabolites Soft-MB allows for solutions with very low magnetization that are generically absent in Hard-MB.

\begin{figure}
\centering
\includegraphics[width=17cm]{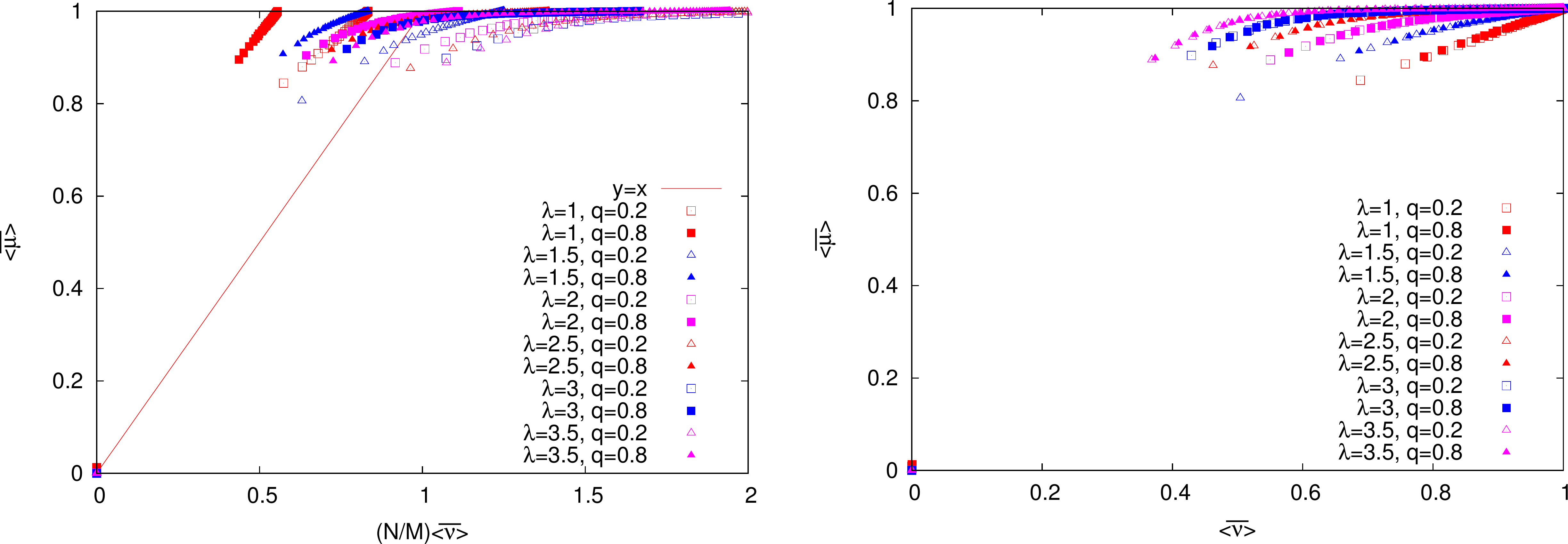}%0.27
\caption{\label{meta_rea_size_comparison}Left: behaviour of $\overline{\avg{\mu}}$ versus $\frac{N}{M}\overline{\avg{\nu}}$ (left) and versus $\overline{\avg{\nu}}$ (right) for $\lambda$ and $q$ as displayed in the legend. In each dataset $\theta$ increases from left to right.}
\end{figure}

Finally, we notice that not all of the solutions to Hard-MB would be able to carry non-vanishing fluxes in the linear problem defined by (\ref{mbe}), which is only possible if the number of available metabolites does not exceed that of active reactions. To see this, one can compare the quantities $M\overline{\avg{\mu}}$ and $N\overline{\avg{\nu}}$, see Figure \ref{meta_rea_size_comparison} (left panel), which are respectively the number of equations and the number of unknowns in the FBA problem. 

It is clear that only for sufficiently large values of $\lambda$ will Boolean configurations correspond to realizable flux states in FBA. This confirms the intuition that redundant network structures (larger $\lambda$'s) confer flexibility (i.e. the possibility of operating in different states) to a reaction network.
What looks counterintuitive in Figure \ref{meta_rea_size_comparison} (left panel) is that small $q$ values are also to be preferred. An explanation to this fact can be obtained by plotting $\overline{\avg{\mu}}$ versus $\overline{\avg{\nu}}$ (right panel in Figure \ref{meta_rea_size_comparison}), and noticing that data with different $q$ values fall on the same curve. Since the data in the left panel of Figure \ref{meta_rea_size_comparison} are obtained by multiplying the $x$ values in the right panel by $N/M=\lambda/(1+q)$, large $q$ data are more keen to cross the line at the boundary of the feasible solutions region.

In the right panel of Figure \ref{meta_rea_size_comparison} we also notice that $\overline{\avg{\mu}}$ spans a rather limited range (roughly $0.8 < \overline{\avg{\mu}} \le 1$) which is mostly independent on the topology (i.e., on $\lambda$), while the range of valid $\overline{\avg{\nu}}$ values becomes very broad for redundant networks (i.e., for large values of $\lambda$). In other words, solutions to the Boolean constrained problem on a RRN do exist only if a very large fraction of metabolites are present, while the fraction of active reactions can be made small only if the topology is redundant enough.

\section{Mean Field Theory: Network Expansion revisited}
\label{sec:MF}
\subsection{The problem}
\label{sec:MF_problem}
The basic idea behind NE is that, given a seed compound (e.g. a nutrient), a reaction can (and will) activate when all its substrates are available (AND-like constraint), whereas a compound will be available if at least one of the reactions that produce it is active (OR-like constraint). The numerical procedure of NE transfers the information about the availability of certain metabolites across the network links, as explained pictorially in Fig. \ref{steps_NE}. We shall term this type of process a Propagation of External Inputs (PEI). 

\begin{figure}
\includegraphics[height=5cm]{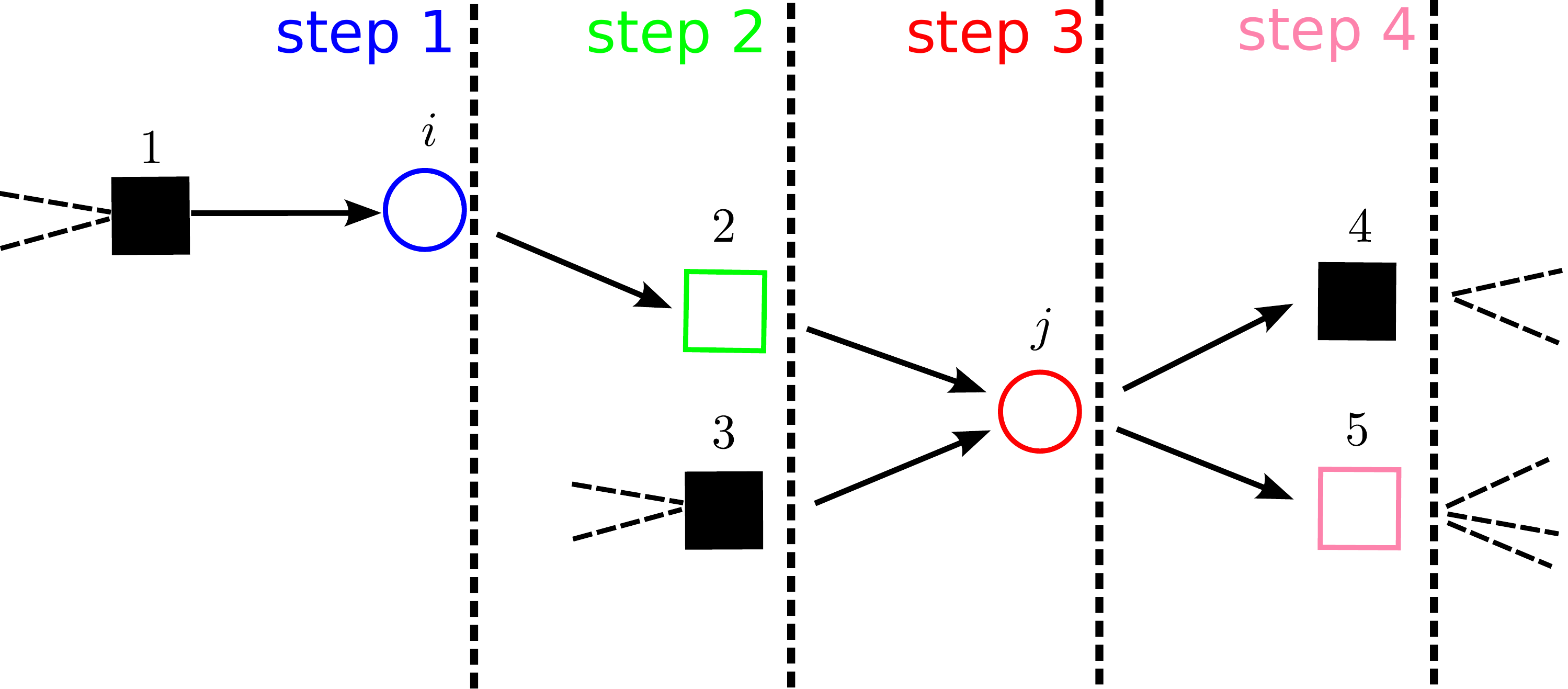}%0.25
\caption{\label{steps_NE}Sketch of four steps of the Propagation of External Inputs (PEI) algorithm (serving as the basis of the Network Expansion method \cite{Ebenhoh}). Black squares represent compounds initially available. In step 1, reaction $i$ is activated by virtue of the availability of compound $1$; in step 2, metabolite $2$ becomes available by virtue of the activation of reaction $i$; in step 3, reaction $j$ activates as both $2$ and $3$ are available; and so on.}
\end{figure}

It is simple to understand that, as soon as the reaction network departs from a linear topological structure, the propagation will likely stop after a small number of steps unless the availability of additional compounds is invoked. Indeed, in Network Expansion PEI is aided by the assumption that highly connected metabolites like water are abundant. Because  of its intuitive appeal, it is useful to analyze briefly the properties of PEI in somewhat more detail.

Within a Mean-Field Approximation, one can write down equations for the probability $\avg{\nu_i}$ that reaction $i$ will be active and for the probability $\avg{\mu_m}$ that metabolite $m$ will be available by simply considering that, under PEI in a given network, a reaction can activate when all of its inputs are available and a metabolite becomes available when at least one reaction is producing it. This implies that
\begin{gather}
\avg{\nu_i}=\prod_{n \in \partial i_{\inn}} \avg{\mu_n}~~, \label{MF_thetaINF_rea} \\
1 - \avg{\mu_m} = \prod_{k \in \partial m_{\inn}}(1-\avg{\nu_k}) \label{MF_thetaINF}~~. 
\end{gather}
where $ \partial i_{\inn}$ and $\partial m_{\inn}$ denote the set of incoming links of nodes $i$ and $m$, respectively. To prove the link between PEI and the CSPs defined above, note that, using the definition (\ref{meas}) one can easily compute the mean values
\begin{gather}
\avg{\nu_i}=\frac{\sum\limits_{\{\mu_n\},\{\nu_i \}} \prod\limits_{n \in \partial i_{\inn}}\Gamma_n\Delta_i e^{\theta \nu_i} \nu_i}{\sum\limits_{\{\mu_n\},\{\nu_i\}}  \prod\limits_{n \in \partial i_{\inn}}\Gamma_n \Delta_i e^{\theta \nu_i}}~~,\\
\avg{\mu_m}=\frac{\sum\limits_{\{\mu_m\},\{\nu_j\}}  \prod\limits_{j \in \partial m_{\inn}}\Delta_j e^{\theta \nu_j} \Gamma_m \mu_m}{\sum\limits_{\{\mu_m\},\{\nu_j\}} \prod\limits_{j \in \partial m_{\inn}}\Delta_j e^{\theta \nu_j} \Gamma_m}~~, 
\end{gather}
Under the Mean Field Approximation, we can set
\begin{gather}
 \Gamma_m(\mu_m,\{\nu_i\})=\Gamma_m(\mu_m,\{\langle \nu_i \rangle \})~~,\\
\Delta_i(\nu_i,\{\mu_m\})=\Delta_i(\nu_i,\{\avg{\mu_m}\})~~,
\end{gather}
which in turn implies
\begin{gather}
\label{punto_fisso_theta_fin}
\langle \nu_i \rangle=\frac{\etheta \prod\limits_{n \in \partial i_{\inn}}\langle\mu_n\rangle}{1+\etheta \prod\limits_{n \in \partial i_{\inn}}\langle\mu_n\rangle} \\
\langle \mu_m \rangle=1-\prod_{k \in \partial m_{\inn}}(1-\langle\nu_j\rangle)~~.
\end{gather}
In the limit $\theta\rightarrow \infty$ we have
\begin{gather}
\avg{\nu_i}= 
\begin{cases}
1 &\text{if }\prod\limits_{n \in \partial i_{\inn}}\langle\mu_n\rangle = 1~~, \\ 
0 &\text{if } \prod\limits_{n \in \partial i_{\inn}}\langle\mu_n\rangle = 0~~.
\end{cases}
\end{gather}
So that equations (\ref{MF_thetaINF_rea}) and (\ref{MF_thetaINF}) are recovered. In other terms, PEI is Mean Field Approximation at $\theta\to\infty$ of the CSPs considered in Section \ref{sec:pop_dyn}.

%It is simple to derive analytically the phase diagram of NE/PEI. Using equation (\ref{MF_thetaINF_rea}) and (\ref{MF_thetaINF}), one can define the probability that a metabolite is available (in the ensemble of RRN defined in Section \ref{sec:pop_dyn} and characterized by parameters $\lambda$, the mean of the Poisson distributed degrees of compounds, and $q$, the probability that a reaction has two input or output componds, as opposed to one compound which occurs with probability $1-q$) as

\begin{figure}
\includegraphics[width=10cm]{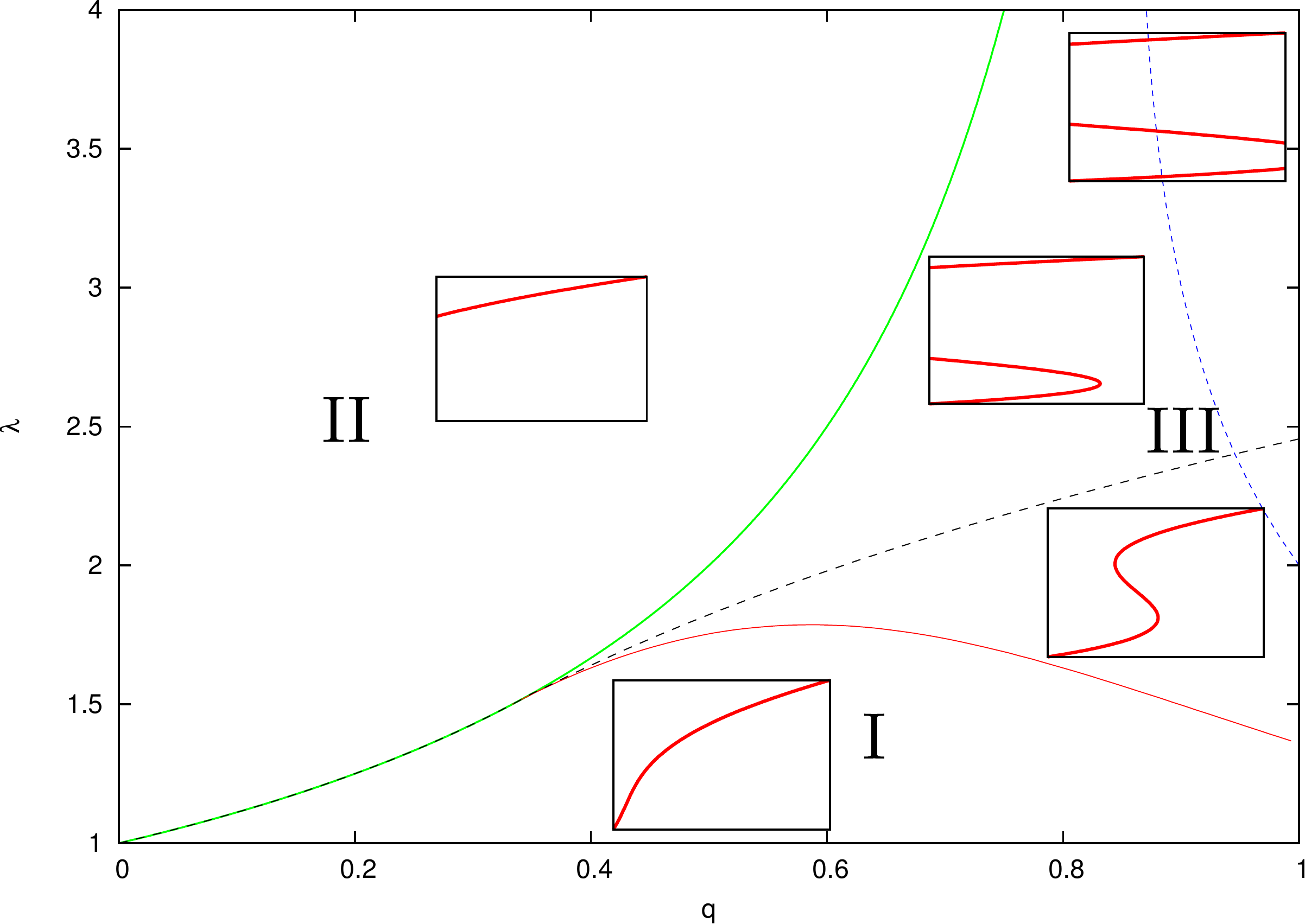}%0.27
\caption{Phase diagram obtained by propagation of external inputs (PEI) in the $(q,\lambda)$ plane. The insets display the curves $\gamma$ versus $\rho_{\inn}$ obtained in the different sectors; all lines are analytical. See text for details.}
\label{diagramma_fase_theta_infty}
\end{figure}

It is simple to derive analytically the phase diagram of PEI in the ensemble of RRN defined in Section \ref{sec:pop_dyn}. In this ensemble, the probability that a metabolite is available is
\begin{equation}
\gamma=\overline{\avg{\mu_m}}~~,
\end{equation}
where the over-bar denotes an average over the network realizations. Using (\ref{MF_thetaINF_rea}) and (\ref{MF_thetaINF}) one sees that
\begin{gather*}
\gamma =e^{-\lambda}\rho_{\inn}+\sum_{k_m \geq 1}\mathit{D}_{M}(k_m)\left(1-\prod_{j=1}^{k_m}(1-\overline{\avg{\nu_j}})\right)~~,
\end{gather*}
where we have assumed that nutrients (fractionally given by roughly $e^{-\lambda}$ nodes) have a fixed probability $\rho_{\inn}$ of being available and where $D_M(k)=e^{-\lambda}\lambda^k/k!$ is the distribution of metabolite in- (and out-)degrees. In turn, this gives
\begin{gather}
\gamma=e^{-\lambda}\rho_{\inn}+1-e^{-\lambda\tau}~~,
\label{eq_punto_fisso_M}
\end{gather}
where $\tau=\overline{\avg{\nu_i}}$ is the probability that a reaction is active, which, recalling that the in- and out-degrees of reactions are distributed according to $D_R(d)=q\delta_{d,2}+(1-q)\delta_{d,1}$, satisfies (within a Mean-Field Approximation)
\begin{gather}
\tau=\overline{\prod_{b\in\partial i_{\inn}} \avg{\mu_n}}=(1-q)\gamma+q\gamma^2~~.
\label{eq_punto_fisso_R}
\end{gather}
Putting things together, $\gamma$ is seen to satisfy the condition
\begin{equation}
\gamma=e^{-\lambda}\rho_{\inn}+1-\exp[-\lambda((1-q)\gamma+q\gamma^2)]~~,
\label{eq_punto_fisso_theta_infty}
\end{equation}
which can be solved for $\gamma$ upon changing the values of $\rho_{\inn}$, $q$ and $\lambda$. The resulting phase diagram, based on the behaviour of the solution $\gamma^\star(\rho_\inn)$, is displayed in Figure \ref{diagramma_fase_theta_infty}.

Three regions can be distinguished in the $(q,\lambda)$ plane. In region I,  Equation (\ref{eq_punto_fisso_theta_infty}) has a unique solution and $\gamma^\star$ is a monotonous function of $\rho_\inn$ (note that $\gamma^\star=0$ is always a solution when $\rho_\inn=0$). Outside region I, the curve $\gamma^\star$ vs $\rho_\inn$ displays an inflection point. If the point lies outside the interval $[0,1]$ (for both $\gamma$ and $\rho_\inn$) then (\ref{eq_punto_fisso_theta_infty}) has a unique non-zero solution for $\rho_\inn>0$ and two different solutions at $\rho_\inn=0$ (region II). In region III, instead, a range of values of $\rho_\inn$ exists where three distinct solutions (with different values of $\gamma$) of (\ref{eq_punto_fisso_theta_infty}) occur. This sector can be further divided according to the number of solutions found for $\rho_\inn=0$ and $\rho_\inn=1$. The black dashed line marks the boundary between phases with, respectively, one and three solutions for $\rho_\inn=0$ while the dashed blue line separates the region with one and three solutions for $\rho_\inn=1$.  

%In Section \ref{percolation_transition} we argue that, from a statistical physics viewpoint, the transitions found above are, in essence, percolation transitions.

\begin{figure}
\includegraphics[width=12cm]{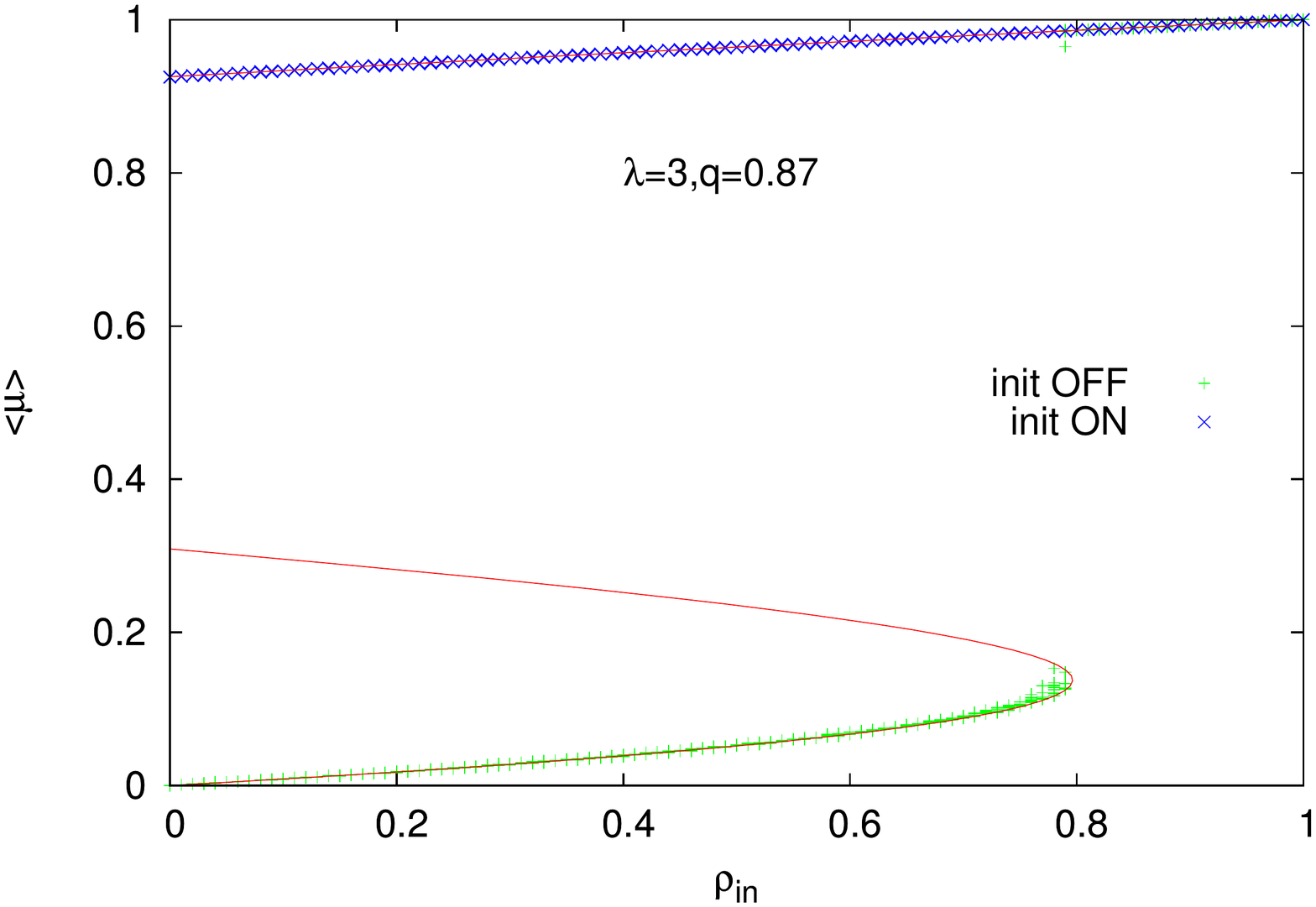}%0.27
\caption{Theoretical solution of Equation (\ref{eq_punto_fisso_theta_infty}) (solid line) versus $\rho_{\inn}$, together with the results obatined by PEI and reverse-PEI.}
\label{PEI_sampling}
\end{figure}

For any fixed $\rho_\inn$, whenever solutions with different values of $\avg{\mu}$ coexist, those with the smallest $\avg{\mu}$ can be retrieved by straightforward PEI starting from a configuration where no metabolite is available except for nutrients. Solutions with larger $\avg{\mu}$, on the other hand, can be found by `reverse-PEI'. In this procedure a configuration where internal metabolites are all available and nutrients are fixed with probability $\rho_{\inn}$ is initially selected, and then a solution is found by enforcing the constraints in an iterative way. The results for both procedures are presented in Figure \ref{PEI_sampling} for $\lambda=3$ and $q=0.87$ (deep into region III in Figure \ref{diagramma_fase_theta_infty}).

\begin{figure}
\includegraphics[width=12cm]{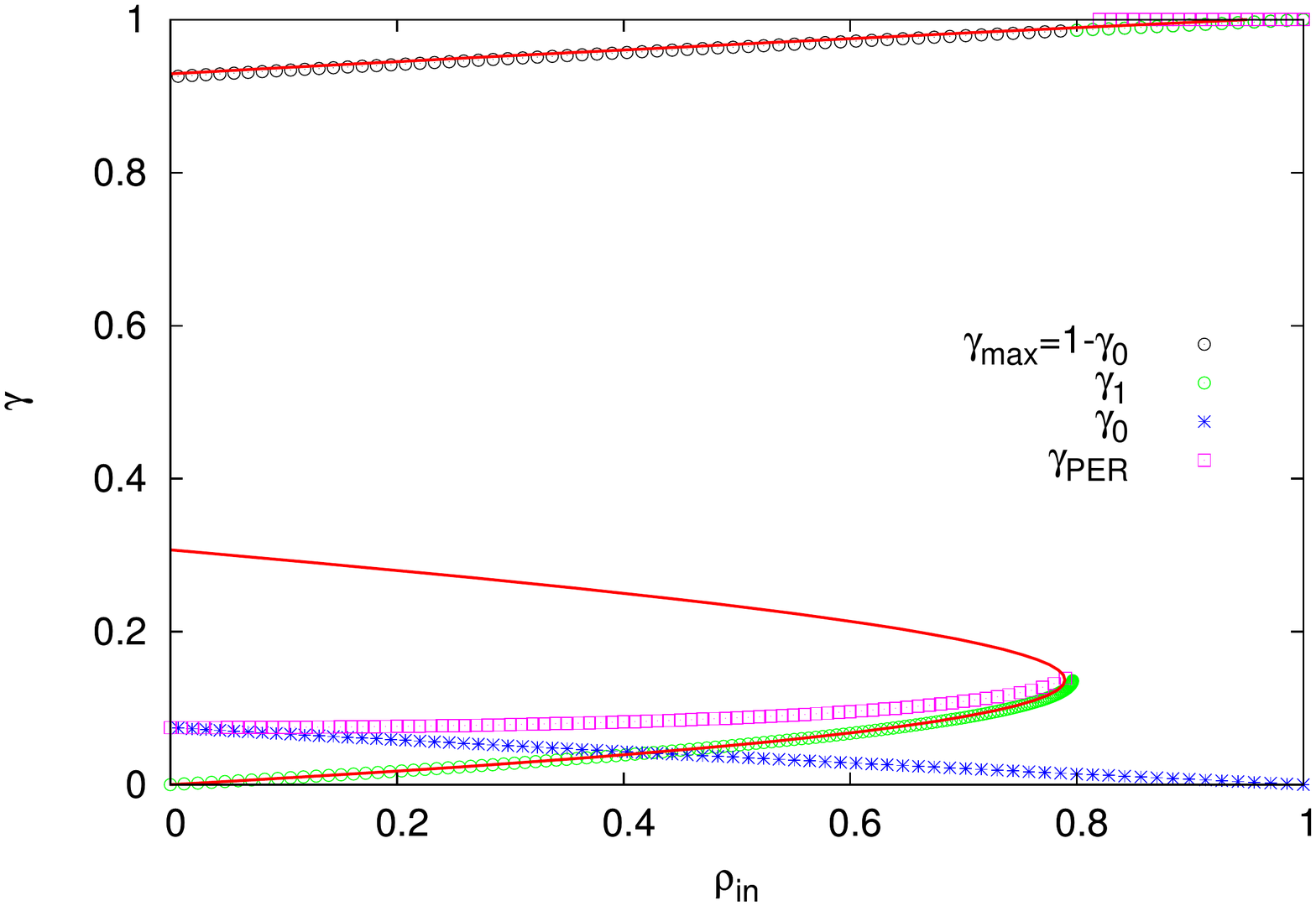}%0.27
\caption{Weights of the different components of the PER core for a graph with $\lambda=3$ and $q=0.87$. The red line corresponds to the solution of equation (\ref{eq_punto_fisso_theta_infty}).\label{PERcore_fig}}
\end{figure}

\subsection{Origin of the phase transition within the Mean-Field Approximation in PEI}
\label{percolation_transition}

We show here that, from a physical viewpoint, the phase transitions occurring in PEI (see Figure \ref{diagramma_fase_theta_infty}) are, in essence, of a percolation type.

To analyze the effectiveness of PEI, we start by identifying the so-called Propagation of External Regulation (PER) Core of the system \cite{Leone_Zecchina}, that is the sub-network obtained by fixing the nutrient availability (with probability $\rho_\inn$) and then propagating this information inside the network. In this way, some variables will get fixed to either $1$ or $0$. At convergence, a fraction $\gamma_1$ (resp. $\gamma_0$) of metabolites will be fixed to $1$ (resp. $0$), while a fraction 
$\tau_1$ (resp. $\tau_0$) of reactions will be fixed to $1$ (resp. $0$). One easily sees that, at the fixed point, the following equations hold:
\begin{gather}
1-\tau_0=q(1-\gamma_0)^2+(1-q)(1-\gamma_0)~~,\\
\gamma_0=\sum\limits_{k\neq 0}\mathit{D}_M(k)\tau_0^k+(1-\rho_{\inn})D_M(0)~~,\\
    \tau_1=q\gamma_1^2+(1-q)\gamma_1~~, \\
    1-\gamma_1=\sum\limits_{k\neq 0}\mathit{D}_M(k)(1-\tau_1)^k+(1-\rho_{\inn})D_M(0)~~.
\end{gather}
In turn, one obtains
\begin{gather}
\tau_0=1-q(1-\gamma_0)^2-(1-q)(1-\gamma_0)~~,\\ 
\gamma_0=e^{-\lambda}e^{\lambda\tau_0}-\rho_{in}e^{-\lambda}~~,\\
    \tau_1=q\gamma_1^2+(1-q)\gamma_1~~,\\ 
    \gamma_1=1-e^{-\lambda\tau_1}+\rho_{in}e^{-\lambda}~~.   
\end{gather}
Unsurprisingly, the equations for $\gamma_1$ and $\tau_1$ take us back to (\ref{eq_punto_fisso_theta_infty}). On the other hand, the fraction of metabolites in the PER core is given by
\begin{equation}
\gamma_{PER}=\gamma_1+\gamma_0. 
\end{equation}
Hence the fraction of metabolites that are not fixed by propagating nutrient availability is given by $1-\gamma_{PER}$, and the maximum achievable magnetization for metabolites is given by $\gamma_{max}=1-\gamma_0$. 
Figure \ref{PERcore_fig} displays the different contributions for a specific choice of the parameters, together with the corresponding solution of Eq. (\ref{eq_punto_fisso_theta_infty}).
\begin{figure}
\includegraphics[width=10cm]{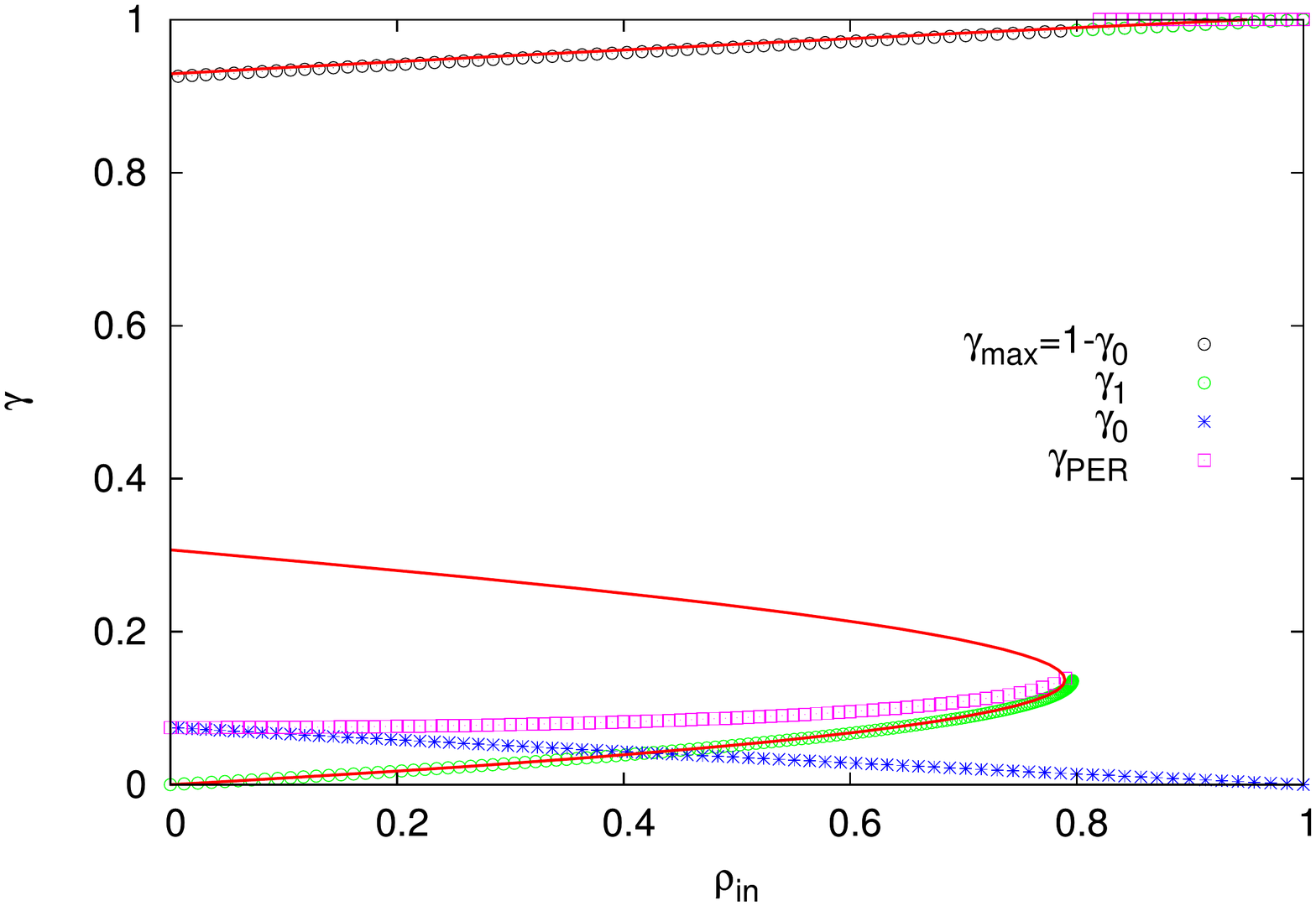}%0.27
\caption{Graph of all the component of the PER core for a graph with $\lambda=3$ and $q=0.87$. The red line is the solution of equation (\ref{eq_punto_fisso_theta_infty}).\label{PERcore_fig}}
\end{figure}
The excellent agreement of $\gamma_1$ with the analytical line for the feasible values of the magnetization suggests that straightforward PEI will be able to recover solutions with lower magnetization when the latter coexist with high-magnetization solutions. On the other hand, the magnetization of the latter coincides, expectedly, with the largest achievable average metabolite availability. Finally, depending on the value of $\lambda$ and $q$, one obtains a single solution when no PER core exists, and two solutions (with magnetizations $\gamma_{max}$ and $\gamma_1$) in presence of a PER core. Hence the transition is a typical percolation transition between a phase in which the internal variables are trivially determined by the nutrients (in absence of a PER core) to one in which the internal are not trivially determined (in presence of a PER core).

\section{Solutions on individual networks by Belief Propagation and decimation}
\label{sec:BP_Decimation}
\begin{figure}
\includegraphics[scale=0.3]{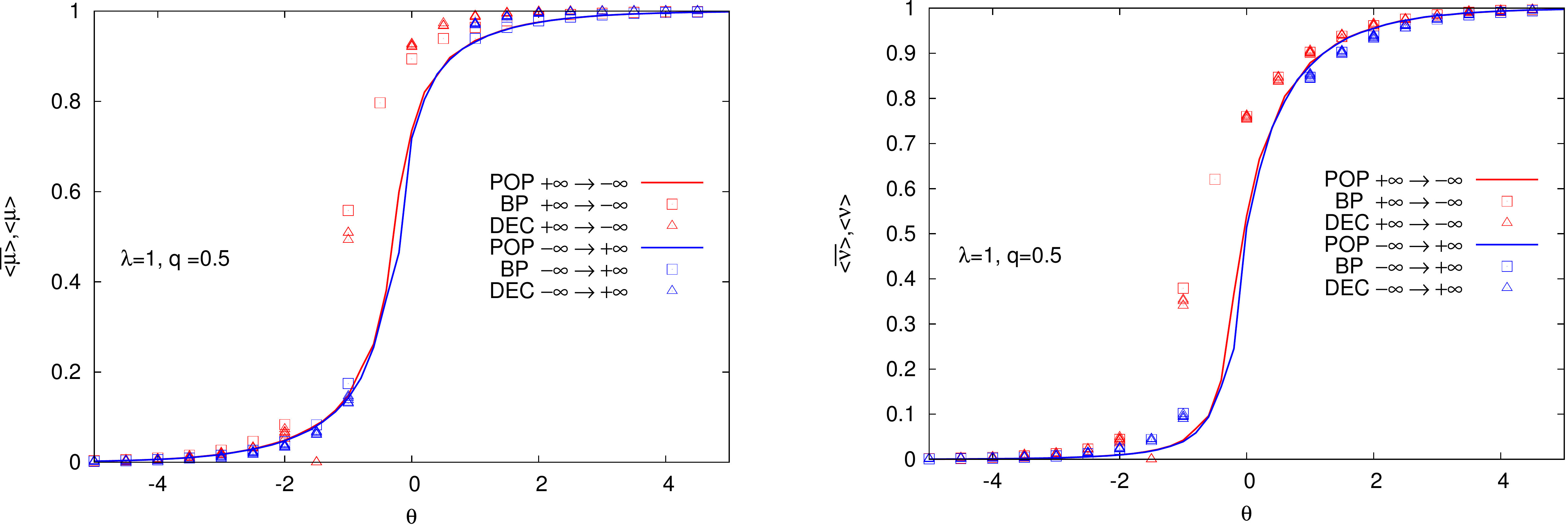}
\caption{{\bf Soft-MB} for $\lambda=1$, $q=0.5$ and $\rho_{\inn}=1$. {\bf Left}: average fraction of available metabolites, $\avg{\mu}$ ($\overline{\avg{\mu}}$ for population dynamics) versus $\theta$. {\bf Right}: average fraction of active reactions, $\avg{\nu}$ ($\overline{\avg{\nu}}$ for population dynamics) versus $\theta$.\label{VN_l1_q05}}
\end{figure}

\begin{figure}
\includegraphics[scale=0.3]{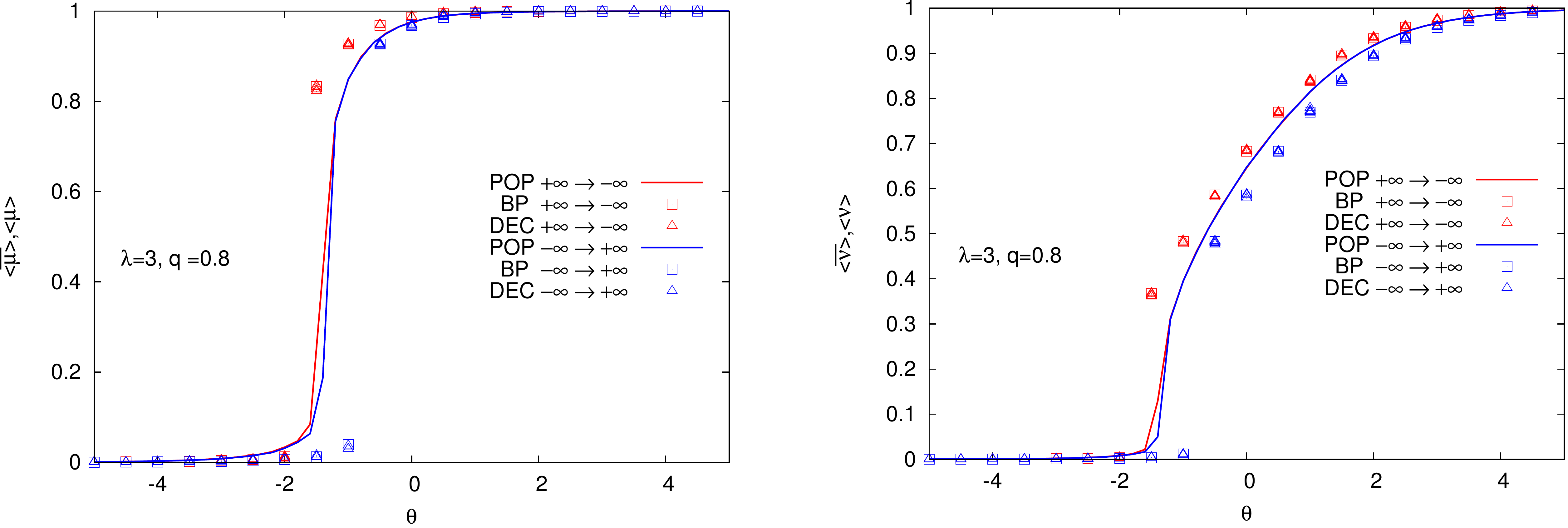}
\caption{{\bf Soft-MB} for $\lambda=3$, $q=0.8$ and $\rho_{\inn}=1$. {\bf Left}: average fraction of available metabolites, $\avg{\mu}$ ($\overline{\avg{\mu}}$ for population dynamics) versus $\theta$. {\bf Right}: average fraction of active reactions, $\avg{\nu}$ ($\overline{\avg{\nu}}$ for population dynamics) versus $\theta$.\label{VN_l3_q08}}
\end{figure}

\begin{figure}
\center
\includegraphics[scale=0.3]{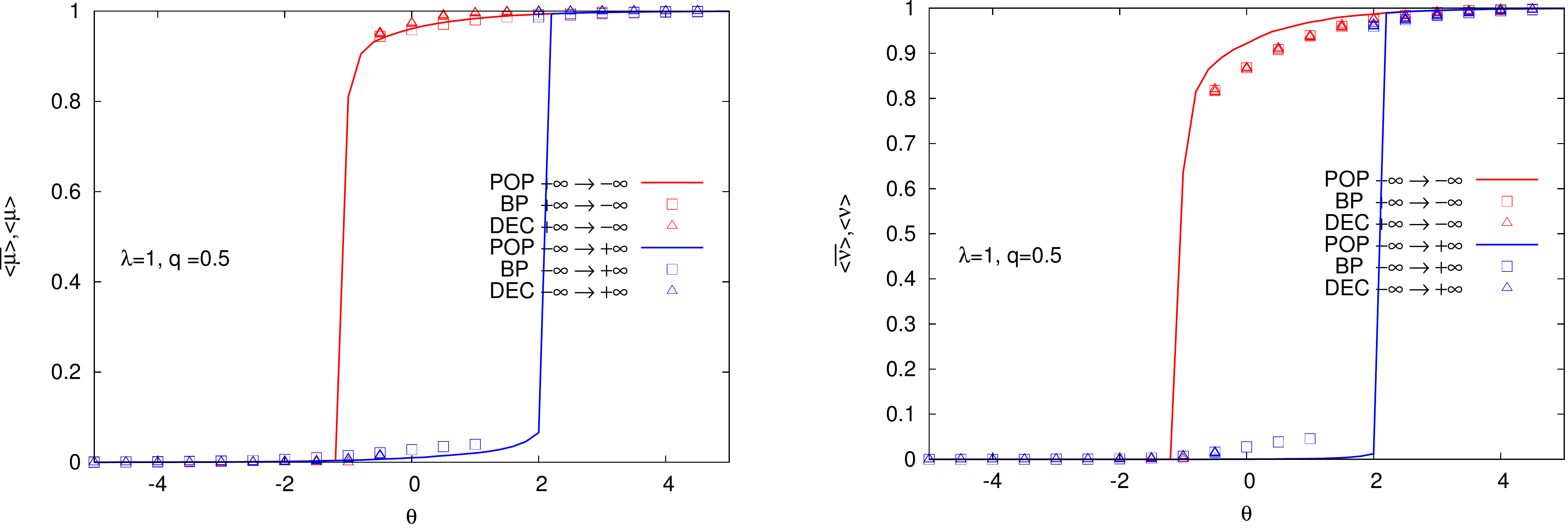}
\caption{{\bf Hard-MB} for $\lambda=1$, $q=0.5$ and $\rho_{\inn}=1$. {\bf Left}: average fraction of available metabolites, $\avg{\mu}$ ($\overline{\avg{\mu}}$ for population dynamics) versus $\theta$. {\bf Right}: average fraction of active reactions, $\avg{\nu}$ ($\overline{\avg{\nu}}$ for population dynamics) versus $\theta$.\label{FBA_l1_q05}}
\end{figure}

\begin{figure}
\center
\includegraphics[scale=0.3]{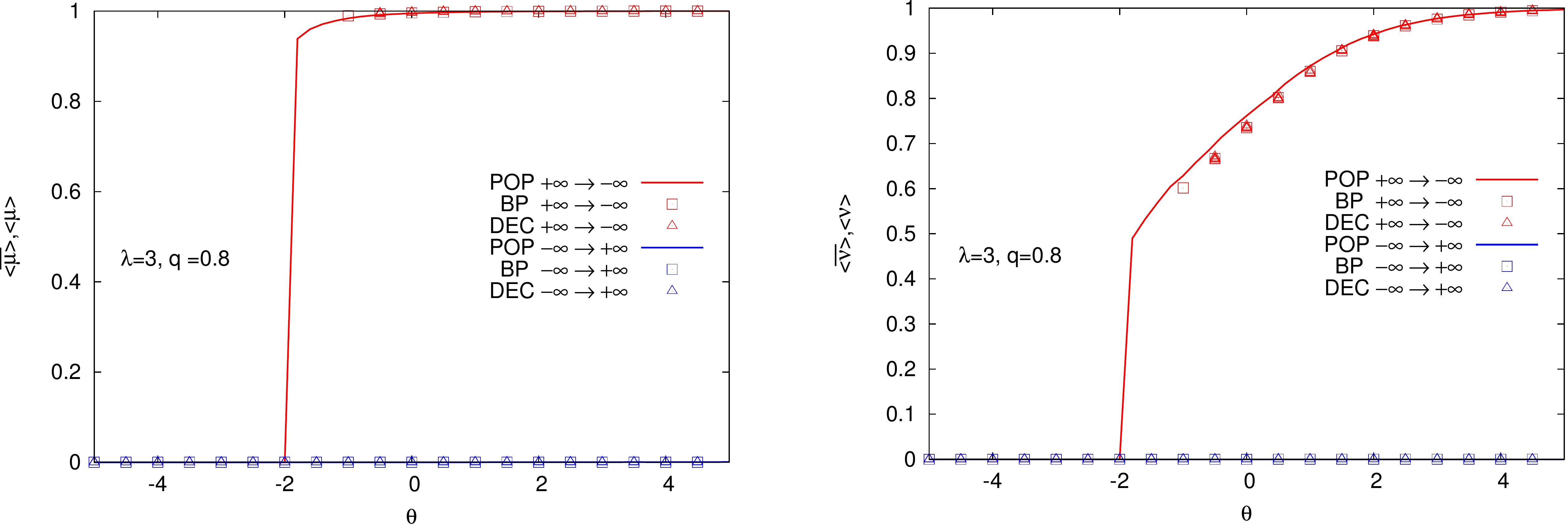}
\caption{{\bf Hard-MB} for $\lambda=3$, $q=0.8$ and $\rho_{\inn}=1$. {\bf Left}: average fraction of available metabolites, $\avg{\mu}$ ($\overline{\avg{\mu}}$ for population dynamics) versus $\theta$. {\bf Right}: average fraction of active reactions, $\avg{\nu}$ ($\overline{\avg{\nu}}$ for population dynamics) versus $\theta$.\label{FBA_l3_q08}}
\end{figure}

We turn now to the analysis of the Soft-MB and Hard-MB CSPs for general $\theta$. In essence, we have derived the cavity equations for the CSPs, presented in Section \ref{cavity_derivation}, and used the Belief Propagation (BP) algorithm discussed in Appendix \ref{sec:BP_explanation} to sample solutions on single instances of RRNs. Next, in order to obtain  \textit{individual configurations} of variables that satisfy our CSPs, we resorted to the decimation scheme presented in Appendix \ref{sec:decimation_explanation}. Results are presented in Figures \ref{VN_l1_q05} and \ref{VN_l3_q08} for Soft-MB ($\alpha=0$) and in Figures \ref{FBA_l1_q05} and \ref{FBA_l3_q08} for Hard-MB ($\alpha=1$). BP results, labeled as `BP', are compared with results retrieved by the population dynamics algorithm developed in Section \ref{sec:pop_dyn} (labeled `POP' and corresponding to the ensemble average) and with the decimation results (labeled `DEC'). In Section \ref{sec:pop_dyn}, the solution space was explored by two different protocols, which we also use here: by reducing $\theta$ starting from a large positive value ($+\infty \rightarrow -\infty$ in the Figure legends) and by doing the reverse ($-\infty \rightarrow +\infty$ in the Figure legends). If the decimation scheme does not converge, the corresponding point is absent.

It is  clear that decimation generically fails to converge close to the transitions both in the Soft-MB and, more severely, in the Hard-MB case. Apart from this, the three methods give results that are in remarkable qualitative agreement, including  the ability to describe discontinuities in $\langle \mu \rangle$ and $\langle \nu \rangle$ upon varying $\theta$. It is noteworthy that many different configurations appear to be feasible. These configurations are spread over a broad range of densities, especially in the Soft-MB case. So our method based on BP and decimation is able to sample the solution space by just varying a single parameter (the chemical potential $\theta$ in the present case), even in cases when only ``extremal'' solutions seem to satisfy the CSP for metabolite nodes (as e.g. in the left panel in Fig.~\ref{FBA_l3_q08}) while the density of active reaction is varying in a more continuous manner (see the right panel in the same figure).

\begin{figure}
\center
\includegraphics[scale=0.3]{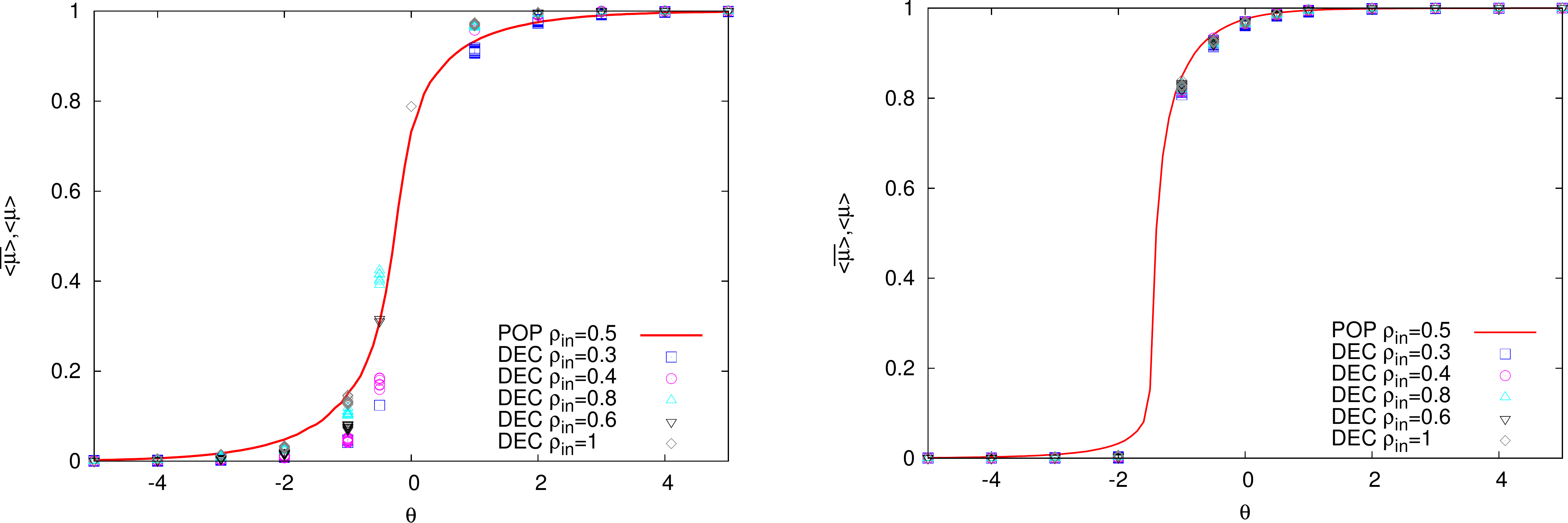}
\caption{{\bf Soft-MB}:  behaviour of the average fraction of available metabolites, $\avg{\mu}$ ($\overline{\avg{\mu}}$ for population dynamics) for $\lambda=1$ and $q=0.5$ (Left) and $\lambda=3$ and $q=0.8$ (Right) at various $\rho_{\inn}$.\label{VN_AllRhoIn}}
\end{figure}

\begin{figure}
\center
\includegraphics[scale=0.3]{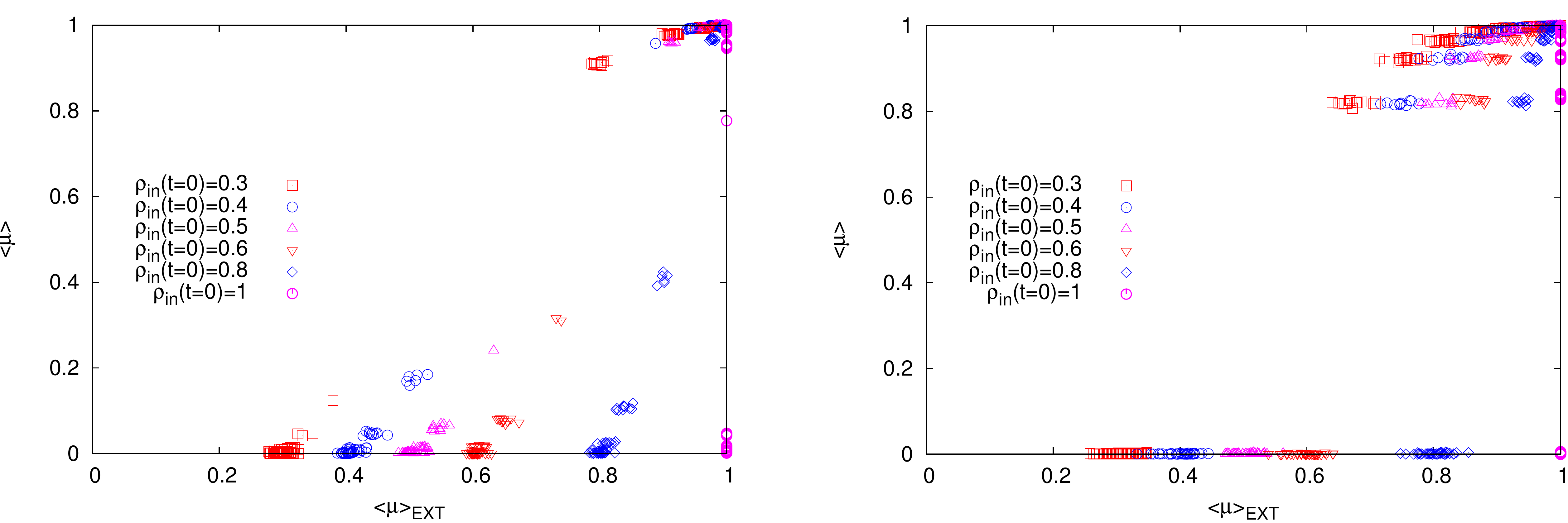}
\caption{{\bf Soft-MB}: Plot of $\avg{\mu}$ vs $\avg{\mu}_{EXT}$ for various $\rho_{\inn}$, for $\theta=(-5,-4.5,..,4.5,5)$ and for $\lambda=1$ and $q=0.5$ (Left) and $\lambda=3$ and $q=0.8$ (Right). \label{VN_rhoInVSMagnIn}}
\end{figure}

As detailed in Appendix \ref{inputs_and_outputs}, during decimation nutrients must be treated with special care. This is because the prior assignment of availability for each nutrient (which, as said above, follows a probabilistic rule with parameter $\rho_{\inn}$) does not always coincide, after decimation, with the frequency with which the nutrient is available in the final assignments (i.e. the actual solutions retrieved), which we denote as $\avg{\mu}_{EXT}$. We analyze the relation between the average magnetization of reactions and metabolites and both $\rho_{\inn}$ and $\avg{\mu}_{EXT}$ in Figures \ref{VN_AllRhoIn} and \ref{VN_rhoInVSMagnIn}. We first note that in this way we are able to obtain solutions at various $\avg{\mu}_{EXT}$ clearly different from the corresponding values of $\rho_{\inn}$. Moreover, solutions are rather stable against changes in $\rho_{\inn}$, as is to be expected expected in random networks, at least for the Soft-MB problem. Hard-MB presents however more difficulties (not shown): because it typically admits solutions with either very high or very low magnetization, it turns out to be hard to obtain solutions with $\avg{\mu}_{EXT}\neq 1$, apart from the trivial case when the whole network is inactive.

\begin{figure}
\center
\includegraphics[scale=0.3]{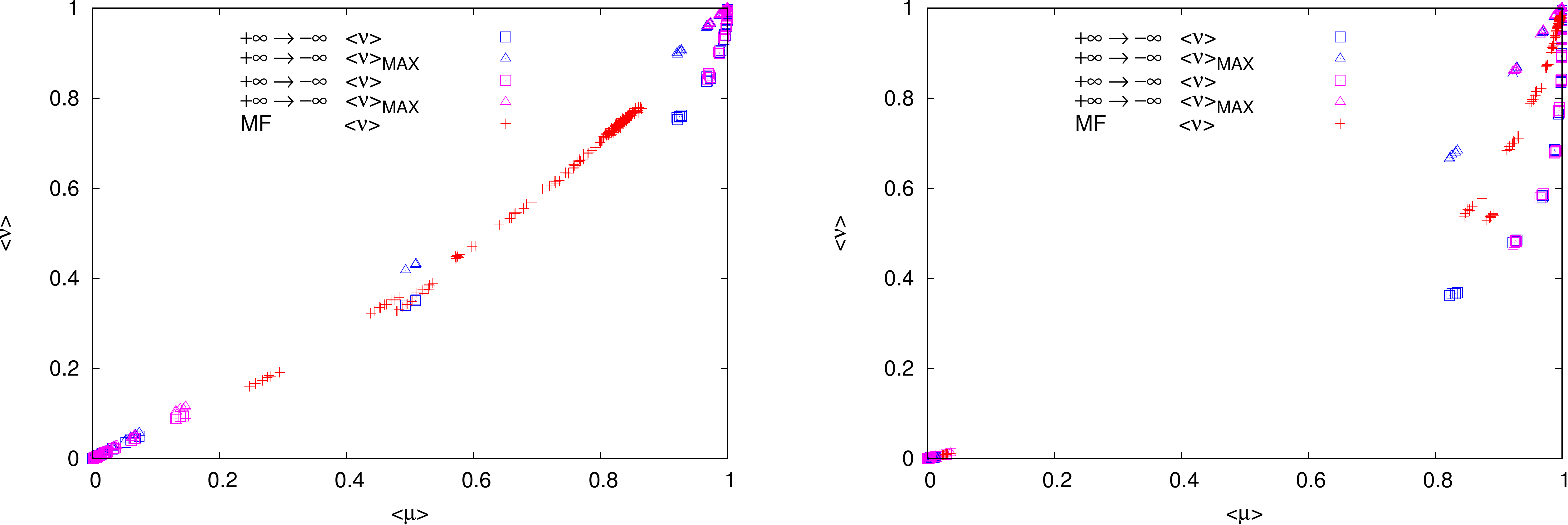}
\caption{Plot of $\avg{\mu}$ versus $\avg{\nu}$ for $\lambda=1$ and $q=0.5$ (Left) and $\lambda=3$ and $q=0.8$ (Right). \label{MF_complete_cmp}}
\end{figure}

Finally we would like to compare the solutions of the complete problem to the solutions obtained using the Mean Field Approximation (MF) presented in the previous Section. Indeed Section \ref{sec:MF_problem} we showed how to obtain solutions for the MF problem at $\theta\rightarrow\infty$ using the PEI or reverse-PEI procedure. However in order to compare the two approaches, the MF solutions at all $\theta$ have to be studied. This is possible by searching for solutions to the MF equations at finite $\theta$ (\ref{punto_fisso_theta_fin}) and then using the same decimation procedure presented in Appendix \ref{sec:decimation_explanation}. It is though important to notice that in the MF case both BP and decimation algorithm can be written in a simpler form as only one message per variable is needed, furthermore for $\theta\rightarrow\infty$ BP and decimation together behave exactly as a Warning Propagation Algorithm \cite{Braunstein_2005_survey}. 

Thus in Figure \ref{MF_complete_cmp} we present the magnetizations $\langle \mu \rangle$ and $\langle \nu \rangle$ for solutions obtained using both MF and the complete problem for various $\theta\in (-\infty,\infty)$. Here MF solutions are reported by crosses while by squares we presented the densities of the solutions obtained by the complete decimation algorithm. The $\avg{\nu}$ density in the latter solutions seems to be always smaller than in MF. However we have to remind that our CSP allows configurations where a reaction is inactive even if all its neighbouring metabolites are present. In these cases the reaction can be switched on without violating any constraint. The data marked $\avg{\nu}_{MAX}$ in Fig.~\ref{MF_complete_cmp} have been obtained by switching on all possible reaction without changing the configuration of metabolites. This is the upper bound for the reaction activity in the complete problem. 

From data shown in Figure \ref{MF_complete_cmp} it is clear that the complete problem allows for a wider variability in the values of $\avg{\mu}$ and $\avg{\nu}$. Moreover the solutions sampled in MF are a subpart of the solutions found in the complete problem. Nevertheless it could be useful on real networks because it is very simple to solve (hence sampling is much faster).

%%% Local Variables: 
%%% mode: latex
%%% TeX-master: "thesis"
%%% End: 

\chapter{Preliminary results on the  metabolic network of E.Coli}
\label{sec:real_case}

\section{Our model in the real case}
In order to use the model we have developed in the latter Chapters on a real metabolic network we have to understand what we are interested in finding. Our main goal is to understand if with our method it is possible to find functional modules of the metabolic network. These modules should be functional in the sense that they do not represent an exact overlap of the biochemical pathways but they should represent ways in which the network can actually function. Another goal is to produce configurations of the network functioning that satisfy FBA requirements and that can grow. By doing this we could actually sample suboptimal configurations of the system and then check, using FBA, whether they are growing or not, thus reducing it to a possible biological space of solutions.

In order to apply the model to real metabolic network, some adjustments have to be made. First of all it is important to understand that there is a huge quantity of external variables in E.Coli network. This is mainly due to the fact that E.Coli is a versatile bacteria that can live using many different substrates as main source of nutrition. For the purpose of this thesis we have decided to use the 7 metabolites that are essential for E.Coli to grow. Other choices could have been made but for the moment this one seemed the most interesting.

It is interesting to note the fact that from an algorithmic point of view applying the model to the real system is a highly non trivial question. This is because in the real metabolic network many loops exist thus violating the independence of the neighbours (see Section \ref{subsec:pure_states}). Another technical difficulty is connected to the fact that some metabolites participate in almost all reactions (as $H$ or $H_2O$) thus resulting in numerical precision errors when computing the messages. Nevertheless we will see that it is possible to overcome these problems and all the algorithms we presented in the previous Chapters function also in the real case.

A very important characteristics of real networks that random network for sure don't possess is the reversibility. It is possible for some reactions to function in either ways while other irreversible reactions are functioning only in one direction. This is clearly a very important aspect and it can be included in our model by adding an additional constraint between every two reversible reactions that makes it impossible for the two to function together. This clearly changes the form of the equations but as we can see in Section \ref{sec:rev_rea} not too much. It is though clear that it is impossible to add this additional constraint in the MF case for the same reason for which it is impossible to do a MF theory of the Hard-MB case: because it introduces too strong correlations to be analyzed in this way.

We expect the solution space that we are sampling using our algorithms to be huge. This not only because the bacteria has evolved during millions of years to develop the behaviours useful for his survival, but also because we expect to sample a space bigger than the biological one. It is thus important to make some assumption on the solutions we will sample to restrict the possible outcome. We already enounced the most important assumption that is the request that the 7 nutrients and that the atpm reaction are present. Furthermore we will request in the following that a solution should also be a single connected component. This is to ensure that we are finding a solution that connects the inputs to the outputs of the system.

\section{Reversible reactions}
\label{sec:rev_rea}
One of the most important features in real networks is the presence of reversible and irreversible reactions. The latter are reactions that can only function in one defined direction while for the others both directions are possible. In the metabolic network of E.Coli almost $40\%$ of reactions are reversible, hence this is a very important characteristic.

To include this behaviour in the model we consider that each reversible reaction is represented in our network as a couple of reactions $(i,-i)$ with reactants and products exchanged and whose states $(\nu_i,\nu_{-i})$ are coupled by a ``reversible'' constraint:
\begin{equation}
 \label{constraint_reversibility}
\Omega(\nu_i,\nu_{-i})=\delta_{\nu_i,0}\delta_{\nu_{-i},0}+\delta_{\nu_i,1}\delta_{\nu_{-i},0}+\delta_{\nu_i,0}\delta_{\nu_{-i},1}=1-\nu_{i}\nu_{-i}.
\end{equation}
Asking for this constraint to be satisfied is equivalent to say: if direct (reversed) reaction is active, reversed (direct) reaction has to be inactive. But if direct (reversed) reaction is inactive, reversed (direct) reaction is not constrained.

Hence the marginal of a reversible couple of reaction is given by 
\begin{align}
  \begin{cases}
    \tilde{p}(0,0)=p_i(0)p_{-i}(0)/W^i\\ \nonumber \\ 
    \tilde{p}(1,0)=p_i(1)p_{-i}(0)/W^i\\ \nonumber \\ 
    \tilde{p}(0,1)=p_i(0)p_{-i}(1)/W^i\\
  \end{cases}
\end{align}
where $W^i=p_i(0)+p_i(1)p_{-i}(0)=p_{-i}(0)+p_{i}(0)p_{-i}(1)$ and $p_i(\nu_i)$ is given by equation (\ref{proba_rel}). This can be written as:
\begin{align}
  \tilde{p}(\nu_i,\nu_{-i})=\left[(1-\nu_i)p_i(0)\left((1-\nu_{-i})p_{-i}(0)+\nu_{-i}p_{-i}(1)\right)+\nu_i(1-\nu_{-i})p_i(1)p_i(0)\right]/W^i
\end{align}
Then the marginal at convergence is given by:
\begin{align}
  \tilde{p}(\nu_i)=\sum_{\nu_{-i}}\tilde{p}(\nu_i,\nu_{-i})=\frac{(1-\nu_i)p_i(0)+ \nu_ip_i(1)p_{-i}(0)}{W^i}
\end{align}
It is thus straightforward to understand that reversible reactions will send a message of the type:
\begin{align}
&  \begin{cases}
\mess{\tilde{\psi}}{i}{a}_{\nu_i}=\left[(1-\nu_i)\mess{\psi}{i}{a}_0+\nu_i \mess{\psi}{i}{a}_1\mess{\psi}{-i}{a}_0\right]/\Zt{i}{a}, \\ \\
    \mess{\tilde{\eta}}{i}{e}_{\nu_i}=\left[(1-\nu_i)\mess{\eta}{i}{e}_0+\nu_i\mess{\eta}{i}{e}_1 \mess{\eta}{-i}{e}_0 \right]/\Zt{i}{e},  \\ \\
  \end{cases}
\end{align}
where messages $\mess{\eta}{(-)i}{e}$ and $\mess{\psi}{(-)i}{a}$ are taken from Section \ref{cavity_derivation}. Therefore it is possible to simulate a metabolic network with reversible reactions by modifying the equations introduced in Section \ref{cavity_derivation} changing messages $\mess{\eta}{(-)i}{e}$ and $\mess{\psi}{(-)i}{a}$.

\section{Mean Field Approximation}
As we already explained the MF case for $\theta\rightarrow\infty$ is a different formulation of a better known problem called Network Expansion (NE) (see Section \ref{sec:NE}). We will thus present in a first part a comparison between the results obtained by our algorithm and the results it is possible to obtain in NE showing that our method and the previous method are consistent. In a second part we will focus on the study of the solution space of the expanded MF model where $\theta$ can be different from $\infty$, showing that by analyzing the solutions, it is possible to recover meaningful dynamical modules of the network.
\subsection{Comparison to Network Expansion}
As already presented in Section \ref{sec:BP_Decimation}, with our method it is possible to recover the probability that a metabolite is present (absent) or that a reaction is functioning (not functioning). Furthermore for $\theta\rightarrow \infty$ the solution at convergence is also a configuration while for $\theta$ finite to obtain a configuration a decimation procedure is used.

\begin{figure}
  \centering
  \includegraphics[scale=0.4]{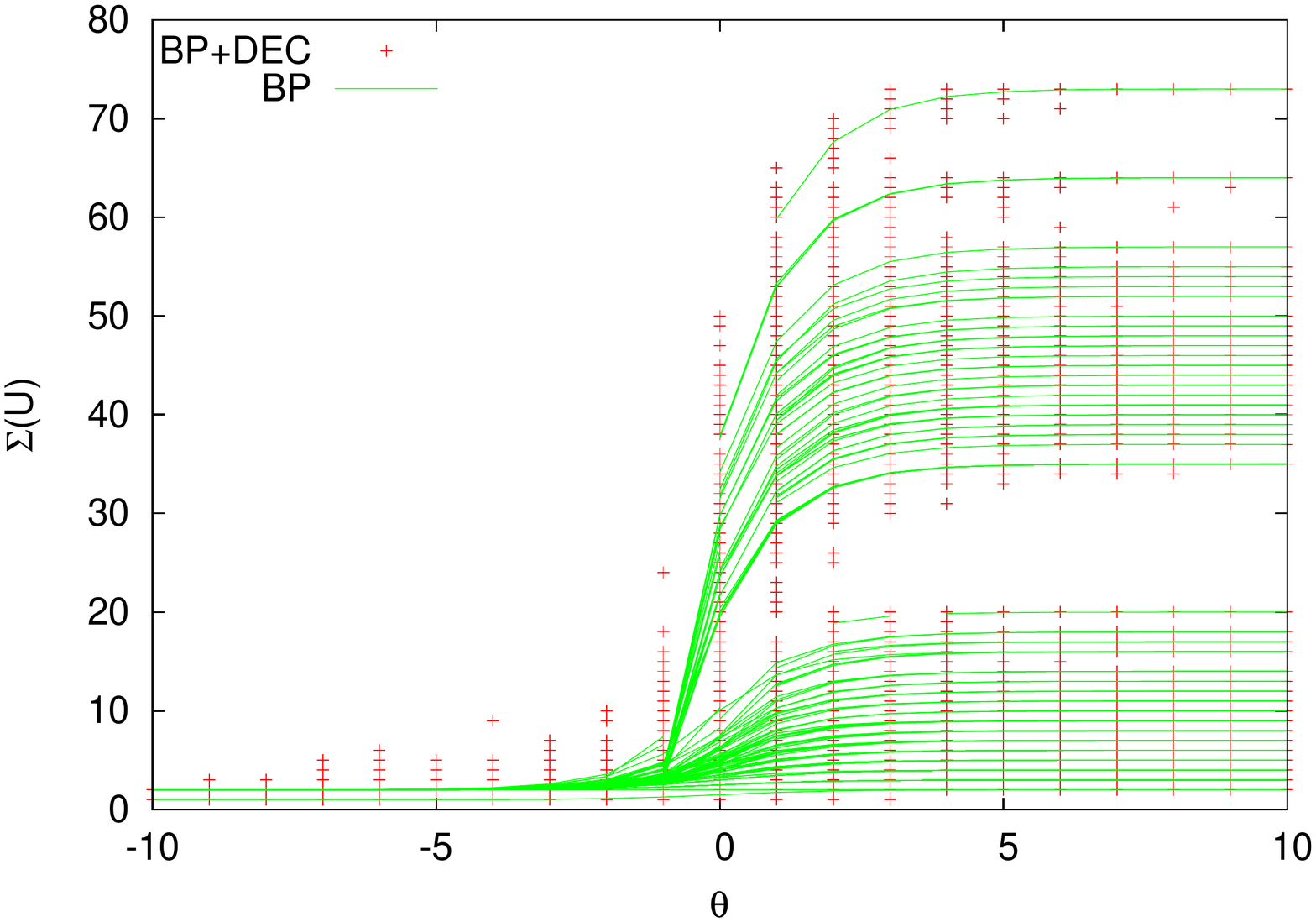}%0.25  
  \caption{Scope of the seed versus $\theta$. The points are after the decimation process, while the lines are given only by the BP procedure}
  \label{NE_VS_MF}
\end{figure}

In Figure \ref{NE_VS_MF} we show the result of the decimation procedure for the network of E.Coli. To obtain the results in this Figure we used the method already presented in Section \ref{sec:NE}. In this method water is always present and at turns each metabolite is used as seed (together with the water). As we can see from this Figure the size of the scopes grow in $\theta$ and reach a maximum at $\theta\rightarrow \infty$. This is as expected as $\theta$ regulates the number of active reactions. It is though interesting to note that the results found at $\theta\rightarrow \infty$ reproduce the one in article \cite{Kruse_2008} thus showing that with our method it is possible to recover known behaviours and generalize it. 

\subsection{The solution space}
\label{sec:solution_space}
We want to explore the solution space of the MF model (equations (\ref{punto_fisso_theta_fin})) and understand if there is some interesting properties of the solution space. As we already explained, a good way to restrain the solution space to meaningful biological solutions is by requiring that the solution found at convergence is connected. Furthermore we will require in this part that the 7 nutrients necessary for E.Coli to grow (see Section \ref{sec:real_net}) are present.

\begin{figure}[h]
\center
 \includegraphics[scale=0.4]{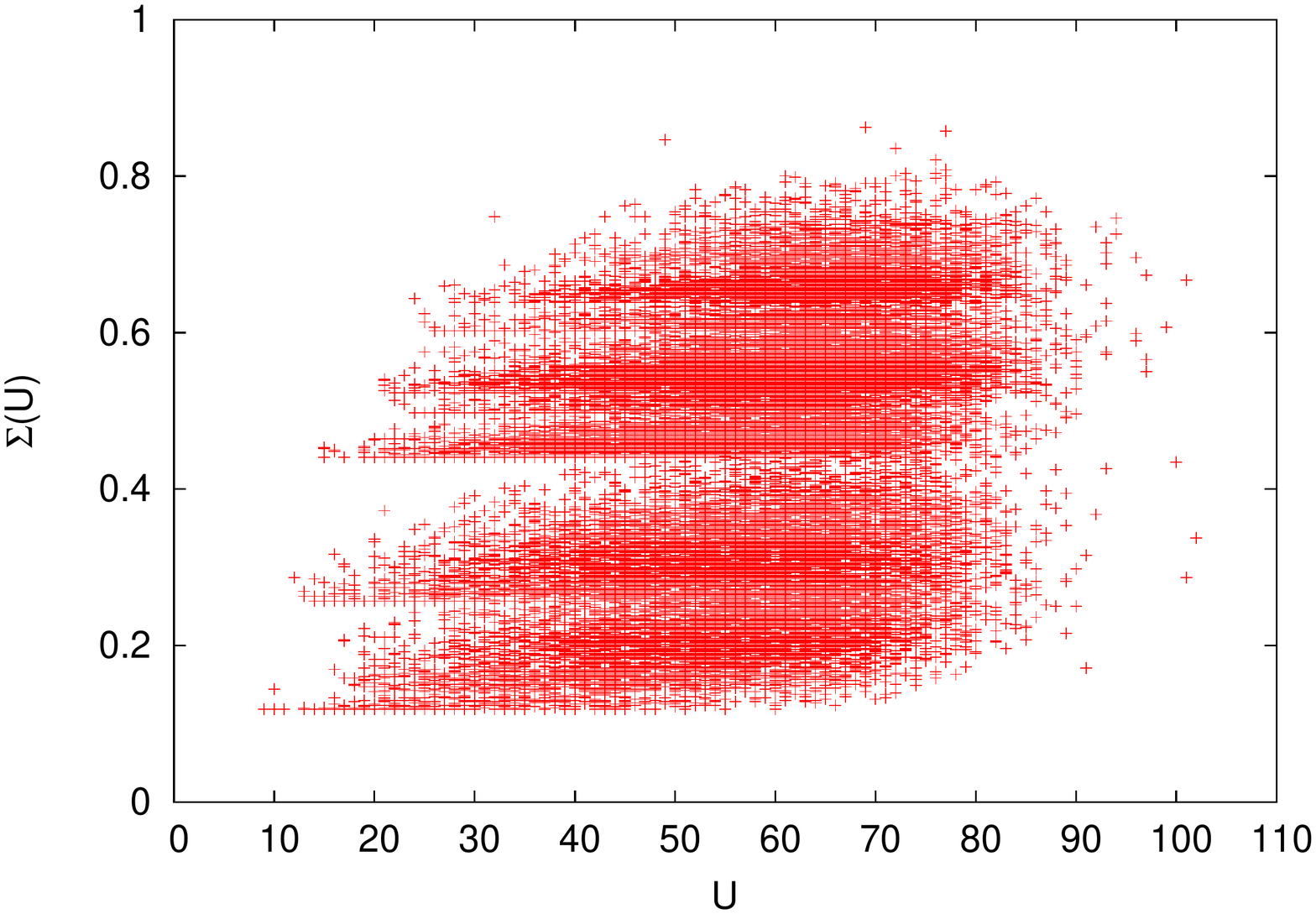}
\caption{Scope $\Sigma(U)$ of the seed $U$ in the region between $p=[0,0.1]$ for $\theta\rightarrow\infty$ with one connected component and 7 nutrients ON, here $N_{rip}=10^4$.}
\label{one_component}
\end{figure}

The protocol that we will use in this part to sample the solutions will be to require that the internal metabolites are present with a probability $p$ little (ranging from $0$ to $0.1$) while the nutrients have to be present. This procedure is repeated $N_{rip}$ times thus the total number of solutions will depend on this parameter. In Figure \ref{one_component} we can see the results of this procedure for the network of E.Coli at $\theta\rightarrow\infty$ where, inspired by NE, we defined the scope $\Sigma(U)$ as the fraction of metabolites present at convergence given the seed $U$ present at time $t=0$. It is interesting to notice that it is possible to arrive to a scope $\Sigma(U)\simeq 0.9$ with as little $100$ metabolites ($14\%$ of the total size) present at the beginning.

\begin{figure}
\center
 \includegraphics[scale=0.3]{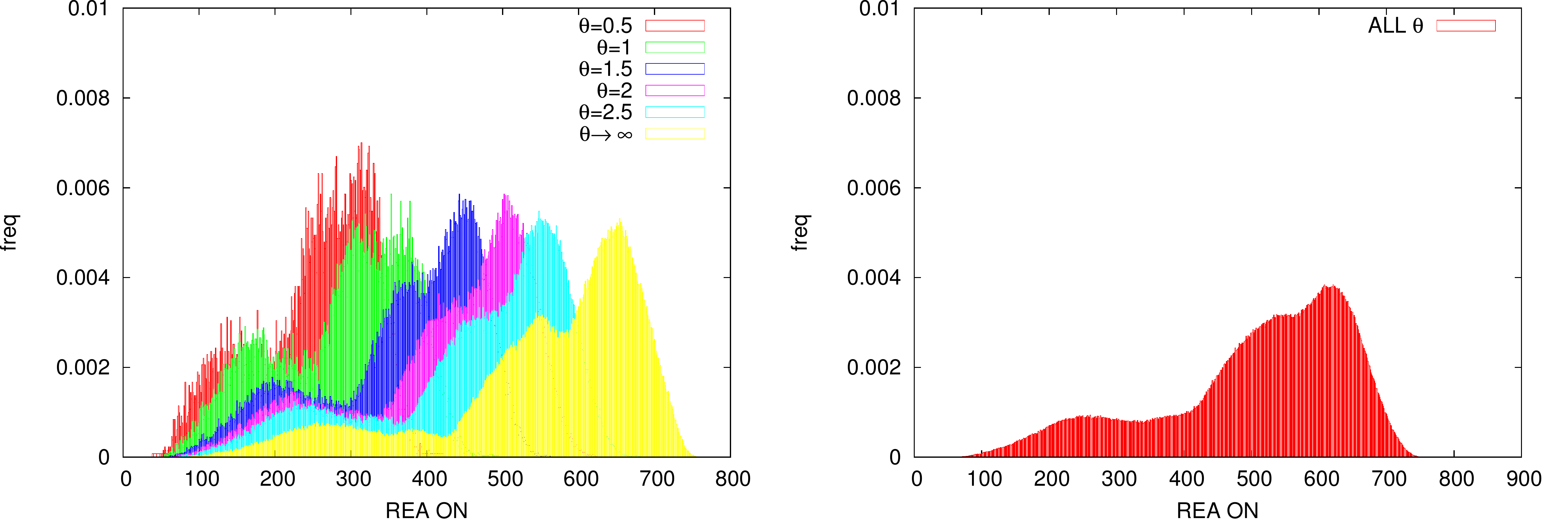}
\caption{Histogram of the number of reactions functioning in the region between $p=[0,0.1]$ for some $\theta$ on the left and the cumulative plot for all $\theta$ on the right. All solutions have one connected component and 7 nutrients ON, here $N_{rip}=2.10^4$.}
\label{histRea_inTheta}
\end{figure}

By changing $\theta$, it is possible to sample solutions with different numbers of reactions functioning obtaining Figure \ref{histRea_inTheta}. Here we see clearly that changing $\theta$ it is possible to sample solutions with bigger number of functioning reactions. Looking at the right part of Figure \ref{histRea_inTheta}, that is the cumulative histogram over all $\theta$, it is clear that there is a part of the solution space for $N_{rea} \in [400,800]$ that it is easier to sample. We expect that these two parts, having different average number of active reactions, will have different properties. Thus in the following we will analyse these two parts independently and we will refer to $\mathcal{R}$ for the right part ($N_{rea} \in [400,800]$) and $\mathcal{L}$ for the ``left'' part ($N_{rea}\in [0,400]$).

In the following we will focus on the reactions as they have already been divided in pathways (see Chapter \ref{chap:biological_back}) thus it will be easier to understand if the modules we are finding have an actual biological interpretation. But in order to find out if there exist an organization of the reactions for the solutions we are sampling, we have to decide in which way we want to analyze the correlation. In a boolean case, the definition of a useful correlation is not trivial at all. This is because with boolean variables it is possible that some variables are frozen thus it is not very interesting to use a correlation in which the fluctuation around the mean is used. Furthermore we are interested in recovering groups of reactions that are working together. Ideally if two reactions are inside the same functional module, they should always be working/not working together. Thus a good correlation matrix is one of the form:
\begin{equation}
C_{ij}=\nu_i \nu_j + (1-\nu_i)(1-\nu_j),
\label{corr_def}
\end{equation}
where $\nu_i=\{0,1\}$ is the state of variable $i$.
Thus we computed the correlation matrix for both $\theta\rightarrow\infty$ and the aggregated data for all $\theta$ together. Then using the correlation matrix, it is possible to cluster the reactions in modules. In order to do this we have first to choose a cutoff, $\tilde{C}$ such that $C_{ij}=1$ if $C_{ij} > \tilde{C}$ and $0$ otherwise. The $C_{ij}\in \{0,1\}$ that we obtain at the end of this process can be considered as an adjacency matrix of a ``correlation graph''. It is thus possible to find the connected components of this graph, obtaining the modules of reactions that are correlated. 
\begin{figure}
\center
\includegraphics[scale=0.3]{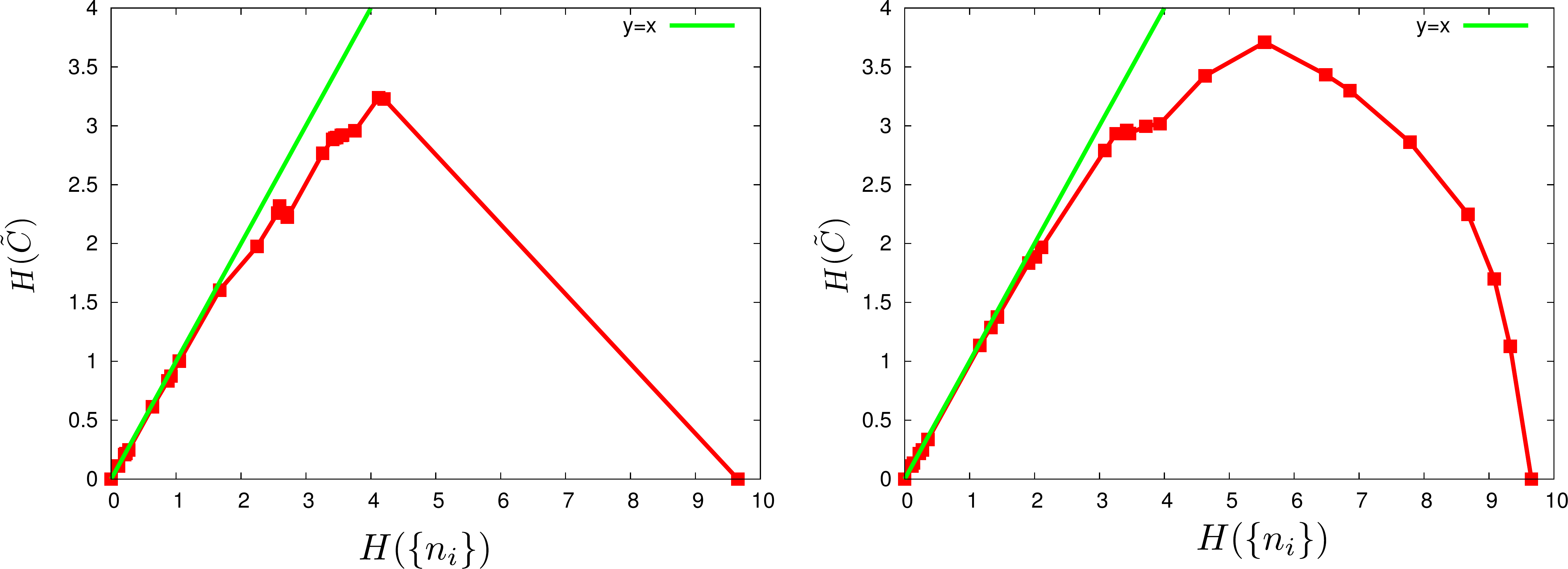}
\caption{$H({\tilde{C}})$ in function of $H(\{n_i\})$ for both $\theta\rightarrow\infty$ (left) and for the cumulative solutions at all $\theta$ on the right. $N_{rip}=2.10^4$ using solutions in $\mathcal{R}$.}
\label{entr_MF_DX}
\end{figure}

Clearly all this process is highly dependent on the choice of $\tilde{C}$ as for $\tilde{C}=0$, all reactions will be in the same module while for $\tilde{C}=1$ all reactions will be in separate modules. A method to choose the right cutoff is given in article \cite{marsili2013sampling}. Here authors consider that the right $\tilde{C}$ can be chosen by finding the maximum of $H(\tilde{C})$ versus $H(\{n_i\})$, where $H(\{n_i\})$ can be seen as the entropy intrinsic in the sample and $H(\tilde{C})$ is the entropy of the particular choice of clustering made. Considering a choice $\tilde{C}$ one can define $N_{mod}$ as the number of modules obtained by clustering and $n_i(\tilde{C})$ as the number of modules with size $i$, thus $N_{mod}=\sum\limits_{i=1}^M n_i$. Using these definitions it is possible to write:
\begin{equation}
  \label{eq:entr_ni}
  H(\{n_i\})=-\sum_{i=1}^M\frac{n_ii}{M}\log\left(\frac{i}{M}\right),
\end{equation}

\begin{figure}
\center
\includegraphics[scale=0.3]{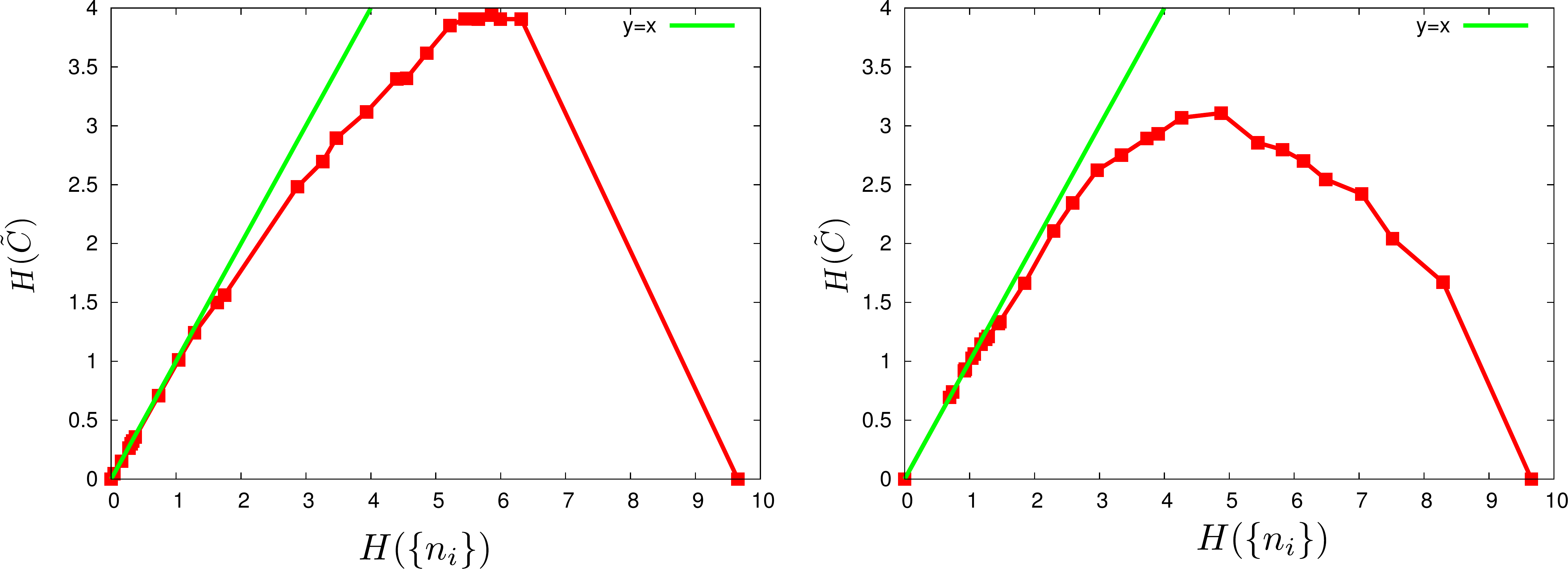}
\caption{$H({\tilde{C}})$ in function of $H(\{n_i\})$ for both $\theta\rightarrow\infty$ (left) and for the cumulative solutions at all $\theta$ on the right. $N_{rip}=2.10^4$ using solutions in $\mathcal{L}$.}
\label{entr_MF_SN}
\end{figure}

where we have written the probability that a reaction is in a cluster with size $i$ as $\frac{i}{M}$. Furthermore $H(\tilde{C})$ can be written as:
\begin{equation}
  \label{eq:entr_C}
  H(\tilde{C})=-\sum_{i=1}^M\frac{n_i(\tilde{C})i}{M}\log\left(\frac{n_i(\tilde{C})i}{M}\right).
\end{equation}
It is thus possible to compute these quantities in our solutions as presented in Figure \ref{entr_MF_DX} and \ref{entr_MF_SN} for both $\theta\rightarrow\infty$ on the left and for the cumulative result on all $\theta$ on the right.

\begin{figure}
\center
\includegraphics[scale=0.3]{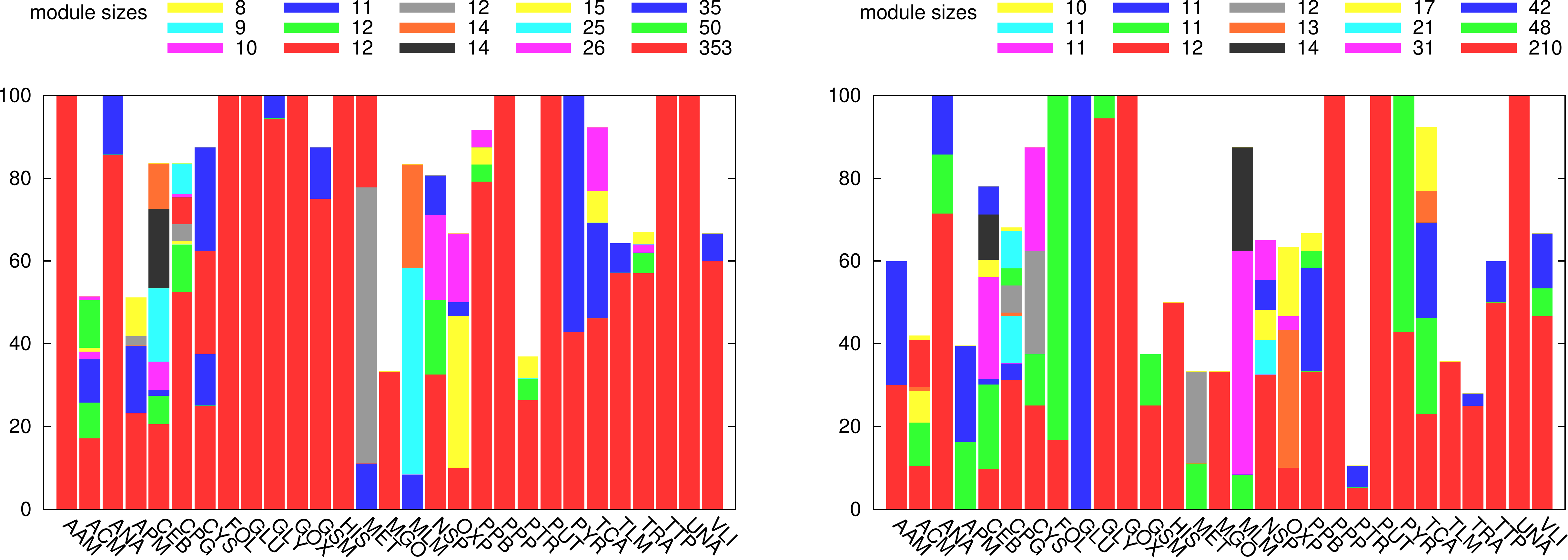}
\caption{On the $x$ axis there is the name of the pathway while on the $y$ axis there is the percentage of this pathway inside the biggest 15 modules for $\theta\rightarrow\infty$ (left) and all $\theta$ together. In the legend there is the size of the module. This plot is for $N_{rip}=2.10^4$ and for the right part of the solution space.}
\label{modules_DX_MF}
\end{figure}

\begin{figure}
\center
\includegraphics[scale=0.3]{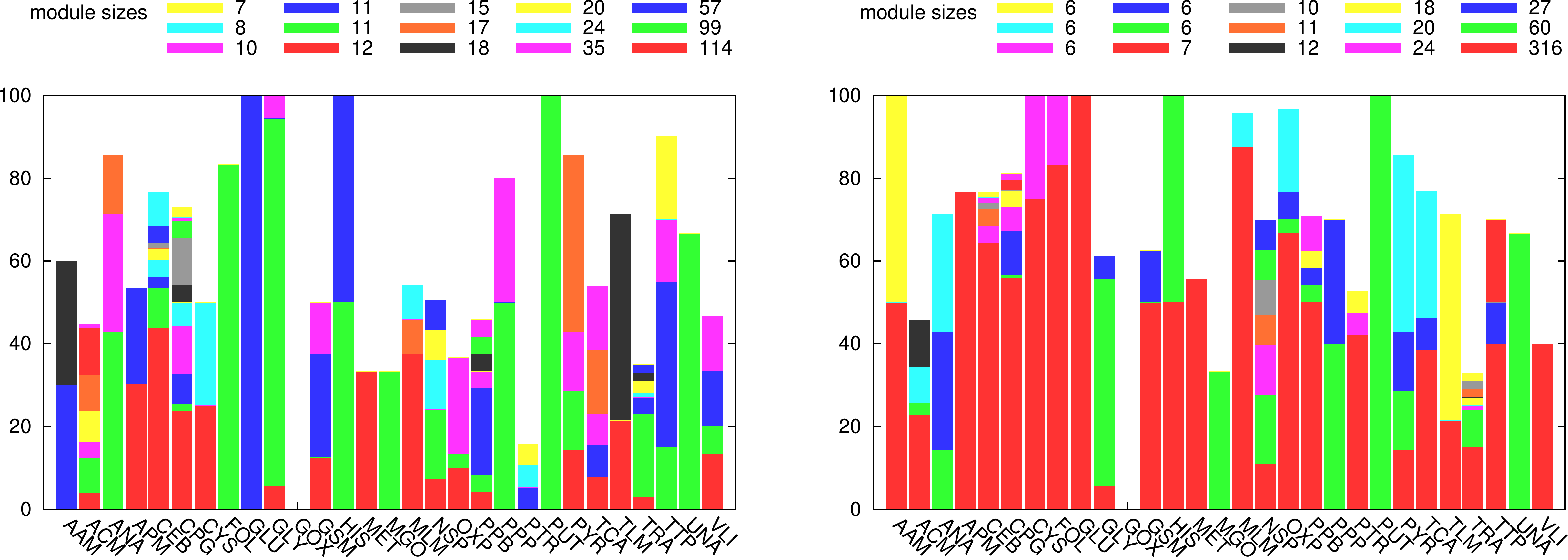}
\caption{On the $x$ axis there is the name of the pathway while on the $y$ axis there is the percentage of this pathway inside the biggest 15 modules for $\theta\rightarrow\infty$ (left) and all $\theta$ together. In the legend there is the size of the module. Plot for $N_{rip}=2.10^4$ and for the left part of the solution space.}
\label{modules_SN_MF}
\end{figure}

\begin{figure}
\begin{center}
\includegraphics[scale=0.3]{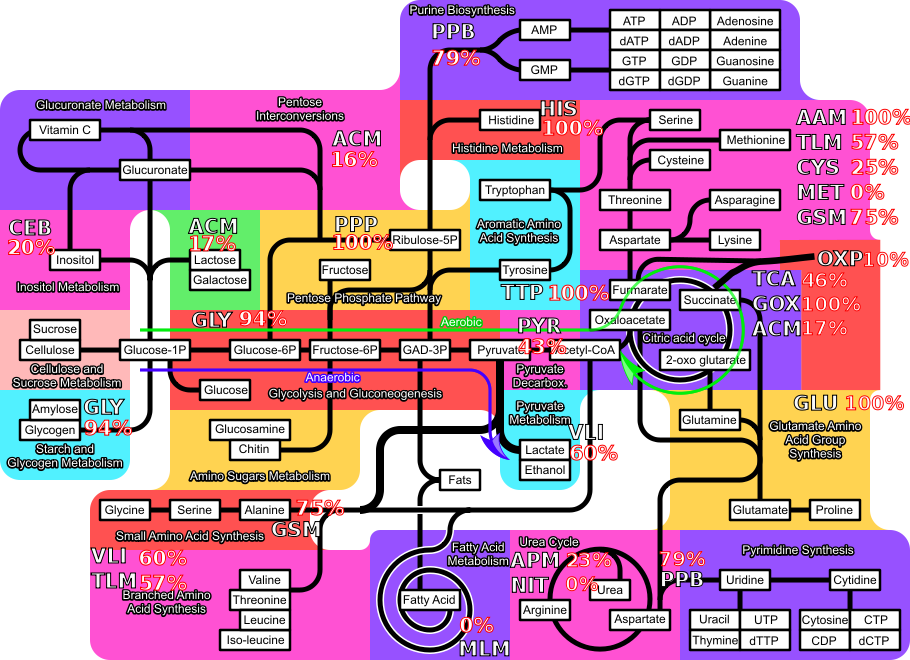}
\caption{Representation of the modules we find divided in pathways for the right part of the solutions. The numbers below the pathways are the percentage of reactions of the pathway switched on. The cutoff is at 0.983.}
\label{pathways_dx}
\end{center}
\end{figure}

Thus using the value of the cutoff at the maximum of $H(\tilde{C})$, it is possible to recover the modules that best represent the reaction interaction for both the ``right'' and the ``left'' part of the solution space. Results are presented in Figures \ref{modules_DX_MF} for the right part and in \ref{modules_SN_MF} for the left part; in each Figure we plotted the percentage of a pathway that is contained in the 15 biggest modules of the network, for both $\theta\rightarrow \infty$ (left) and all together (right). Thus if a pathway do not reach $100\%$ in this plot, this means that its reactions are not inside one of the biggest module. In Figure \ref{modules_DX_MF} we see that the modules on the right for $\theta\rightarrow\infty$ contains all the necessary pathways for the aerobic respiration, considering then that the uptakes of the $O_2$ is always functioning (as it is one of the nutrients), we can be positive that this is the case. This is better represented in Figure \ref{pathways_dx} where the biggest pathway on the right is superposed to the metabolic map, to stress what this module function is. Clearly the whole respiration is a composition of many modules, but the biggest module is already taking most of the behaviour. The same is not true on the left part, where we have still not a good interpretation of what the function of these modules could be. It is interesting the fact that in each case, the modules that it is possible to find using all $\theta$'s or only $\theta\rightarrow \infty$ can differ in some components but generally seem to represent the same biological function as the most important pathways are present. Furthermore, the fact that in Figure \ref{modules_SN_MF} the biggest module using all $\theta$ is much bigger than the one only at $\theta\rightarrow\infty$ seems to suggest that part of the accuracy of the reconstruction of the modules is dependent on how many solutions are sampled in this part. It is also interesting that some modules take exactly a pathway and that some pathways are divided into two or three modules only.

\section{Soft and Hard Mass Balance}

\begin{figure}
\center
\includegraphics[scale=0.3]{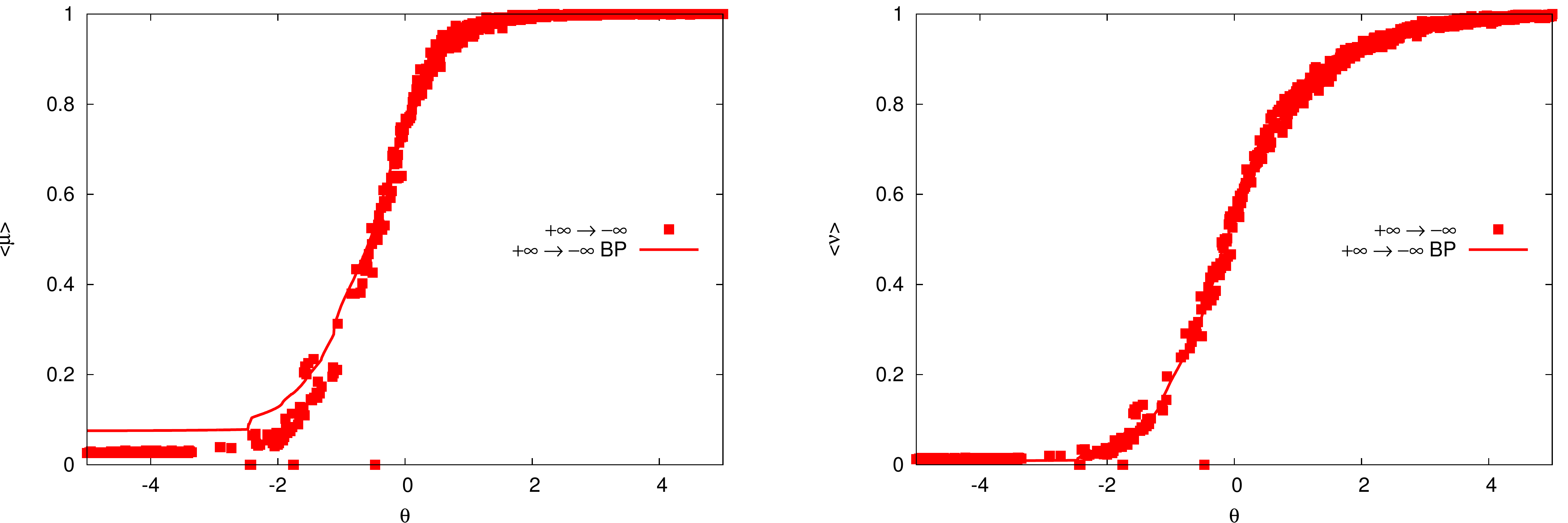}
\caption{Soft-MB solutions for the reactions (right) and the metabolites (left) in the irreversible case. Solutions are found using the BP algorithm (BP) and the decimation (DEC) algorithm for various $\theta$. Here we present only the results for the protocol $+\infty\rightarrow-\infty$ because the other protocol gives the same results.}
\label{fig:BP_DEC_VN}
\end{figure}

\begin{figure}
\center
\includegraphics[scale=0.3]{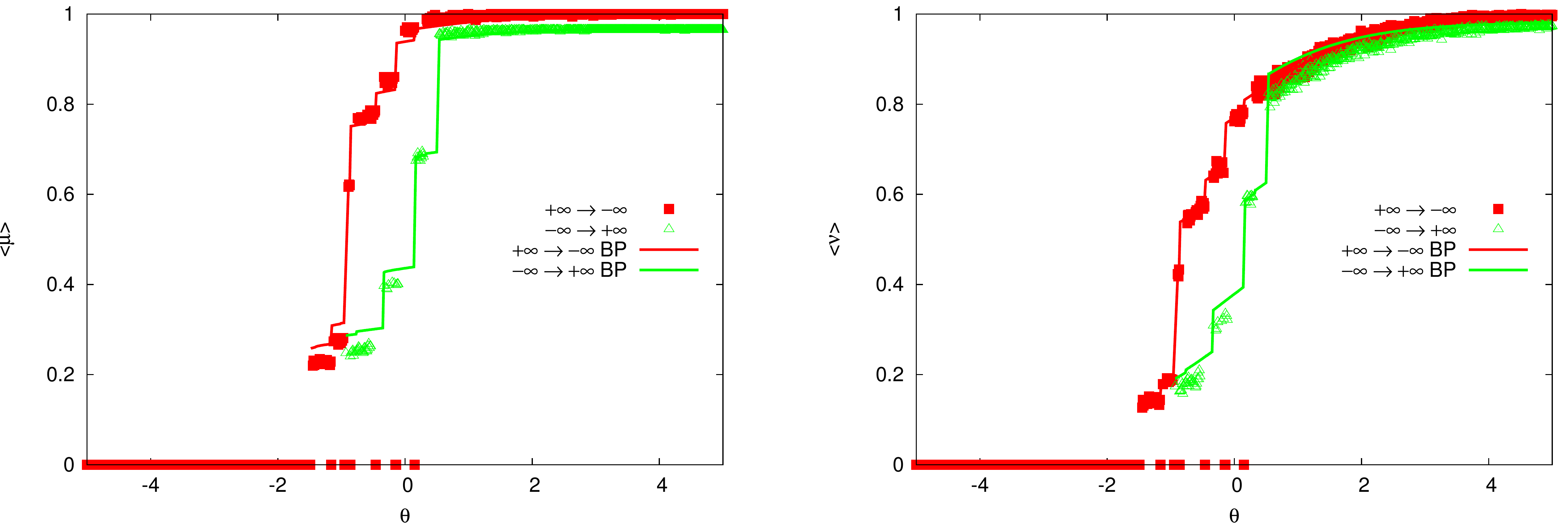}
\caption{Hard-MB solutions for the reactions (right) and the metabolites (left) in the irreversible case. Solutions are found using the BP algorithm (BP) and the decimation algorithm for various $\theta$.}
\label{fig:BP_DEC_FBA}
\end{figure}

In this section we want to review the main results obtained so far in the Soft- and Hard-MB case. First of all we would like to understand if the properties of these CSPs observed in random networks hold also in real networks. In order to do this, we developed a BP algorithm solving the cavity equations derived in Section \ref{cavity_derivation} for the non reversible case. Furthermore it is also possible to apply the decimation procedure presented in Appendix \ref{chap:algorithms} to obtain configurations of metabolites and reactions that satisfy our constraints. It is thus possible to sample the phase space using the same procedure presented in Chapter \ref{sec:random_case} finding the results presented in Figure \ref{fig:BP_DEC_VN} for Soft-MB and in Figure \ref{fig:BP_DEC_FBA} for Hard-MB. In this plots we present the results for $\rho_{in}=1$ and using the E.Coli network in which all reversible reactions are written once.

As already explained in Section \ref{sec:BP_Decimation}, in order to obtain configurations of the system using the decimation, the nutrients are decimated according to the internal marginal. This means that, as opposed to the MF case, it is possible that in the complete problem, not all of the 7 nutrients are used together in a solution as the only compulsory requirement is that the solution is connected and that the ATPM is functioning. It is then possible to verify which solution is using which nutrient, to understand the organization of the solution space (work in progress).

\begin{figure}
\center
\includegraphics[scale=0.4]{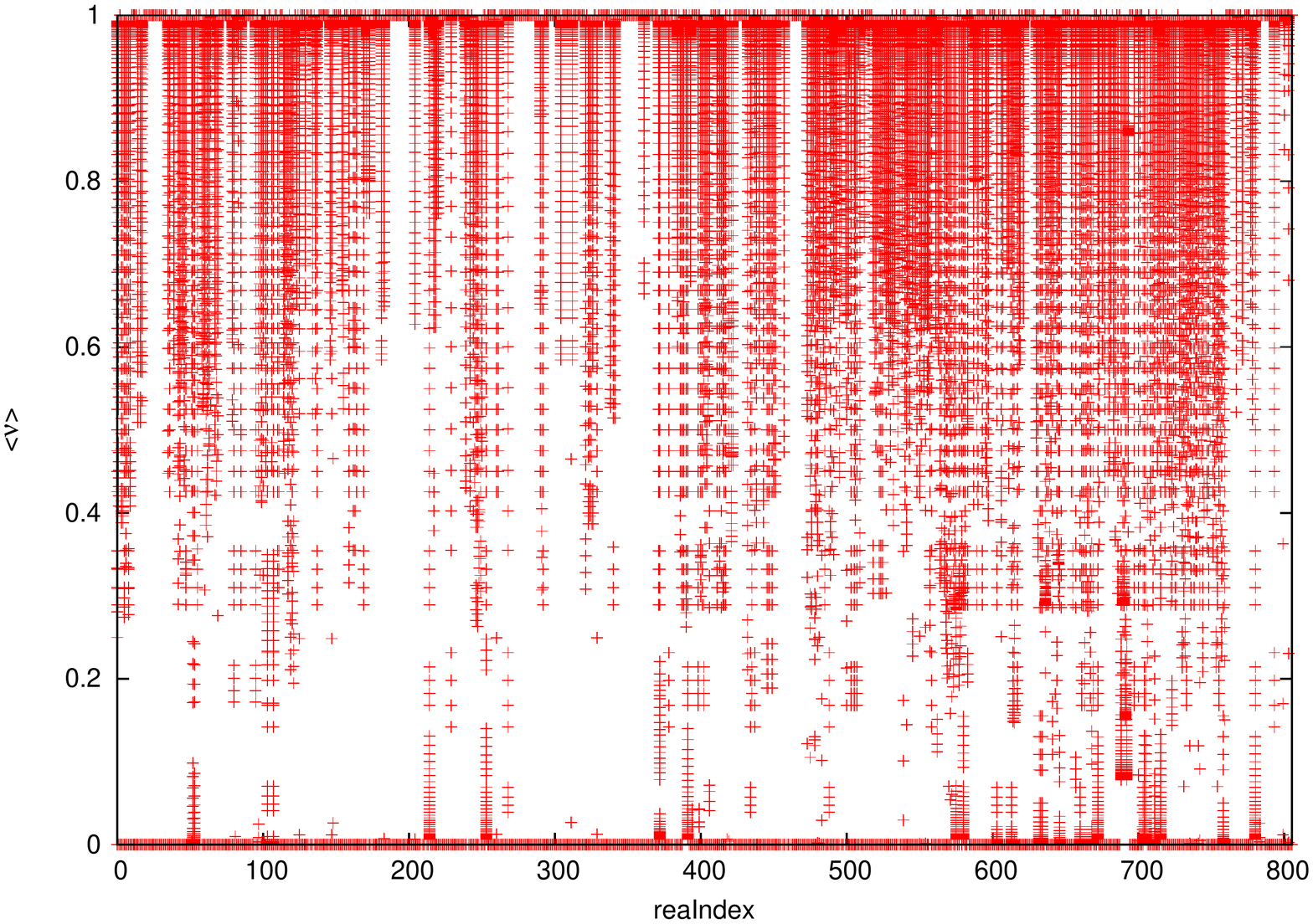}
\caption{Marginal of the reactions in Hard-MB, for the case $+\infty\rightarrow\-\infty$ and $\rho_{in}=1$.}
\label{fig:FBA_MARG}
\end{figure}

It is thus clear from these results that exactly as we found on the random network, Soft-MB seems to  undergo a continuous transition while in Hard-MB the transition is discontinuous. In the random network, due to the presence of the giant components, in the Hard-MB case the only values of the magnetization possible where $1$ or $0$. Instead here many possible states are reached during the transition, showing the complexity of the real network. We expect that each plateau of the magnetization represents a state of the system and that each time there is a transition, the system is switching from one state to another. This is seen in Figure \ref{fig:FBA_MARG} where we see the marginal of the reactions for all $\theta$'s. As we can see, the marginals are polarized and not all values of the marginal are possible.

%%% Local Variables: 
%%% mode: latex
%%% TeX-master: "thesis"
%%% End: 

\chapter{Conclusions and Outlook}
In this Thesis we presented a work on the metabolic network from both a theoretical and a practical point of view. We first recalled basic biological and statistical mechanics concept and theories in Chapters \ref{chap:biological_back} and \ref{sec:stat_mech}. Then we defined in Section \ref{sec:boolean_problem} two novel types of Constraint Satisfaction Problems representing known biological approaches in the study of metabolic networks (FBA \ref{bio:FBA} and VN \ref{sec:VN}). In order to study this problem we presented the representation of the real metabolic network and its random equivalent the Random Reaction Network in Sections \ref{sec:real_net} and \ref{sec:RRN}. We thus studied the problem in the random system by deriving the cavity equations for the problem \ref{cavity_derivation} finding a convenient notation to write both CSP problems in a common way. Then we studied the random problem in depth by studying both the RRN ensemble (using population dynamics) and the single RRNs (using BP) in Sections \ref{sec:pop_dyn} and \ref{sec:BP_Decimation}. We also showed that the Mean Field approximation of our model, Section ~\ref{sec:MF}, correspond to the Network Expansion approach for metabolic network (see Section~\ref{sec:NE}). Finally in Section \ref{sec:real_case} we showed preliminary results of applying our algorithm on real metabolic networks.

We wanted by this study to understand the possibilities behind a theoretical approach in metabolic network analysis. Most of this Thesis was devoted to understand how to effectively model known biological approaches on random reaction networks. It was difficult to find how to formalize these approaches in terms of CSPs but we believe to have shown that the study of the properties of the Soft-MB (both MF and complete) and Hard-MB shows a highly non trivial organization of the phase space on random networks. Furthermore, we think that it is overall interesting to understand how the various approaches developed by biologists are related to them.

If on the one hand, we believe our work to be a quite comprehensive study of the properties of the Soft-MB and Hard-MB problems on random networks, on the other hand the results presented on the real network are only a sketch of the potentiality of our method. Indeed in the real network we have encountered many difficulties as long as the algorithms are concerned. It is though conforting that the irreversible Soft-MB and Hard-MB problems give a interesting insights on the properties of the real network. Especially the behaviour of the magnetization on the Hard-MB problem show that with our approach it is possible to highlight properties of the structure of the real network. Furthermore we believe that the results on the functional modules using the MF approximation could be interesting from a practical point of view and could be used by biologist to better understand the metabolic network and its properties.

In the near future, we will try to takle the most important issues. The first step will be to develop the reversible BP and decimation algorithm, applying the considerations of Section \ref{sec:rev_rea}. Then we would like to analyze the Soft-MB solutions by dividing them in classes of input and output usage. Thus it shoule be possible to divide these groups of solutions in modules of reactions, using the method presented in Section \ref{sec:solution_space}, and understand which biological functions we are actually sampling. Hopefully the analysis in this case should extend the results on functional modules of the MF case. The second step will be to select the reactions in the Hard-MB case that have a polarized marginal. It should then be possible to use these informations about the reactions to sample solutions of FBA (by running a linear programming algorithm) that are growing less than the maximal growth rate.

%%% Local Variables: 
%%% mode: latex
%%% TeX-master: "thesis"
%%% End: 

\appendix

\chapter{Nutrients and Outputs}
\label{inputs_and_outputs}

\section{Random case}
\begin{figure}[h]
  \centering
  \includegraphics[scale=0.35]{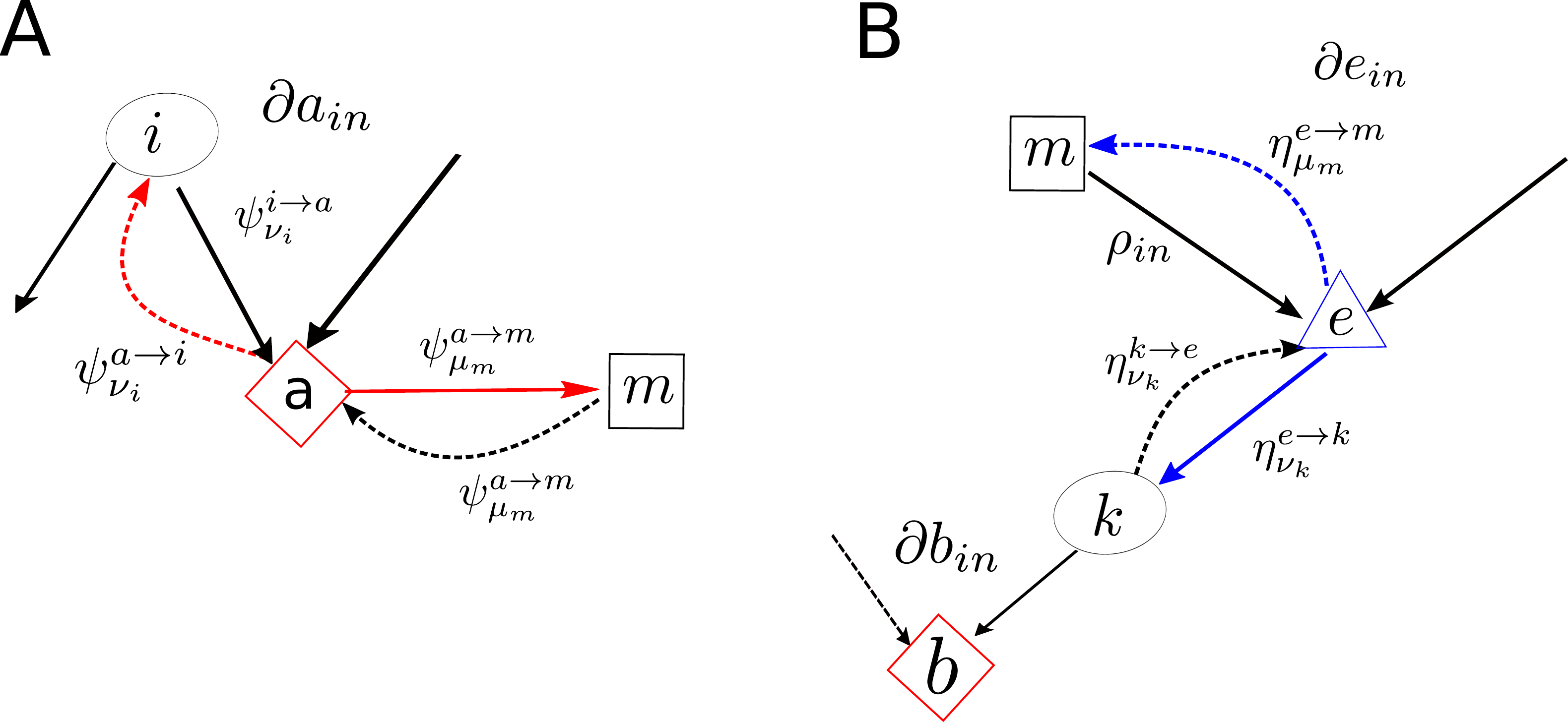}
  \caption{Representation of the external metabolites in our network. \textbf{A} is the product while \textbf{B} is the nutrient.}
  \label{nutrOut_schema}
\end{figure}

How to deal with the nutrients (metabolites with in-degree 0 and out-degree larger than 0) and the outputs (metabolites with in-degree larger than 0 and out degree 0) is probably the trickiest part of the network analysis.  Indeed looking at the cavity equations derived in \ref{cavity_derivation}, we immediately see that nutrients and outputs are automatically switched off because in these cases $\mess{\psi}{a}{m}_{\mu_m}=\delta_{\mu_m,0}$, while in real systems these variables are usually active, as they represent the interaction with the environment. To overcome this limitation we will consider in the following that nutrients are \textit{external} variables fixed by the environment and thus have a probability of being present $\rho_{in}$. Furthermore these variables send a message to the neighbouring reaction-constraint of the type:
\begin{gather}
\mess{\eta}{m}{e}_{\mu_m}=  (1-\rho_{in})\delta_{\mu_m,0} + \rho_{in}\delta_{\mu_m,1}.
\end{gather}
On the other hand the products are \textit{internal} variables with no reaction constraint node associated, a probability of being present $p(\mu_{m})=\mess{\psi}{a}{m}_{\mu_m}$ (taken from \ref{proba_rel}) and send a message:
\begin{gather}
\label{output_message}
\mess{\psi}{m}{a}_{\mu_m}=\mess{\psi}{a}{m}_{\mu_m},
\end{gather}
to the neighbouring metabolite-constraint node. Furthermore the metabolite-constraint has to be a Soft-MB constraint otherwise the outputs would be always off in the Hard-MB case. A schematic view of the form of the network for the external metabolites is presented in Figure \ref{nutrOut_schema}.

\begin{figure}
  \centering
  \includegraphics[scale=0.45]{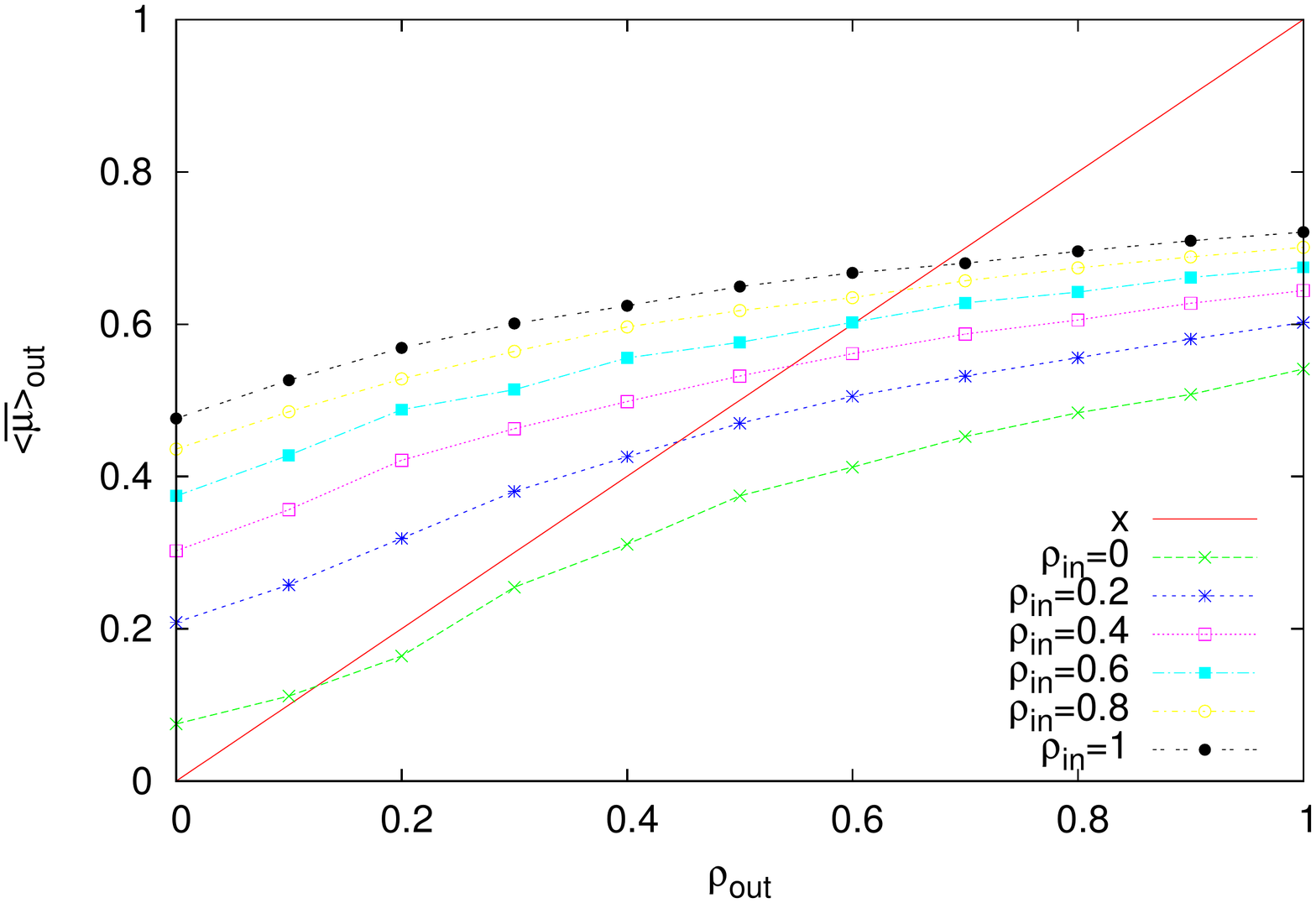}
  \caption{Plot of $\overline{\avg{\mu}}_{out}$ versus $\rho_{out}$ for $q=0.4$, $\lambda=1.5$ and various values of $\rho_{in}$.}
  \label{inRHOOUT}
\end{figure}

In principle for the outputs it is possible to define another parameter $\rho_{out}$ as:
\begin{gather}
\mess{\psi}{m}{a}_{\mu_m}=   (1-\rho_{out})\delta_{\mu_m,0} + \rho_{out}\delta_{\mu_m,1},
\end{gather}
to force the network to switch on a fraction of outputs. Nevertheless it is then required to check at convergence of the algorithm that this value is consistent with the value of $p(\mu_m)$. A simple way to check this is by measuring $\overline{\avg{\mu}_{out}}=\overline{\sum\limits_{i \in outputs} \mu_i/N_{out}}$ and checking if this value is consistent with the value of $\rho_{out}$ given as a parameter. In Figure \ref{inRHOOUT} this check is done for a particular case. In this Figure we see that there is only one value of $\rho_{out}$ consistent with $\overline{\avg{\mu}}_{out}$ (this result holds similarly for Soft-MB and Hard-MB, and for any $q$ and $\lambda$). We then verified that this solution is exactly the same as the one obtained by using equation (\ref{output_message}), hence showing that the parameter $\rho_{out}$ is not necessary to explore all the possible solutions of the outputs.

As a consequence of this setting on the inputs and outputs, the presence of the nutrients is determined by the parameter $\rho_{in}$ while the presence of the outputs is determined at convergence depending on the state of the network.

\section{Real network}
In the real network the relationship nutrient and metabolite with in degree $0$ is not always satisfied. Furthermore in the real network it is possible to have reactions with in degree $0$ (the uptakes) and with out degree $0$ (outtakes). The simplest way to adapt the program functioning on random network for this case is by considering that
\begin{itemize}
\item Metabolites with in degree $0$ do not have a metabolite constraint attached and that are not nutrients have $\mess{\psi}{a}{m}_1=\frac{1}{2}$.
\item Nutrients have $\mess{\psi}{a}{m}_1=\mess{\psi}{m}{a}_1=\rho_{in}$ either if they have in degree $0$ or not.
\item Reactions with in degree $0$ do not have a reaction constraint attached and have $\mess{\eta}{e}{i}=\frac{1}{2}$.
\item Reactions with out degree $0$ are not a problem as the equations derived in Section \ref{cavity_derivation} are still valid.
\end{itemize}

\chapter{Algorithms}
\label{chap:algorithms}
\section{Belief Propagation Algorithm}
\label{sec:BP_explanation}
In a nutshell, in BP it is considered that each variable sends a message to its neighbours. This message represents the belief that the variables has about the state of its neighbours. The outcome of this algorithm is the BP-marginal for variables, $\mu$ and $\nu$.

It is worth noting that while in the complete case many different messages exist between the variables (see Section \ref{cavity_derivation}), in PEI, that is in a Mean Field approximation, the messages are the same for all neighbours and correspond to $\avg{\mu_m}$ and $\avg{\nu_i}$. Nevertheless the functioning of the algorithm is similar in the two cases: first we generate a RRN with a given $q$ and $\lambda$, then we initialize the messages (to a random value or to the last value computed) and we iterate the equations until convergence. Finally for the complete problem (in PEI the BP-marginal is equal to the marginal) at convergence it is possible to recover the marginals using equations (\ref{proba_rel}) and (\ref{normalizations}). All networks used in this thesis have $M=10^4$ while $N=\lambda M / (1+q)$.

The simplest way to sample the solutions is by fixing one of the two free variables remained: $\theta$ or $\rho_{in}$. By changing $\rho_{in}$ we can see how the configuration of the solutions changes when the nutrients have a probability $\rho_{in}$ of functioning. Whereas by changing $\theta$ we can observe what happens if we constrain the system to switch on (or off) the reactions. Each behaviour is interesting to understand how the system is organized. In each case the mean over the metabolites,$\avg{\mu}$, and the reactions, $\avg{\nu}$ ($\avg{x}$ is the average over the measure $P(\mu,\nu)$, (\ref{meas})) has been computed.

\section{Population dynamics}
\label{sec:PopDyn_explanation}

BP equations presented in Section \ref{cavity_derivation} are meant for inferring the marginal probabilities on a specific graph. However, when one is interested in the behavior of typical samples of the RRNs with given parameters $q$ and $\lambda$, then the equations presented in Section \ref{cavity_derivation} can be solved using {\it population dynamics} \cite{Parisi_cavity}. The idea behind this approach is that, instead of computing the messages on a given graph, one considers the probabilities, $P(\psi)$ and $Q(\eta)$, of having a message $\psi$ or $\eta$ in the system. 
Self-consistency equations for these probabilities can be written as follows:
\begin{align}
P(\psi)=E_{\lambda,q}\left[\prod \int d\eta \;Q(\eta)\; d\psi'\; P(\psi') \delta(\psi-F(\eta,\psi'))\right], \\ \nonumber \\
Q(\eta)=E_{\lambda,q}\left[\prod \int d\eta' \;Q(\eta')\; d\psi\; P(\psi) \delta(\eta-G(\eta',\psi))\right],
\end{align}
where the product is over the neighbours and the functions $F(\eta,\psi)$ and $G(\eta,\psi)$ are given by the equations in Section \ref{cavity_derivation}.
These population dynamics equations can be solved iteratively and once the fixed point has been reached, averages over the RRN {\it ensemble} can be directly computed.

In the population dynamics algorithm, we start by initializing the system with a random {\it population} of messages and by fixing the parameters of the RRN, $q$ and $\lambda$. Then we iterate using the equations of Section \ref{cavity_derivation} where the neighbours are extracted at random, using the distributions (\ref{degree_M}) and (\ref{degree_R}). This is done until convergence of the mean of the messages in the system. At convergence we can compute the mean value of the metabolites, $\overline{\avg{\mu}}$, and reactions, $\overline{\avg{\nu}}$, with respect to the ensemble of RRNs and over the measure (\ref{meas}).

All the data shown in the present work have been obtained with a population with $M=10^4$ and $N$ derived from equation (\ref{rel_N_M_q_lam}). We have checked that doubling the population has no relevant effects on the results obtained.

\section{Decimation Procedure}
\label{sec:decimation_explanation}
The BP algorithm is an efficient way for obtaining the {\it probability} that a variable take a certain state. Nevertheless one is generally confronted with the problem of obtaining actual \textit{configurations} of variables that satisfy a CSP. In order to find it, we resorted to a decimation procedure already used in other cases \cite{RicciT_Semerjian}.

In decimation, first BP is run and then the BP marginal is used as the real marginal of the variable, thus setting the variable to $0$ or $1$ \textit{according to the marginal}. Hence during decimation, variables are set one at a time, starting from the most polarized (with BP-marginal near $0$ or $1$) then running BP to make sure that the constraints are satisfied and that no contradiction occurs. This procedure is then iterated until all variables are decimated or until some constraint is violated.

Using this procedure it is thus possible to obtain a Boolean configuration that is a solution of the CSP problem under study. It is important to note that while BP is an unbiased way of sampling the solution space (at least for problems on random graphs), the decimation process is highly dependent on the procedure used to decimate. Nevertheless, if the procedure converges, the configuration found will be a solution of the CSP. Furthermore assuming BP marginals are unbiased for a RRN it is possible to understand whether we are sampling fairly well the solution space with decimation.

In order to reproduce the behaviour presented in Section \ref{sec:pop_dyn}, the algorithm that we used to obtain the results presented in Figures \ref{VN_l1_q05}-~\ref{FBA_l3_q08} is an extension of the standard decimation procedure presented above. In our algorithm, for a given $\theta$, first a BP solution is found and stored, then the system is decimated $N_{dec}$ times each time starting from the same BP solution stored. Finally BP solution for the next $\theta$ is obtained by initializing the messages with the last stored BP solution. For each system under study we applied this procedure following the two protocols ($+\infty \rightarrow -\infty$ or $-\infty \rightarrow +\infty$) presented in Section \ref{sec:pop_dyn}. All the results in this article have been obtained with $N_{dec}=5$ for the complete problem and $N_{dec}=10$ for MF.

%%% Local Variables: 
%%% mode: latex
%%% TeX-master: "thesis"
%%% End: 

\backmatter
% bibliography 
\cleardoublepage 
\phantomsection 
\bibliographystyle{unsrt}
\bibliography{bib_complete.bib}

\end{document}